\documentclass[preprint]{aastex631}
\usepackage{longtable}
\usepackage{gensymb}
\usepackage{natbib}

\shorttitle{TTV search in the TrES-2 system in TESS era}
\shortauthors{Biswas et al.}

\begin{document}

\title{Probing the Possible Causes of the Transit Timing Variation for TrES-2b in TESS Era}

\author{Shraddha Biswas}
\affiliation{Indian Centre For Space Physics \\
466 Barakhola, Singabari road, Netai Nagar, Kolkata, West Bengal 700099}
\email{hiyabiswas12@gmail.com}

\author[0000-0002-8075-5598]{D. Bisht}
\affiliation{Indian Centre For Space Physics \\
466 Barakhola, Singabari road, Netai Nagar, Kolkata, West Bengal 700099}
\email{devendrabisht297@gmail.com}

\author[0000-0001-7359-3300]{Ing-Guey Jiang}
\affiliation{Department of Physics and Institute of Astronomy \\
National Tsing-Hua University, Hsinchu 30013, Taiwan}
\email{jiang@phys.nthu.edu.tw}

\author[0000-0001-8452-7667]{Devesh P. Sariya}
\affiliation{Department of Physics and Institute of Astronomy \\
National Tsing-Hua University, Hsinchu 30013, Taiwan}
\email{deveshpathsariya@gmail.com}

\author{Kaviya Parthasarathy}
\affiliation{Department of Physics and Institute of Astronomy \\
National Tsing-Hua University, Hsinchu 30013, Taiwan}
\email{kaviyasarathy1998@gmail.com}
%% Note that the \and command from previous versions of AASTeX is now
%% depreciated in this version as it is no longer necessary. AASTeX 
%% automatically takes care of all commas and "and"s between authors names.

%% AASTeX 6.31 has the new \collaboration and \nocollaboration commands to
%% provide the collaboration status of a group of authors. These commands 
%% can be used either before or after the list of corresponding authors. The
%% argument for \collaboration is the collaboration identifier. Authors are
%% encouraged to surround collaboration identifiers with ()s. The 
%% \nocollaboration command takes no argument and exists to indicate that
%% the nearby authors are not part of surrounding collaborations.

%% Mark off the abstract in the ``abstract'' environment. 
\begin{abstract}
Nowadays, transit timing variations (TTVs) are proving to be a very valuable tool in exoplanetary science to detect exoplanets by observing variations in transit times. To study the transit timing variation of the hot Jupiter, TrES-2b, we have combined 64 high-quality transit light curves from all seven sectors of  NASA’s Transiting Exoplanet Survey Satellite (TESS) along with 60 best-quality light curves from the ground-based facility Exoplanet Transit Database (ETD) and 106 mid-transit times from the previous works. From the precise transit timing analysis, we have observed a significant improvement in the orbital ephemerides, but we did not detect any short period TTVs that might result from an additional body. The inability to detect short-term TTVs further motivates us to investigate long-term TTVs, which might be caused by orbital decay, apsidal precession, Applegate mechanism, and ${R\phi}$mer effect and the orbital decay appeared to be a better explanation for the observed TTV with $ \bigtriangleup\mathrm{BIC} = 4.32$. The orbital period of the hot Jupiter TrES-2b appears to be shrinking at a rate of $\sim - 5.58 \pm 1.81$~ms~yr$^{-1}$. Assuming this decay is primarily caused by tidal dissipation within the host star, we have subsequently calculated the stellar tidal quality factor value to be $ \sim 9.9 \times {10}^{3}$, which is 2-3 orders of magnitude smaller than the theoretically predicted values for other hot-Jupiter systems and its low value indicates more efficient tidal dissipation within the host star. Additional precise photometric and radial velocity observations are required to pinpoint the cause of the change in the orbital period.
 
\end{abstract}

%% Keywords should appear after the \end{abstract} command. 
%% The AAS Journals now uses Unified Astronomy Thesaurus concepts:
%% https://astrothesaurus.org
%% You will be asked to selected these concepts during the submission process
%% but this old "keyword" functionality is maintained in case authors want
%% to include these concepts in their preprints.

%% From the front matter, we move on to the body of the paper.
%% Sections are demarcated by \section and \subsection, respectively.
%% Observe the use of the LaTeX \label
%% command after the \subsection to give a symbolic KEY to the
%% subsection for cross-referencing in a \ref command.
%% You can use LaTeX's \ref and \label commands to keep track of
%% cross-references to sections, equations, tables, and figures.
%% That way, if you change the order of any elements, LaTeX will
%% automatically renumber them.
%%
%% We recommend that authors also use the natbib \citep
%% and \citet commands to identify citations.  The citations are
%% tied to the reference list via symbolic KEYs. The KEY corresponds
%% to the KEY in the \bibitem in the reference list below. 

\section{Introduction} \label{sec:intro}

Since the first detection of the hot Jupiter, 51 Pegasi b \citep{1995Natur.378..355M} orbiting a solar-like main sequence star, the field of exoplanetary science has emerged as one of the most rapidly growing branches of astrophysics due to continuous exciting discoveries. 
Hot Jupiters are easier to detect and observe due to their short orbital periods ($P < 10\,~$days) and the close proximity to their host stars ($a < 0.1\,~$au). There are several techniques to detect these planets, namely the radial velocity method (\citealt{2005PThPS.158...24M,2005A&A...439..367M}), transit photometry method (\citealt{2003ApJ...597.1076K, 2005A&A...444L..15B, 2004A&A...426L..15P, 2005ApJ...626..523C, 2005ApJ...624..372K}), astrometry method (\citealt{Ford_2003, 2004ApJ...617.1323P, 2005tvnv.conf..455G}), planetary microlensing method (\citealt{2000ApJ...535..176A, 10.1046/j.1365-8711.2003.06720.x}) and so on. Till now, according to NASA Exoplanet Archive\footnote{https://exoplanetarchive.ipac.caltech.edu/}, a total of 5602 exoplanets have been discovered, among which 4168 have been discovered using the transit photometry method, and for this reason, the photometric observation using transit method is considered as one of the most powerful techniques of planet detection. 
The Long-term transit-monitoring observations of hot-Jupiter systems help to perform transit time analyses of individual systems and to refine the ephemeris information of transiting planetary systems (e.g., \citealt{2009ApJ...691.1145S, 2012MNRAS.427.2757M, 2013ApJ...770...36K, 2013A&A...551A.108M, 2017AJ....153...78C}). Furthermore, \citet{2022ApJS..258...40K} and \citet{2023ApJS..264...37S} have recently conducted homogeneous timing analyses with a large dataset of hot Jupiters. In addition to this, the photometric study also provides us a golden opportunity to find out any periodic variation in the planet’s orbit caused by the gravitational influence of an additional body through the temporal shifts (either delays or advancements), occurring at the midpoint of transit events across successive orbits, commonly referred to as Transit Timing Variations (TTVs; \citealt{2013AJ....145...68J, 2013A&A...551A.108M, 2018MNRAS.480..291S, 2021RAA....21...97S, 2022AJ....163...77A}). 
                                    
However, with long-term high-precision transit timing data, 
we can also explore the several possible causes of TTV, e.g. when the tightly bound hot Jupiters encounter a strong tidal
interaction (\citealt{2014ARA&A..52..171O, 2019psce.confE..38B}) in between the planet and the host star, it leads to a shrinkage in planet’s orbit, 
known as orbital decay through the tidal dissipation of energy
in their host stars and exchange of orbital angular momentum 
occurring from the planetary orbit to a star's spin 
(\citealt{1966Icar....5..375G, 1973ApJ...180..307C, 1996ApJ...470.1187R, 1999ssd..book.....M, 2003ApJ...596.1327S, 2009ApJ...698.1778R, 2009ApJ...692L...9L, 2010ApJ...725.1995M, 2019MNRAS.490.4230S, 2020ApJ...888L...5Y} and references therein). 
As a result, the exoplanets eventually collide with their host stars and from the orbital decay rate measurement, we can directly estimate the stellar tidal quality factor
$Q^{'}_{\ast}$
(\citealt{2014MNRAS.440.1470B, 2018AcA....68..371M, 2017AJ....154....4P,2020AJ....159..150P}). This orbital decay can lead to a shift in transit times when observed for more than a decade. Previously, orbital decay was studied in several cases; for example, for WASP-43 (\citealt{Gillon_2012}), \citet{2016AJ....151...17J} preferred an orbital decay scenario by estimating an orbital decay rate of $\dot{P_q}=-\,28.9~ \pm ~7.7$~ms~yr$^{-1}$, whereas \citet{2016AJ....151..137H} observed a period change of $ \dot{P_q}=-\,0.0~ \pm ~6.6$~ms~yr$^{-1}$, indicating consistency with a constant orbital period and consequently ruling out the aspect of orbital decay phenomenon. Currently, the most promising candidate to observe orbital decay through transit observations is WASP-12b, a planet around a main-sequence F star (\citealt{2009ApJ...693.1920H}) with a $ 1.09\,d$  orbital period. \citet{2016A&A...588L...6M} first reported an orbital decay for WASP-12b with a decay rate of $ -\,25.60~ \pm ~4.0$~ms~yr$^{-1}$ , which was later confirmed by \citet{2017AJ....154....4P}, but they also predicted the possibility of apsidal precession. Later, \citet{2020ApJ...888L...5Y} provided additional compelling evidence supporting the phenomenon of orbital decay instead of apsidal precession and $ R\phi $mer effect. \citet{2021AJ....161...72T} conducted the initial analysis of the TESS light curves pertaining to this system; subsequently, \citet{2022AJ....163..175W} augmented their findings with additional TESS data, unveiling distinct evidence of orbital decay in WASP-12b. More recently, \citet{Yeh_2024} validated the presence of orbital decay. On the other hand, in the case of WASP-4b, although a notable change in the transit times seems to have been observed (\citealt{2019AJ....157..217B, 2019MNRAS.490.4230S}), the interpretation remains ambiguous, possibly attributable to line-of-sight acceleration (\citealt{Baluev_2020, Bouma_2020,Turner_2022, Harre_2023}). 

In addition to the tidal orbital decay, there could be other possible causes of TTVs, like the apsidal precession (\citealt{2002ApJ...564.1019M, Heyl_2007, Jordán_2008, 2009ApJ...698.1778R, 2017AJ....154....4P}) due to the rotational and tidal bulges on the planet, the Applegate mechanism (\citealt{1992ApJ...385..621A, 2010MNRAS.405.2037W}),  the line-of-sight acceleration in hot-Jupiter systems (e.g., \citealt{2019AJ....157..217B,Bouma_2020, 2019MNRAS.490.4230S, 2020ApJ...888L...5Y, 2021AJ....161...72T,Turner_2022, 2022AJ....163..175W, 2024MNRAS.tmp..965Y}), starspots or star-spot crossing events(\citealt{2004MNRAS.351..110W, 2009A&A...508.1011R, Sanchis-Ojeda_2013, 2016AJ....151..150M}); the presence of additional bodies in close-in orbit to the hot Jupiters (see \citealt{2008ApJ...682..586M, 2009A&A...506..369B, Gibson_2009, 2009A&A...508.1011R} 
and references therein).

To investigate these potential sources of Transit Timing Variations (TTVs), observations conducted from ground-based telescopes by amateur astronomers were not sufficient. The multi-sector Transiting Exoplanet Survey Satelite (TESS; \citealt{2014SPIE.9143E..20R}), a space-based survey telescope, was primarily launched in 2018 by NASA to discover and characterize new exoplanets orbiting bright stars as well as to conduct follow-up observations of previously discovered hot Jupiters via full-sky observations. It provides unprecedented high-quality and high-precision photometric transit data and is well-equipped to reduce the noise components that arise due to the Earth’s atmospheric effects. It also provides a better opportunity to get longer uninterrupted time-series data compared to ground-based telescopes, and by combining the TESS data with previous works, we can obtain refined mid-transit timings with an extended observational baseline, enhancing our capability to investigate Transit Timing Variations (TTVs).

For our work, we have selected the most massive gas-giant exoplanet TrES-2b (M$_P$ = 1.198\, M$_J$, R$_P$ = 1.222\, R$_J$), discovered by the Trans-Atlantic Exoplanet Survey (TrES, \citealt{2006ApJ...651L..61O}), orbiting a G0V main-sequence dwarf GSC 03549-02811 (V = 11.41 Mag, M$_\ast$ = 0.98\ M$_{\sun}$, R$_{\ast}$= 1.003 \ R$_{\sun}$) in a close-in orbit ($a = 0.0367 $~au) with an orbital period of 2.47 \, days. Because of the following reasons, we have chosen this object for our transit timing analysis:- (1) The availability of transit timing data for more than a decade (2006-2024), (2) Availability of TESS data along with follow-up observations by previous workers, (3) Some contradictory findings of previous authors based on the presence of additional planets. Some early publications suggested the evidence of Short-term TTVs caused by a third body or an exo-moon in the system (\citealt{Mislis_2009, 2009A&A...508.1011R}), whereas the later studies neither found evidence for short-term TTVs nor for long-term TTVs for TrES-2b (\citealt{2009AN....330..475R, Kipping_2011, Schr_ter_2012, Raetz_2014}). For several years, there were no further TTV studies for this system, but recently, \citet{2022AJ....164..220H} found the existence of long-term TTVs in TrES-2b, which is consistent with the findings of \citet{2007ApJ...664.1185H}. So, these contradictory findings motivated us to explore the exoplanetary system TrES-2b. 

Along with TESS light curves, we have collected all available transit light curves from the ETD\footnote{http://var2.astro.cz/ETD/predictions.php} (\citealt {Poddan__2010}) and from the literature, to increase the transit timing observation baseline. In total, we have employed 230 transit light curves of TrES-2b in this work. ETD is a user-friendly online portal (established in 2008) by the Variable Star and Exoplanet Section of the Czech Astronomical Society. It has over 400 planets and 12,000 contributed observations spanning 15 years.

We have organized the remainder of the paper as follows. Section \ref{sec:observation} presents the data processing from TESS observations and details data handling from the ETD and the literature. Section \ref{sec:light_curve_analysis} describes the Transit light curve analysis procedure, and the transit timing analyses using three different timing models are described in Section \ref{sec:timing_analysis}. The results of this analysis are discussed in Section \ref{sec:results}, and the concluding remarks are given in Section \ref{sec:remarks}.

\section{Observational Data} \label{sec:observation}
\subsection{Extraction of Detrended Time-series Data from TESS Observations}

    During the TESS observing runs, 64 transits of TrES-2b were observed in total seven sectors 26, 40, 41, 54, 55, 74 and 75 with 120\,s cadence during the time interval 09-06-2020 to 30-01-2024 and the star, TrES-2 is designated as TIC 39986044, in the TESS input catalog (\citealt{2018AJ....156..102S}). For our work, we directly downloaded the TESS light curve files of TrES-2b from the Barbara A. Mikulski Archive for Space Telescopes\footnote{https://exo.mast.stsci.edu}, (MAST), a public data archive, using the JULIET package (\citealt{2019MNRAS.490.2262E}) and accessed into Presearch Data Conditioning Simple Aperture Photometry (PDCSAP; \citealt{2012PASP..124.1000S, 2012PASP..124..985S, 2014PASP..126..100S, Caldwell_2020}) light curves. These calibrated light curves are produced by the TESS Science Processing Operations Center (SPOC; \citealt{Caldwell_2020}) and by NASA Ames Research Center (\citealt{2016SPIE.9913E..3EJ}) and are optimized to remove instrumental systematic artifacts (see \citealt{2012PASP..124.1000S} and \citealt{2012PASP..124..985S, 2014PASP..126..100S}) and to characterize transiting planets using PDCSAP pipeline. Instead of simple aperture photometry (SAP) observations, we preferred the PDCSAP light curves here due to their lower scatter.

The library, JULIET\footnote{https://github.com/nespinoza/Juliet}, written in Python, is publicly available on GitHub. Since it discards the data points with nonzero quality flags, that’s why during extraction of the time series data (times, fluxes, and flux errors), we retained data points only with a quality flag zero for time series analysis (e.g., \citealt{2022AJ....163...79H}). We converted the format of the time of the observations to TESS BJD$_{\rm TDB}$ by adding 2,457,000\footnote{https://archive.stsci.edu/missions/tess/doc/EXP-TESS-ARC-ICD-TM-0014.pdf}.  Within Juliet, we have incorporated nested sampling schemes through the MultiNest algorithm (\citealt{2009MNRAS.398.1601F, Feroz_2019}) via the PyMultiNest\footnote{https://github.com/JohannesBuchner/PyMultiNest} package (\citealt{Buchner_2014}). JULIET conventionally accepts input data in the form of text files or arrays containing time stamps, flux values, uncertainties, and any additional parameters required for the analysis. We further used the JULIET package to remove the trends appearing in the time series data of TrES-2b by masking the in-transit points using the transit ephemeris of \citet{2019MNRAS.486.2290O}. Next, we fit a Gaussian Process  (GP) to our data by employing a simple (approximate) Mat\'ern kernel (\citealt{2019MNRAS.490.2262E}) implemented by CELERITE. To use this kernel within Juliet, we have to give the priors for some parameters. We have given a normal prior for the mean out-of-transit flux (mflux), fixed the dilution factor for the photometric noise (mdilution) to unity, and we have assumed wide log-uniform priors for all the other parameters, the jitter (in parts per million) added in quadrature to the errorbars of the instrument (${\bf {\sigma}_{\omega}}$), the amplitude of the GP (${\sigma}_{mGP}$), and the timescale of the Mat\'ern part of the GP ($\rho_{mGP}$).

\begin{figure}
\centering
  \begin{tabular}{@{}c@{}}
    \includegraphics[width=\columnwidth]{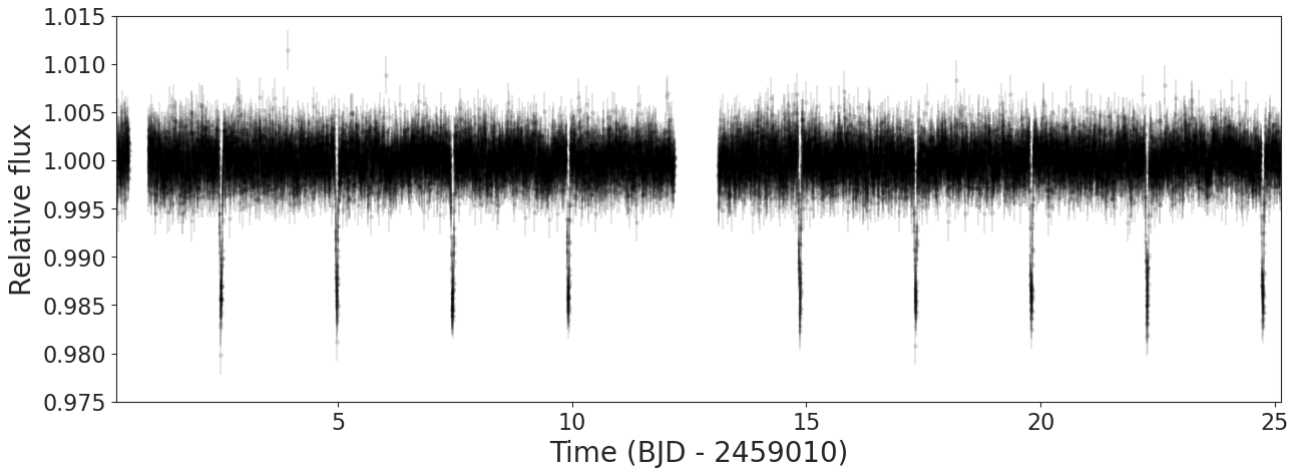}\\
    \includegraphics[width=\columnwidth]{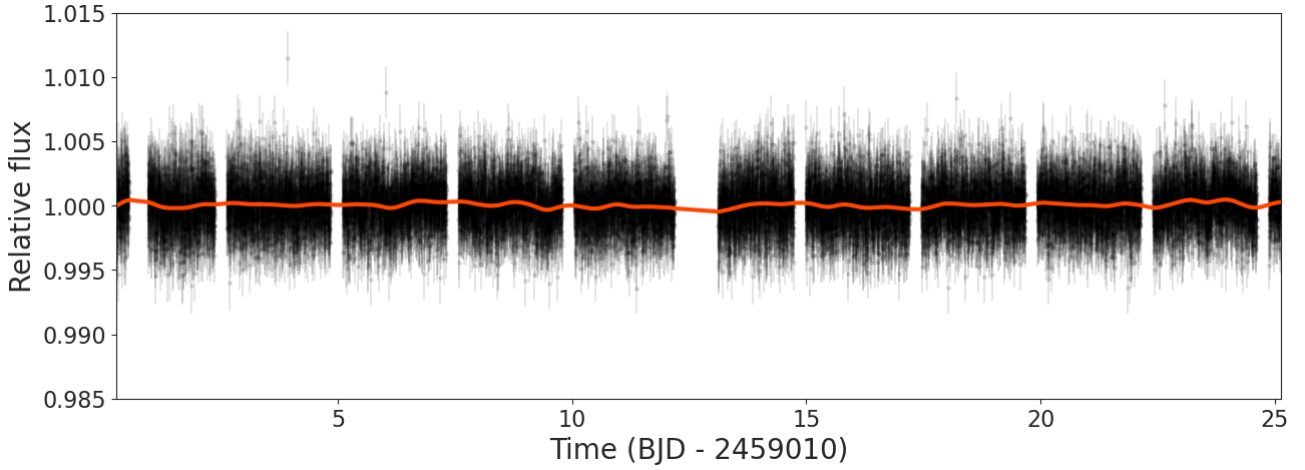}\\
    \includegraphics[width=\columnwidth]{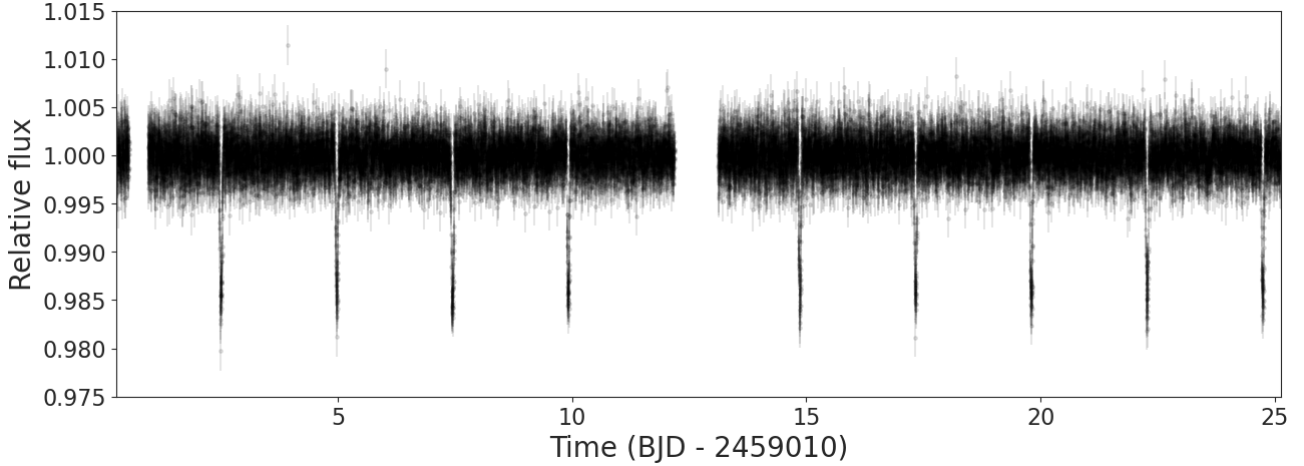} \\
  \end{tabular}
  \caption{Top panel: time series data of TrES-2b observed by TESS within sector 26. Middle panel: corresponding OOT time series data along with the best-fit GP model (red curve). Bottom panel: detrended time series data corresponding to the top panel.}
  \label{fig:ind_transits_sec26_1}
\end{figure}

In order to obtain the GP-fitted detrended normalized light curves corresponding to the time series data, we divided the flux and its corresponding error by the GP model of each sector. In Figure \ref{fig:ind_transits_sec26_1}, we have presented the trend of the time series data of TrES-2b (top panel), the OOT (out-of-transit) time-series data with the best-fitting GP model (middle panel), and the detrended time series data (bottom panel) of TrES-2 for sector 26. The same figures for other sectors are shown in Figures \ref{fig:ind_transits_sec40_2}--\ref{fig:ind_transits_sec40_7} (see the Appendix). To construct each transit time series, we extracted sections of the detrended normalized light curve within $\pm ~$0.1 day of the predicted mid-transit times.

\subsection{Observational Data from ETD and the Literature}
In addition to the TESS data, we have also included 60 transit light curves of TrES-2b with the best quality data (quality index, DQ $<3$) from the largest database ETD, observed during 2013 to 2023 by several observers at different observatories all over the globe. Now, before doing light curve analysis, to remove the atmospheric effects that arose due to interference from the Earth’s atmosphere, we normalized the OOT flux values of the light curves close to unity by fitting a linear function of time to the OOT part. By following the methodology adopted by \citet{2016AJ....151...17J}, we also employed a third-degree polynomial to model the airmass effect and a linear function to model the seeing effect. We can express the original light curve $F_{o}(t)$ as,
\begin{equation}
F_{o}(t) = F(t) \mathcal{P}(t) \mathcal{Q}(s),
\end{equation}
 
In the above equation, $F(t)$ represents the corrected light curve. The function P(t) is defined as a polynomial expression, specifically $\mathcal{P}(t) = a_0 + a_1 t + a_2 t^2+ a_3 t^3$. Additionally, $\mathcal{Q}(s) = 1 + c_0 s$, where '${\bf s}$' denotes the seeing associated with each individual image. The seeing conditions are unknown for all 60 light curves obtained from ETD. Consequently, the implementation of any seeing correction is not feasible. As a result, we uniformly set $\mathcal{Q}(s)=1$ across all light curves sourced from ETD. We conducted a numerical search to determine the best values for the four parameters: $a_0, a_1, a_2, a_3$. This optimization aimed to minimize the standard deviations and ensure that the out-of-transit portion of $F(t)$ closely approximates unity. Subsequently, this normalization process was applied to all the light curves.

After the normalization process, we converted all the time stamps of our mid-transit times from Julian Day (JD) or Heliocentric Julian Day (HJD) into Barycentric Julian Day (BJD) on the Barycentric Dynamical Time (TDB) timescale, i.e., TDB-based BJD) to account for the Earth’s movement, using an online converter tool provided by \citet{2010PASP..122..935E}\footnote{astroutils.astronomy.ohio-state.edu/time/hjd2bjd.html}. 

\begin{figure}
    \centering
    \includegraphics[width=1.1\linewidth]{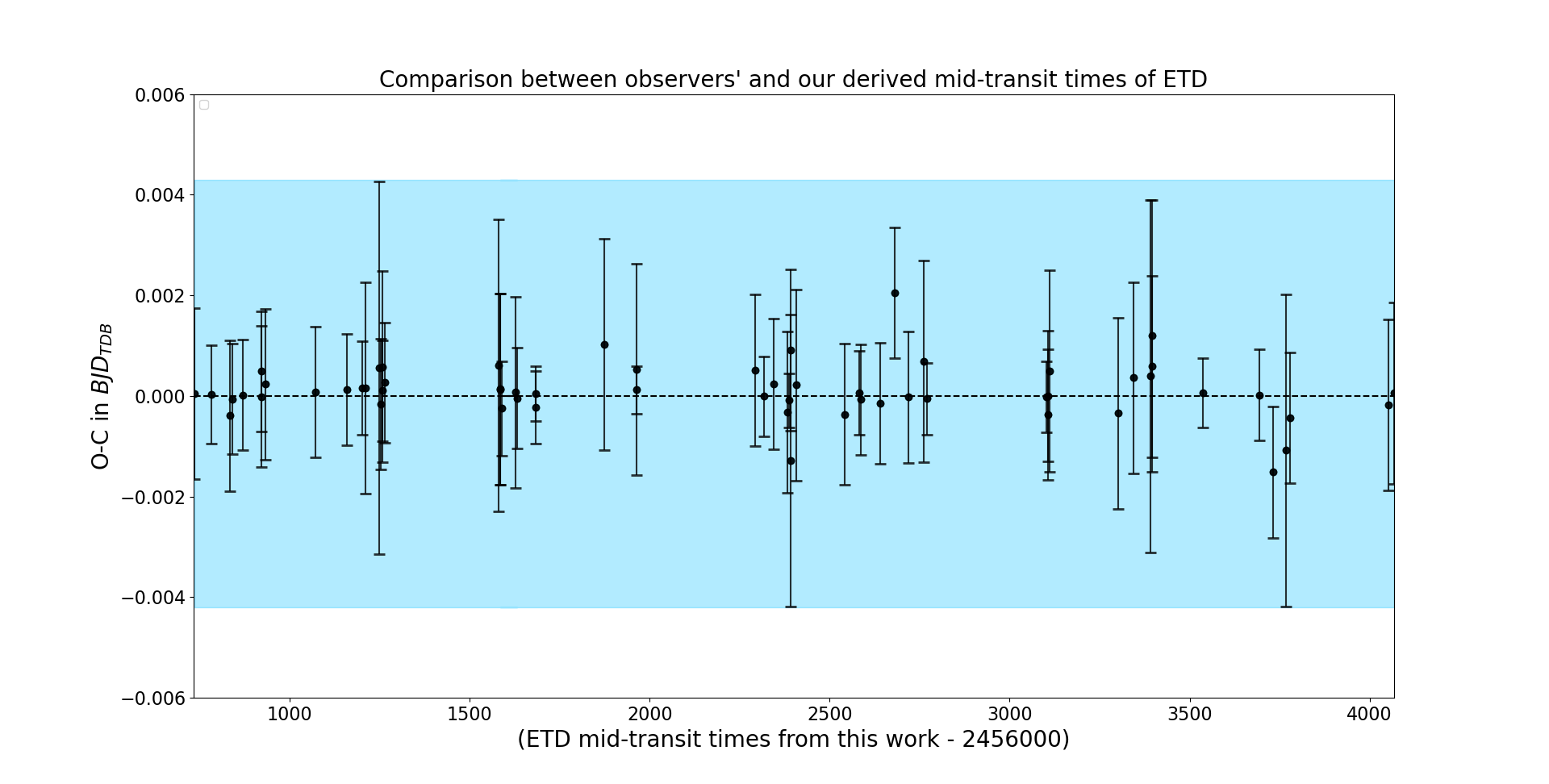}
    \caption{A comparative analysis of mid-transit times derived from both observer data and our calculations, where O-C represents the deviation of the observed mid-transit times from our derived mid-transit times. The sky-blue shaded region shows the maximum deviation range from the central line.}
    \label{fig:comparison_in_ETD_data}
\end{figure}

We conducted a graphical comparison (see Fig. \ref{fig:comparison_in_ETD_data}) between our calculated mid-transit times and those recorded by observers in ETD, focusing on the deviation between our derived values and the observer-reported times. In our analysis, we observed a strong agreement between the mid-transit times we derived from all 60 light curves of ETD and those reported by the observers.

In addition to these, we have also considered all the published light curves of TrES-2b available in the literature. The details of all 230 transit light curves taken from TESS, ETD, and the literature employed in this paper are given in Table \ref{tab:4}.

\begin{table}
\begin{center}
\caption{Details of all 230 transit light curves of TrES-2b considered in this work.}
\label{tab:4}
\begin{tabular}{lclc}
%\bottomrule
%\bottomrule
\hline
Object & Number of light  & Sources & Total number of \\
name &  curves taken & & light curves \\
\hline
 & 64 & TESS&\\
& 60 & ETD \citep{Poddan__2010} &\\
& 82 & \citet{Schr_ter_2012} &\\
& ~6 & \citet{refId0} &\\
TrES-2b & ~9 & \citet{2011ApJ...726...94C} &230\\
& ~4 & \citet{2019MNRAS.486.2290O} &\\
& ~3 & \citet{2007ApJ...664.1185H} &\\
& ~1 & \citet{Mislis_2009} &\\
& ~1 & \citet{2006ApJ...651L..61O} &\\

\hline
\end{tabular}
\end{center}
\end{table}

\section{\textbf{The Analysis of Transit Light Curves}} \label{sec:light_curve_analysis}

To determine the individual mid-transit times, ${T}_{mid}$ of all 124 transit light curves obtained from TESS and ETD and simultaneously to refine the stellar and planetary parameters of the TrES-2 system, we modeled all the light curves by utilizing the Transit Light Curve Analysis (TAP; \citealt{2012AdAst2012E..30G}), an interface-driven software package. We loaded all the light curves of TrES-2b individually into TAP to derive the transit parameters, such as the ratio of the planet to the star radius (${R_p /R_\ast}$), the mid-transit time (${T_m}$), the orbital inclination ({\it i}), the semi-major axis scaled in star radii ($a/R_\ast$), and the linear and quadratic limb-darkening (LD) coefficients ($u_1$, $u_2$). The TAP software employs the Markov Chain Monte Carlo (MCMC; \citealt{JSSv035i04, 2019A&A...623A.107C}) technique, driven by the Metropolis-Hastings algorithm and a Gibbs sampler, to fit every transit data with the model light curves of \citet{Mandel_2002}, derived from a star-planet system and to account for correlated noise while estimating the parameter uncertainties, they used the wavelet-based likelihood function developed by \citet{2009ApJ...704...51C}. To get the most reliable results for model parameters along with error estimation, we took a standard approach and computed five Markov Chain Monte Carlo (MCMC) chains with lengths of $10^5$ links for each transit light-curve analysis. The process for assessing the convergence of Markov Chain Monte Carlo (MCMC) chains utilized by the TAP involves employing Gelman–Rubin statistics (referred to as G-R statistics) to evaluate the likelihood that multiple chains have reached convergence within the same parameter space (\citealt{2006ApJ...642..505F, 2012AdAst2012E..30G}). This assessment entails computing G-R statistics (denoted as $\hat{R}(z)$) for each model parameter and estimating the effective number of independent samples (denoted as $\hat{T}(z)$). The chains are iteratively extended until both $\hat{R}(z)$ and $\hat{T}(z)$ meet specific criteria, $ \hat{R}(z)$ $\leqslant 1.01$ and $\hat{T}(z)$ $\leqslant 1000$ (see \citealt{2006ApJ...642..505F}). Upon fulfilment of these convergence criteria, the MCMC chains are deemed to have achieved convergence and are regarded as sufficiently mixed (\citealt{2006ApJ...642..505F}).

To start the Transit Light Curve analysis, we followed the same approach as adopted by \citet{2013AJ....145...68J}, except incorporating a Gaussian prior on the orbital period. First, during the MCMC calculations in TAP, we set the ratio of the planet to star radius (${R_p /R_\ast}$) and the mid-transit time (${T_m}$) as completely free parameters. As suggested by \citet{2009ApJ...704...51C}, \citet{2010ApJ...718..920C} and \citet{2010ApJ...711..374F}, we adopted the eccentricity of orbit (e) and longitude of periastron ($\omega$) values to be set to zero (a circular orbit with e =  0), the remaining parameters, such as the orbital period (P), scaled semi-major axis ($a/R_\ast$), orbital inclination (i) were fitted under Gaussian penalties and they are allowed to vary only between the $\pm 1\sigma$ values reported in the literature. For all 60 transit light curves obtained from ETD, the linear ($u_1$) and quadratic ($u_2$) limb-darkening coefficients were also fitted using the above mentioned method. The initial values of the LD coefficients used for TrES-2 for different filters are listed in Table \ref{tb:LD_coefficients}, and the initial parameter setting is shown in Table \ref{tb:initial_parameter}. We calculated the initial values of the linear and quadratic LD coefficients ($u_1$, $u_2$) for the TESS light curves of TrES-2b by interpolating the coefficients effective temperature (${T}_{\rm eff}$), surface gravity ($log_{g} $), metallicity ([Fe/H]) and microturbulence velocity ($V_t$ )from the tables of \citet{2017A&A...600A..30C}. 

The initial values of the parameters $a/R_\ast$, i, and ${R_p /R_\ast}$ were taken from \citet{Kipping_2010}, the initial orbital period value was taken from \citet{2006ApJ...651L..61O}(see Table \ref{tb:initial_parameter}), and the initial value of ${T_m}$ was set automatically for TrES-2b. 

For the ETD transit light curves observed in the clear, V, I, and R filters, following \citet{Su_2021}, we linearly interpolated the values of $u_1$ and $u_2$ from the tables of \citet{2011A&A...529A..75C}, using the EXOFAST \footnote{https://astroutils.astronomy.osu.edu/exofast/limbdark.shtml} package (\citealt{2013PASP..125...83E}) with the stellar parameters of effective temperature (${T}_{\rm eff}=5850\,$ K), surface gravity ($log_{g} = 4.55$ cm $s^{-2}$), metallicity ([Fe/H] $= -\,0.15$) and microturbulence velocity ($V_t$ = 2.0 km $s^{-1}$ ) (values taken from \citealt{Sozzetti_2007}). The LD coefficients for the light curves of TrES-2b observed through a clear filter were calculated by taking the average of their values in the {\it V} and {\it R} filters (see \citealt{2013IBVS.6082....1M}). The 50.0 percentile, i.e., median level, is considered as the best-fit value, as well as the 15.9 and 84.1 percentile levels (i.e., 68\% credible intervals) of the marginalized posterior probability distribution for each model parameter are considered as its lower and upper $1\sigma$ uncertainties or the error bars, respectively. The best-fit values of the model parameters ${P}$,  ${T_m}$, i, $a/R_\ast$, ${R_p /R_\ast}$, $u_1$ and $u_2$ along with their $1\sigma$ uncertainties derived from 64 TESS Transit Light Curves of TrES-2b, are given in Table \ref{tb:lighcurve_model_TESS1}, whereas, the same best-fit model parameter values of all 60 Transit light curves of ETD can be found in Table \ref{tb:lighcurve_model_TESS}. Furthermore, the graphical representations of all modelled light curves from TESS and ETD are shown in Figures \ref{fig:6}--\ref{fig:8} and in Figures \ref{fig:ETD_light_curves1}-\ref{fig:ETD_light_curves3}, respectively. A more detailed description of TAP can be found in the study by \citet{Fulton_2011} and \citet{2012AdAst2012E..30G}.

\begin{table*}
\begin{center}
\caption {The theoretical limb-darkening coefficients}
\label{tab:3}
\begin{tabular}{clcc}
%\bottomrule light curves
%\bottomrule
\hline
Object Name & Filter & $u_1$ & $u_2$\\
\hline
{TrES-2} & {\it V} $^{a}$ & 0.4378 & 0.2933\\
& {\it R} $^{a}$ & 0.3404 & 0.3190\\
& {\it I} $^{a}$ & 0.2576 & 0.3186\\
& clear $^{b}$ & 0.3891 & 0.3062\\
& TESS$^{c}$ & 0.5283 & 0.3621\\
\hline

\end{tabular}\\
\label{tb:LD_coefficients}
\end{center}
{Notes:
$^a$ Calculated using {\sc exofast} with ${T}_{\rm eff}=5850$~K, $\log{g}=4.40$ and [Fe/H] $=0.2$.\\
$^b$ Calculated as the average of their value in the {\it V} and {\it R} filters.\\
$^c$ $u_1$ and $u_2$ taken from the tables of \citet{2017A&A...600A..30C}.}\\
\end{table*}

\begin{table}
\begin{center}
\caption{The Initial Parameter Setting}
\label{tab:2}
\begin{tabular}{ccc}
%\bottomrule
%\bottomrule
\hline
\hline
Parameter & Initial Value & During MCMC Chain \\
\hline
\hline
P (days) & 2.47063 & A Gaussian Prior with $\sigma$ = 0.00001 \\
i (deg) & 83.71 & A Gaussian Prior with $\sigma$ = 0.42\\
$a/R_\ast$ & 7.969 & A Gaussian Prior with $\sigma$ = 0.055\\
${R_p /R_\ast}$ & 0.1278 & Free\\
${T_m}$ & Set by eye & Free\\
$u_1$ & According to filter &   A Gaussian Prior with $\sigma$ = 0.21\\
$u_2$ & According to filter &   A Gaussian Prior with $\sigma$ = 0.23\\
\hline
\end{tabular}
\label{tb:initial_parameter}
\end{center}

\end{table}

\begin{center}
\small\addtolength{\tabcolsep}{2pt}
\begin{longtable*}{cp{2.8cm}cp{1.8cm}cp{1.8cm}cp{1.8cm}cp{1.5cm}cp{1.5cm}cp{1cm}c}
\caption{The Best-ﬁt Values of Parameters $P$, T$_{m}$, i, a/R$_\ast$, R$_p$/R$_\ast$, $u_1$, and $u_2$ for 64 TESS Transit Light Curves of TrES-2b using \texttt{TAP} \label{tb:lighcurve_model_TESS1}}\\
\hline
\hline \multicolumn{1}{c}{Epoch} & \multicolumn{1}{c}{$P$}&\multicolumn{1}{c}{${T}_{m}$} & \multicolumn{1}{c}{$\it i$} & \multicolumn{1}{l}{$a/{R}_{\ast}$} & \multicolumn{1}{c}{R$_p$/R$_\ast$} & \multicolumn{1}{c}{u$_1$} & \multicolumn{1}{c}{u$_2$}\\
(E) & \ \ \ \ \ \ \ \ \ (days) & (BJD$_{\rm TDB}$) & \ \ \ \ (deg) &  &  &   \\
\hline
\endfirsthead
		2046 & $2.4706301^{+0.0000099}_{-0.000010}$&$2459012.51027^{+0.00083}_{-0.00082}$ & $83.91^{+0.15}_{-0.14}$ & $7.801^{+0.055}_{-0.054}$ & $0.1262^{+0.0044}_{-0.0043}$ &
$0.43^{+0.18}_{-0.18}$ & $0.22^{+0.19}_{-0.20}$\\
		2047 & $2.470630^{+0.000010}_{-0.000010}$&$2459014.98072^{+0.00092}_{-0.00089}$ & $83.87^{+0.16}_{-0.16}$ & $7.812^{+0.054}_{-0.054}$ & $0.1282^{+0.0054}_{-0.0048}$ &
$0.50^{+0.17}_{-0.18}$ & $0.25^{+0.18}_{-0.19}$\\
...&    ...&...&...&...&...&...&...&\\
\hline
\end{longtable*}
Note. This table is available in its entirety in machine-readable form. A portion is shown here for guidance regarding its form and content.
\end{center}

\section{\textbf{Transit Timing Analysis}}    \label{sec:timing_analysis}
For the precise Transit Timing Analysis of the hot Jupiter TrES-2b, we fitted three different timing models (\citealt{2017AJ....154....4P, 2020ApJ...888L...5Y, 2021AJ....161...72T, Turner_2022}) to our observational transit timing data by using the Markov Chain Monte Carlo (MCMC) algorithms in emcee package (\citealt{2013PASP..125..306F}). For these three models, we consider the cases of (1) The ﬁrst model is the linear model, which assumes a circular orbit with a constant period; (2) The second one, the orbital decay model, also assumes a circular orbit but here the orbital period changes at a steady rate, dP/dE and (3) The third model is the apsidal precession model, which assumes the orbit of the planet is slightly eccentric, with the argument of periastron, $\omega$ uniformly precessing.

\subsection{The Linear Ephemeris}       \label{sec:linear}

After assembling all mid-transit times from TESS, ETD and previously published studies (106 mid-transit times directly taken from \citealt{2022AJ....164..220H}), to study the period variation and to estimate new linear ephemerides for the orbital period P and mid-transit time $T_0$  of the hot Jupiter TrES-2b, covering a baseline of 2594 transit epochs, we ﬁtted a linear model, 
\begin{equation}
T_{c} (E) = T_0 + EP,
\end{equation}
to the precise 230 mid-transit times, $T_m$ as a function of epoch E, where $T_c (E)$, E, P, and $T_0$ are the calculated mid-transit time, epoch or sequence of orbits (an integer) counted from the reference epoch, orbital period, and the mid-transit time at the reference epoch ($E = 0$)\footnote{The first transit of TrES-2b observed by \citet{O’Donovan_2006} was considered as $E=0$}.

To refine the best-fit parameter values derived from linear model fitting in parameter space, we assume a Gaussian likelihood by imposing uniform priors on P and $T_0$ and sampled the posterior probability distribution, following the algorithm proposed by \citet{2010CAMCS...5...65G}. To evaluate the convergence and sampling efficacy of the Markov Chain Monte Carlo (MCMC) chains, we computed the mean acceptance fraction ($a_f$), the integrated autocorrelation time ($\tau$), and the effective number of independent samples ($N_{\rm eff}$). These metrics signify robust convergence and effective sampling of the MCMC chains, evidenced by the observed $a_f$ falling within the optimal range of 0.2–0.5, and $N_{\rm eff}$ surpassing the minimum threshold of 50 per walker, as prescribed in our MCMC methodology (refer to \citealt{2013PASP..125..306F, 2016MNRAS.462.4018C, 2017ApJ...848....9S}). We estimated the mean acceptance fraction ($a_f$) value of $\sim 0.44$, which lies within the ideal range of 0.2-0.5, ensuring good sampling of the MCMC chains. The models may show unstable behavior due to non-convergence of the MCMC chain (\citealt{2022AJ....164..220H}); that’s why we computed the integrated autocorrelation time ($\tau$) to check the convergence of the MCMC chain (\citealt{2010CAMCS...5...65G, 2013PASP..125..306F}) and found its value to be $\sim 19$ steps. We also determined the effective number of independent samples to be $\sim 1052$, which was found to exceed the minimum threshold value of 50 per walker established in the MCMC analysis, as advised by the emcee group.

Using the procedure mentioned above, we derived the value of orbital period, $P = (2.470 613 55 \pm 0.000000041)$~days, which is shorter by $\sim 1.42$\,s than the discovery paper (\citealt{2006ApJ...651L..61O}). As the uncertainty value in the orbital period is inversely proportional to the number of epochs over time (\citealt{2019AJ....157..242E}), a longer observational time baseline of 18 years (2006\,-\,2024) will result in a highly precise orbital period, and we noted a significant improvement in the uncertainty value of orbital period ($\pm $ 0.000000048\,days) than in the discovery paper (± 0.00001\,days). The resulting transit time, $T_0$ = 2453957.635381  $\pm  0.0000195$ BJD$_{\rm TDB}$ is $\sim  4.5\sigma$ times more precise than the previously published ephemeris (\citealt{2019MNRAS.486.2290O}) under the constant period assumption.

These results are produced by taking 100 MCMC walkers, each walker running 300 steps and discarding the initial 37 steps as a final burn-in to avoid the strongly correlated parameters (\citealt{2016A&A...595L...5A}). The derived values of the linear ephemerides are fully consistent with the results available in the literature.  The median value and the 68\% credible intervals of the posterior probability distribution are considered as the best-fit value and its lower and upper $1 \sigma$ uncertainties, respectively.  

\begin{figure*}
\centering

\begin{tabular}{@{}c@{}}
    \includegraphics[width=1.2\textwidth]{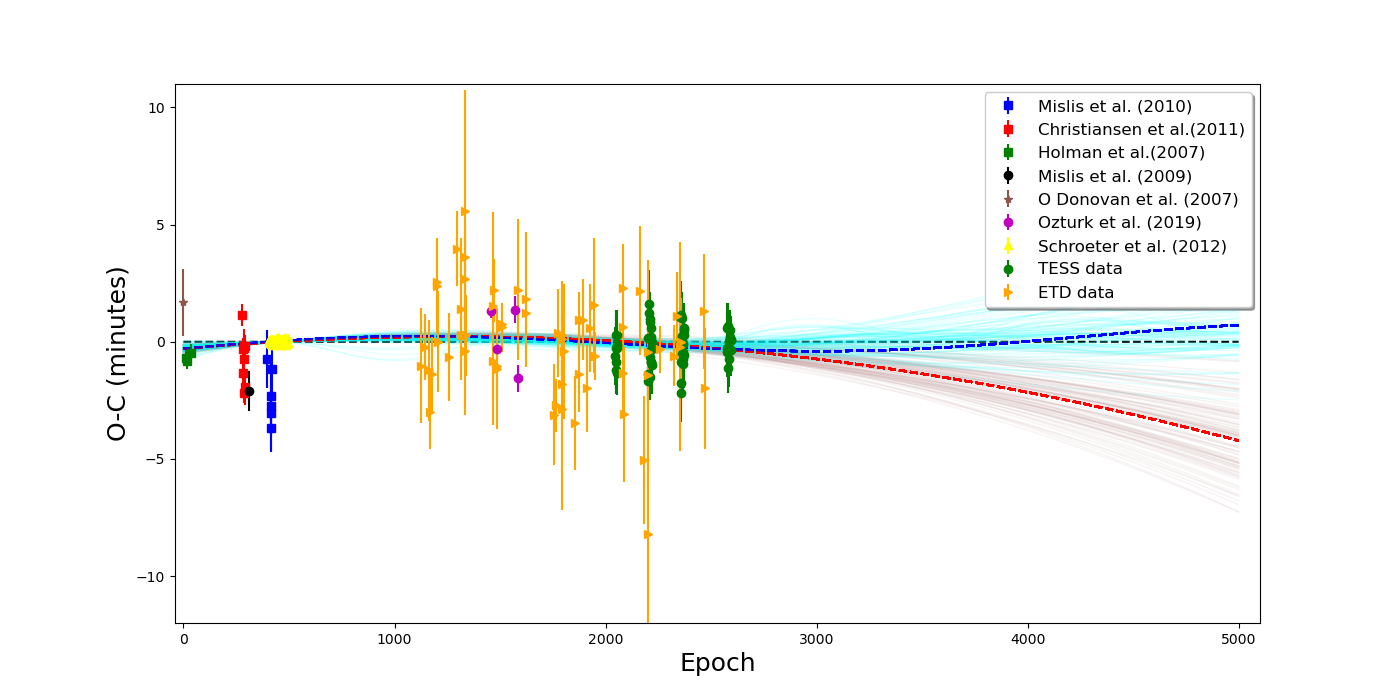}\\

\end{tabular}

  \caption{$TTV$ diagram for analysing 230 mid-transit times of TrES-2b. The dashed black line and red and blue curves represent the linear, orbital decay and apsidal precession models. The lines are drawn for 100 randomly chosen sets of parameters from the Markov chains of posteriors of the orbital decay (brown) and apsidal precession (cyan) models. The models are extrapolated for the next $\sim 17$ years to illustrate the broad spectrum of possible solutions.}
  \label{fig:10}
\end{figure*}

To search for TTVs in the TrES-2b system, we found out the value of timing residuals (O-C, pronounced as “O minus C”; \citealt{2005ASPC..335....3S} ) by subtracting the calculated (based on a fixed ephemeris) mid-transit time, $T_{c}$ from the observed mid-transit time, $T_m$, for each observed epoch E, considered for timing analysis. If there is no sign of TTVs, then we expect the derived values of O-C to be consistent with zero. But, here, we observed significant timing deviations on both sides of zero after linear model fitting, which may be due to some timing anomalies caused by additional planets or moons. We prepared the TTV diagram to analyze the possible shift in transit times (see Fig.  \ref{fig:10}) by plotting the timing residuals, i.e., O-C values concerning the observation epoch. Our estimated timing residuals (O-C) along with their corresponding epochs and original mid-transit times ($T_m$) are shown in Table \ref {tab:1}. 

In Table \ref{tb:timing_models}, we represent the best-fit values of all parameters, their lower and upper $1\sigma$  uncertainties adopted from the median value, and the 68\% credible intervals of the posterior probability distribution during linear model fitting for this transiting planet.

\begin{center}
\small\addtolength{\tabcolsep}{-3pt}
\begin{longtable*}{cp{3.8cm}cp{4.0cm}cp{3.5cm}lll}
\caption{Mid-transit Times $({T}_{m})$ and Timing Residuals (O-C) for 230 Transit Light Curves of TrES-2b} 
\label{tab:1}\\
\hline
\hline 
Transit Number & \ \ ${T}_{m}$  (BJD$_{\rm TDB}$) & $O-C$ (days) & \ \ \ \ Transit Source & Timing Source\\
\hline
\endfirsthead
0 & 2453957.63655000 & 0.0011691 & \citet{2006ApJ...651L..61O} & \citet{2022AJ....164..220H}\\
 13 & 2453989.75288000 & -0.0004771 & \citet{2007ApJ...664.1185H}& \citet{2022AJ....164..220H} \\
 ...&...&...&...&...&\\
\hline
\end{longtable*}
Note. This table is available in its entirety in machine-readable form.\\
References. \citet{2006ApJ...651L..61O, 2011ApJ...726...94C, Mislis_2009, refId0, 2022AJ....164..220H, Schr_ter_2012, 2019MNRAS.486.2290O, 2007ApJ...664.1185H}.
\end{center}

\subsection{The Search for Periodic Additional Planets through Frequency Analysis} \label{sec:additional}
\begin{figure}
    \centering
    \includegraphics[width=1.1\linewidth]{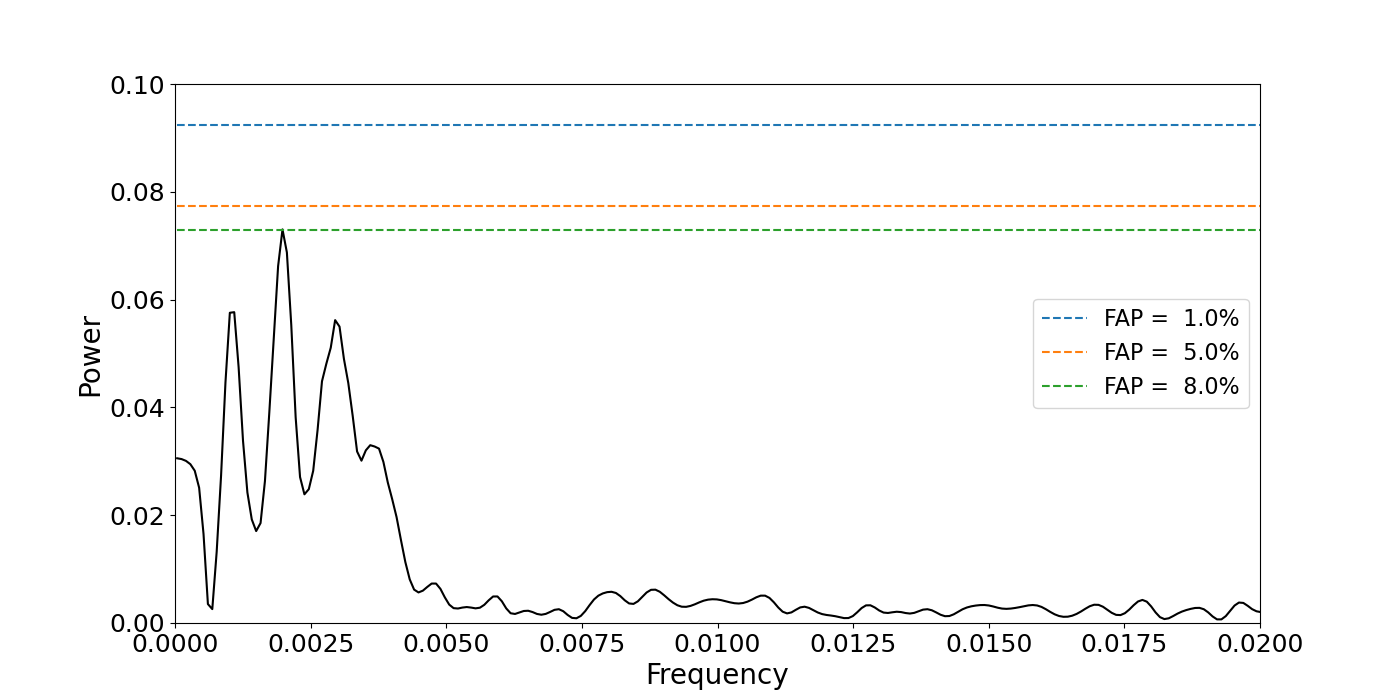}
    \caption{Generalized Lomb-Scargle periodogram computed for the timing residuals of TrES-2b.}
    \label{fig:periodogram}
\end{figure}

Since the discovery of TrES-2b (\citealt{2006ApJ...651L..61O}), there have been several debates regarding the presence of short-term TTV in this system. To search for a short-term TTV, we need to probe periodicity in the timing residuals (O-C) of all transit times of TrES-2b by computing a generalized Lomb–Scargle periodogram (GLS; \citealt{2009A&A...496..577Z}) on a large number of observational datasets in the frequency domain and we learn that the timing residuals help us to determine whether there is any presence of strict orbital periodicity in our system caused by a potential additional planet (not necessarily transiting) or an exo-moon, in an orbit close to the hot Jupiter, TrES-2b. To accomplish this task, we begin by computing the generalized Lomb-Scargle periodogram. Subsequently, we determine the index of the highest peak within the periodogram, indicating the frequency exhibiting the greatest power, which signifies the presence of the most significant signal. Following this, we employ PyAstronomy (\citealt{2019A&A...623A.107C}) to compute the False Alarm Probability (FAP) associated with the identified highest powered peak and the value of FAP for the highest peak of power 0.071969 at a frequency of 0.0015625 cycle/epoch, was found to be 8\%, which lies below the threshold level of FAP = 5\% and 1\% (\citealt{2021A&A...656A..88M, 10.1093/mnras/stad248}). That means there is no signature of statistically significant periodicity in the timing residuals, and due to the lack of periodicity,  we discarded the existence of short-term TTVs in the system. This finding is fully consistent with the previous conclusions available in the literature (e.g., \citealt{Kipping_2011, Schr_ter_2012, Raetz_2014}). The resulting periodogram with spectral power as a function of frequency is shown in Fig. \ref{fig:periodogram}.   The failure of detection for short-term TTVs next motivates us to search for long-term TTVs (which need a long timescale for more than a decade) to model O-C residuals, which may be caused by other mechanisms like orbital decay and apsidal precession.

\subsection{A Search for Orbital Decay Phenomenon} \label{sec:decay}
While no compelling evidence of short-term transit timing variations (TTV) attributable to an adjacent planet in a close-in orbit around the exoplanet TrES-2b was detected, the potential for transit timing variations (TTV) resulting from orbital decay can also be a viable explanation. We know tidal interactions between short-period exoplanets and their host stars over a longer time baseline, spanning more than a decade, usually result in the engulfment of planets by their host stars and shortening of orbital period leading to orbital decay. Thus, to explore whether the observed shift in transit times in TrES-2b is produced by orbital decay or tidal inspiralling (\citealt{1973ApJ...180..307C, 2009ApJ...692L...9L, 2010ApJ...725.1995M, 2018AJ....155..165P}), we followed \citet{2017AJ....154....4P} and fitted the following quadratic orbital decay model (Equation \ref{lab:equation}) to the timing data of the hot Jupiter, which includes a dimensionless additional curvature term dP/dE:
\begin{equation}
T_{q} (E) = {T}_{0} + P E + \ \frac{1}{2} \ \frac{dP}{dE} \ {E}^2,
\label{lab:equation}
\end{equation}

here, the three free parameters are the ${T}_{0}$ (the mid-transit time at E = 0), P (the orbital period), and $\frac{dP}{dE}$ (the change in the orbital period between succeeding transits. ${T}_{q} (E)$  is the calculated mid-transit time, and we assume that the orbital period P decreases at a constant rate.  
For the orbital model fitting, we adopted the same procedure as employed in the linear model fitting (see Section \ref{sec:linear}), except for running 32,000 steps per walker with the MCMC sampling of the posterior distribution in parameter space, for determining the best-fit ephemeris for the three free parameters of the orbital decay model (i.e., P, ${T}_{0}$, $\frac{dP}{dE}$)  and their corresponding uncertainties. We discarded the initial 60 steps (i.e., nearly two times the estimated value of integrated autocorrelation time) as a ﬁnal burn-in from the 32,000 steps per walker of MCMC (see \citealt{2016A&A...595L...5A, Hogg_2018}) to avoid the strongly correlated parameters. The prior corresponding to the changing orbital period in the decay model was allowed to vary between positive and negative numbers. 
Table \ref{tb:timing_models} gives the results of best-fitting orbital decay ephemerides. We find that our result for the quadratic term in the orbital ephemeris of TrES-2b (see Table \ref{tb:timing_models}) agrees well with the value of $ \frac{dP}{dE}$ = -(9.9 $\pm 1.9$) × $ 10^{-10}$ days/epoch from \citet{2022AJ....164..220H} to a $2.9 \sigma$ significance level and the negative value of this curvature term can be attributed to tidally induced orbital decay. 
Now, by substituting the derived values of P and $\frac{dP}{dE}$ for TrES-2b in the below equation (equation (4) of \citealt{2017AJ....154....4P}),
      
\begin{equation}
\ \frac{dP}{dt} = \ \frac{1}{P} \ \frac{dP}{dE},
\end{equation}
we converted the value of the quadratic coefficient of the best-fitting parabola, $\frac{dP}{dE}$ into the period derivative, $\dot {P}$  to be $\sim - 5.58 \pm 1.81$~ms~yr$^{-1}$, which displays that the orbital period of TrES-2b appears to be shrinking based on the full set of available transit-timing measurements spanning 18 yr and this $\dot {P}$ value is almost six times that for WASP-12 b. Some recent analyses of the TESS data for TrES-2b have been conducted; two notably relevant studies include \citet{2022ApJS..259...62I} and \citet{2024arXiv240407339A}. \citet{2024arXiv240407339A} determined the period derivative to be $\sim - 5.9 \pm 4.1$~ms~yr$^{-1}$, which is consistent with our findings. \citet{2024arXiv240407339A} conducted a series of omit-one tests where each midtime was sequentially excluded, and the resulting Bayesian Information Criterion ($\bigtriangleup\mathrm{BIC}$) was recalculated. They found that inaccurate mid-transit times, especially those with very small error bars that strongly constrain a specific model can substantially influence $\bigtriangleup\mathrm{BIC}$ values. \citet{2024arXiv240407339A} also computed a rescaled $\bigtriangleup\mathrm{BIC}$ to account for these influential data points. They noted that certain transit midtimes display unrealistically small error bars, the increased uncertainty in the measured rate of period change compared to our study implies that these discrepancies could originate from underestimated uncertainties related to transit time measurements. Our derived period derivative ($ \sim - 5.58 \pm 1.81$~ms~yr$^{-1}$) lies within $ 3.9\sigma$ confidence interval of the reported rate of period decrease by \citet{2022AJ....164..220H} ($ - 12.6 \pm 2.4$~ms~yr$^{-1}$). This discrepancy in $ \dot {P}$ value between our work and \citet{2022AJ....164..220H} may arise due to the inclusion of data both from the TESS and existing literature in this present work, but still preference is given to the orbital decay scenario. 

Furthermore, our calculation of the period derivative using exclusively the ETD data yielded a value of approximately $\dot {P}$ $ \sim - 10.4 \pm 2.1$~ms~yr$^{-1}$, which falls within the $1\sigma$ confidence interval of the rate of period decrease reported by \citet{2022AJ....164..220H} ($- 12.6 \pm 2.4$~ms~yr$^{-1}$). Further investigation and analysis are likely necessary to reconcile these differences and better understand the system's behavior.

Similar to the linear model, here we also calculated the timing residuals, i.e., O-C values by subtracting the mid-transit times calculated using the linear ephemeris $T_c (E)$ from those derived after fitting orbital decay model, ${T}_{q} (E)$  to analyze the possible shift in transit times and the TTV diagram (O-C diagram), we represented these timing residuals as a function of epoch with the red dashed curve showing the quadratic ephemeris model (see Fig. \ref{fig:10}). Here, the brown solid lines, consisting of 100 random draws from the Markov chains, are extrapolated for the next $\sim$ 17~yr  to illustrate the future trend of the quadratic trend scenario. As in the O-C diagram, we can see the orbital decay model shows a significant deviation from the linear ephemeris, we can further enhance the indication of apparent shortening of the orbital period, leading to a potential characteristic decay timescale of $T_d$ = {P}/{$\dot {P}$} = 37.59 Myr, longer than the value derived by \citet{2022AJ....164..220H} of 16.94 Myr, due to gradually decreasing decay rate. This decay time is approximately 0.74 \% of the derived age for the host star, TrES-2.  
\subsection{Apsidal Precession Study for TrES-2 System}      \label{sec:apsidal}
From some previous theoretical works (see \citealt{2016A&A...588L...6M, 2017AJ....154....4P, 2019AJ....157..217B, 2020ApJ...888L...5Y, 2022AJ....163...77A}), we came to know the apsidal precession phenomenon may also be a plausible reason for observing long-term TTV in hot-Jupiter systems. Apsidal precession is a gradual rotation of the apsidal line (the line connecting the aphelion and perihelion) of a planet within the orbital plane (\citealt{1995Ap&SS.226...99G}) due to various gravitational perturbations caused by other celestial bodies present in the orbital system or due to relativistic effect (General Relativity, GR). According to \citet{2009ApJ...698.1778R}, the hot Jupiter, TrES-2b, is already a potential candidate for studying apsidal precession with a slightly eccentric orbit, e = 0.021. That’s why, to probe the possibility of this phenomenon in our system, we fitted our observational data using the model from \citet{1995Ap&SS.226...99G}, used previously by \citet{2017AJ....154....4P} and \citet{2020ApJ...888L...5Y}:
\begin{equation}
T_{ap} (E) = T_{ap0} + P_{s} E -  \frac{e {P_s} \cos{({\omega}_0 + E \frac{d\omega}{dE})}}{\pi(1-\frac{\frac{d\omega}{dE}}{2 \pi})},
\label{eq:1}
\end{equation}
In this model, the five free parameters ($T_{ap0}$, $P_s$, e, $\omega_0$, $\frac{d\omega}{dE}$) denote the mid-transit time at E = 0,  the sidereal period, the orbital eccentricity, the argument of periastron at reference epoch (E = 0) and the precession rate of periastron. In equation \ref{eq:1}, $E$ is the epoch, $T_{ap} (E)$ is the calculated mid-transit time, $\omega$ is the angle between the ascending node on the plane of the sky and the periastron of the orbit (\citealt{Ballard_2010}). To determine the best-fit ephemeris of the apsidal precession model, we adopted here the same technique as mentioned in sections \ref{sec:linear} and \ref{sec:decay}, except for running $10^5$ steps per walker of MCMC.
The best-ﬁt parameter values of e, $\omega_0$ and $\frac{d\omega}{dE}$, obtained from the MCMC posterior probability distribution are, $e = 0.00145^{+0.00088}_{-0.00046}$, indicating a nearly circular orbit with $\omega_0 = 1.71^ {+0.77}_{-1.08}$ and $\frac{d\omega}{dE}) =  0.00040^ {+0.00031}_ {-0.00018} $~rad~epoch$^{-1}$. As the estimated $1\sigma$ uncertainty values of these model parameters are found to be large, these parameters appear to be statistically less significant. These statistically insignificant results might have originated due to the nearly circular orbit of TrES-2b, or they might have appeared due to the presence of a strong correlation between the model parameters, or it might not be the correct model to explain the long-term transit timing studies, which is fully consistent with the previous result (\citealt{2022AJ....164..220H}). Though there is currently no evidence for strong periodic TTVs caused by a nearby perturber (see section \ref{sec:additional}) but, a wider range of perturbers may still be able to excite this low eccentricity of the inner planet in the system(\citealt{10.1111/j.1365-2966.2008.14164.x}). So, we can not exclude the possibility of the apsidal precession phenomenon. We again plotted the O-C diagram with the timing residuals $T_{ap} (E)$ - $T_c (E)$ as a function of the epoch, where  $T_{ap} (E)$ and $T_c (E)$ values are derived using the best-fit ephemeris for the apsidal precession and linear models. In the O-C diagram, the blue dashed curve depicts the apsidal precession model. To see the future trend for TrES-2b, we randomly drew a sample of 100 parameter sets from the posteriors of the apsidal precession model (shown by a cyan-coloured line, see Figure \ref{fig:10}) and extrapolated them for the next $\sim \, 17\, \rm yrs$. The O-C diagram of TrES-2b shows that the apsidal precession model agrees well with the linear model. Observing the true nature of the O-C variations will be interesting by collecting further long-term high-precision follow-up observations of the transits and occultations.

\begin{table*}[!htb]
\caption{Best-Fit Model Parameters for TrES-2b}
    \centering
    \begin{tabular}{lcccc}
     \hline
     Parameter                          &  Symbol &   units    & value  & 1 $\sigma$ uncertainty  \\
     \hline
     \multicolumn{5}{c}{\textbf{Constant Period Model}} \\
        Period                          & P$_{\text{orb}}$          & days         & 2.47061355 & $^{+0.000000041}_{-0.000000041}$    \\
        Mid-transit time                & T$_{0}$   &BJD$_{\rm TDB}$     & 2453957.635381  &$^{+0.000019}_{-0.000020}$ \\
        \hline                                                              
        $a_f$, $\tau$, $(N_{\rm eff})^{a}$  &                 &           &$\sim 0.44, \sim 19, \sim 1052$ &\\
        N$_{dof}$                         &            &              &228         & \\
        $\chi^{2}$, $\chi^{2}_{red}$	                    &            &              & 348.84, 1.53       &       \\
        AIC                             &           &               & 352.41 \\
        BIC                             &           &               & 359.29 \\
     \hline
     \hline 
    \multicolumn{5}{c}{ \textbf{Orbital Decay Model}}\\  
        Period                         & P$_{\text{orb}}$        &  days         & 2.47061411 & $^{+0.00000018}_{-0.00000019}$ \\
        Mid-transit time               & T$_{q0}$    &BJD$_{\rm TDB}$      & 2453957.635172 & $^{+0.000070}_{-0.000069}$ \\
        Decay Rate                      & dP/dE     &days/epoch     & -4.45$\times$10$^{-10}$ &  $^{+1.4168\times10^{-10}}_{-1.4185\times10^{-10}}$  \\
        Decay Rate                      & dP/dt     & msec/yr       & -5.68                   & 1.81      \\
        \hline
        $a_f$, $\tau$, $(N_{\rm eff})$  &                 &           &$\sim 0.35, \sim 30, \sim 1067$ &\\
        N$_{dof}$                         &            &              &227         & \\
        $\chi^{2}$, $\chi^{2}_{red}$	                    &            &              & 338.23, 1.49       &       \\
        AIC                             &           &               & 344.66 \\
        BIC                             &           &               & 354.97 \\
        \hline 
        \hline
       \multicolumn{5}{c}{  \textbf{Apsidal Precession Model}} \\
        Sidereal Period                  & P$_{s}$    & days          &  2.47061364      &  $^{+0.00000021}_{-0.00000014}$             \\
        Mid-transit time                 & T$_{ap0}$    & BJD$_{\rm TDB}$     & 2453957.635134     & $^{+0.00027}_{-0.00036}$                  \\
        Eccentricity                    & e         &               &  0.00145          & $^{+0.00087}_{-0.00046}$ \\
        Argument of Periastron          & $\omega$$_{0}$& rad           &  1.707             &$^{+0.77}_{-1.08}$               \\
        Precession Rate                 & d$\omega$/dE& rad/epoch     &  0.000401          & $^{+0.00031}_{-0.00018}$                 \\
       \hline                                                          
        $a_f$, $\tau$, $(N_{\rm eff})$  &                 &           &$\sim 0.24, \sim 814, \sim 246$ &\\
        N$_{dof}$                         &            &              &225         & \\
        $\chi^{2}$, $\chi^{2}_{red}$	                    &            &              & 348.75, 1.55       &       \\
        AIC                             &           &               & 358.51 \\
        BIC                             &           &               & 375.70 \\
        \hline 
        \hline
    \end{tabular}
   
    \label{tb:timing_models}
     {Note: $^a$ $N_{\rm eff}$ is the effective number of independent samples.
}
\end{table*}

\section{Discussion}     \label{sec:results}
\subsection{Applegate Mechanism}
Fluctuations in the quadrupole moment of the low-mass host star, induced by spin-orbit coupling (\citealt{2002ApJ...564.1019M}), generate a new effect known as Applegate mechanism (\citealt{1987ApJ...322L..99A} and \citealt{1992ApJ...385..621A}) and this Applegate effect can produce variations in the transit times of close stellar binary systems on quasi-periodic timescales over the length of the stellar activity cycle. \citet{2010MNRAS.405.2037W} calculated the TTV amplitudes for many stars, including TrES-2, by considering an orbital period modulation time scale of 11, 22, and 50 yr for magnetic activity cycles. The largest TTV amplitude or the O-C variations expected by them for TrES-2b due to the Applegate mechanism was $ \delta{t} \sim 3.7$~s for a 50 yr activity cycle, which is much smaller than the maximum deviation in transit timing found here. Therefore, no clear signature of the Applegate effect explains the observed TTV in the TrES-2 system.

\subsection{The $R\phi$mer Effect}
In the past, we observed that the transiting planets, with a very short orbital period ($P \sim 1-20\,d$), are very good candidates for radial velocity observations (\citealt{10.1111/j.1365-2966.2012.20839.x}) and an accelerating motion of the center of mass of the binary system along the observer’s line of sight due to the force of wide-orbiting stellar companions can also cause a change in the measured transit timings of an extrasolar planet (see WASP-4b: \citealt{2019AJ....157..217B, 2019MNRAS.490.4230S, Bouma_2020, Turner_2022}; XO-2N b: \citealt{Damasso_2015, 2024MNRAS.tmp..965Y}; Kepler-1658b: \citealt{2022ApJ...941L..31V}; TrES-5b: \citealt{2021A&A...656A..88M}). Using the transit data, we have already computed the value of the period derivative to be $\dot {P}$ = $- 5.58 \pm 1.81$~ms~yr$^{-1}$, and because of the detection of the decreased orbital period, we assume that the system, TrES-2 is accelerating toward us along our line of sight. To probe whether this assumption is correct or not, we collected all the RV data available in the literature  (TrES-2: \citealt{2006ApJ...651L..61O, 10.1111/j.1365-2966.2012.20839.x, 2014ApJ...785..126K, 2017A&A...602A.107B}) and modeled them using the Radvel package (\citealt{2018PASP..130d4504F}). For the RV data modeling, we have fitted the RV semi-amplitude ($K_b$), the instrument zero point velocity ($\gamma$), the linear acceleration in RV ($\dot{v}_{RV}$), and the jitter term ($\sigma$) specific to each spectrograph freely. We have set the Gaussian priors to the parameters of the orbital period and mid-transit time by taking the $1 \sigma$ uncertainties from \citet{2019MNRAS.486.2290O}, and we have fixed the remaining parameters, such as eccentricity (e), argument of periastron ($\omega$), and quadratic trend in RV ($\ddot{\gamma}$) to zero. We reported a positive radial acceleration from the RV dataset analysis equal to $ 0.0636 \pm 0.0081$ ~m~sec$^{-1}$~day$^{-1}$, and by substituting the derived value of this radial acceleration (slope) and P (from Section \ref{sec:decay} ) in the following equation (\citealt{2021A&A...656A..88M}):

\begin{equation}
\dot{v}_{RV}=\frac{\dot{P_q}}{P_q} {c},
\end{equation}
we can derive the value of $\dot {P}$. Here, $\dot{v}_{RV}$ represents the line-of-sight acceleration of the RV of the host star, TrES-2, $\dot {P}$ represents the orbital period derivative, P is the orbital period, and c is the speed of light. We observed a turnover in the value of the period derivative, as the associated Doppler effect led to the shrinkage of the orbital period at a rate of $ 16.52 $~ms~yr$^{-1}$ and an order of magnitude greater than the decay rate observed using transit data. This increased orbital period further indicates that the center of mass of the star-planet system is accelerating away from us, clearly contradicting the result obtained using transit data, i.e., $\dot {P}$ = $- 5.58 $~ms~yr$^{-1}$. So, we conclude that the line-of-sight acceleration phenomenon is not responsible for the observed TTV in the TrES-2 system, and we also excluded the presence of an unseen distant companion in a wider orbit than the hot Jupiter system. As we are observing a limited RV dataset, further continued high-precision RV observations of TrES-2 would be helpful to strengthen this conclusion.

\subsection{Planetary Love Number ($k_p$)}
The planetary love number, $k_p$ (\citealt{1911spge.book.....L, 1977Icar...32..443G, 1978ppi..book.....Z}) is a very useful dimensionless parameter to study the effect of the interior density distribution of the hot Jupiter on the star-planet orbit through the gravitational quadrupole tidal bulges. The tidal evolution theory expected the orbits of the hot Jupiters to be circularized on a very short timescale than the ages of their systems (\citealt{2007A&A...462L...5L, 2018ARA&A..56..175D}). Using Equation (25) of \citet{1966Icar....5..375G} and by assuming the value of planetary quality factor to be ${Q}^{'}_{p}=10^6$, the estimated timescale for the tidal orbital circularization of TrES-2b is $\sim 67.40\,\rm Myr$, which is several orders of magnitude smaller than the age of the host star (TrES-2: $5.1^{+2.7}_{-2.3} $Gyr; \citealt{Sozzetti_2007}). So, the presence of an orbital eccentricity is inconsistent with tidal theory. To continuously excite the eccentricity, a process must be present there (see \citealt{10.1111/j.1365-2966.2007.12500.x, 2011Natur.473..187N, 2017AJ....154....4P, 2018ARA&A..56..175D, 2019AJ....157..217B, 2021A&A...656A..88M, 2022ApJ...941L..31V}), so that the apsidal precession can take place. In this context, \citet{2009ApJ...698.1778R} showed that the planetary interiors of very hot Jupiters could be a dominant component of the apsidal precession. We estimated the value of $k_p$ =  $30.09 \pm 2.58$ for TrES-2b by using Equation (16) of \citet{2017AJ....154....4P}:
\begin{equation}
\frac{d\omega}{dE}={15}{\pi}{k_p}\left(\frac{M_\ast}{M_p}\right)\left(\frac{R_p}{a}\right)^5,
\end{equation}
where the value $\frac{d\omega}{dE}$ is taken from Table \ref{tb:timing_models}, and the other relevant parameters are taken from \citet{Sozzetti_2007} and \citet{2012A&A...540A..99E}. Here, it is seen that the rate at which apsidal precession occurs is directly proportional to the planetary tidal Love number, $k_p$ (\citealt{2009ApJ...698.1778R, 2017AJ....154....4P}) and the higher value of $k_p$ leads to a higher precession rate with smaller centrally condensed interior structure. Despite the precession model having more free parameters and presenting an anomaly in the Love number with considerable uncertainty, indicating it may exceed that of Jupiter by two orders of magnitude ($k_p$ = 0.59: \citealt{2016ApJ...831...14W}), we cannot entirely dismiss apsidal precession as a plausible explanation for the observed long-term trend of transit timing variations (TTVs). Previous findings suggest that a broader range of perturbers could potentially excite the low eccentricity of the inner planet in the system (see Section \ref {sec:apsidal}). Further additional follow-up timing observations of both transits and occultations will be required to strengthen this finding.

\subsection{Statistical Approach for Finding Plausible Causes for Observed TTV}
\subsubsection{Goodness of the Fit}
 To estimate the quality of the statistical model and to check whether the observed data fits the model created by the previous data well, we performed the chi-squared test of the observed transit midpoint and the expected transit midpoint (see Table \ref{tb:timing_models}).  For the chi-square values, the orbital decay model (${\chi}^{2}$ = 338.23) provides a much better fit in comparison to constant period model (${\chi}^{2}$ = 348.84) and apsidal precession model (${\chi}^{2}$ = 348.75), which is fully consistent with the previous result of \citet{2022AJ....164..220H}.
Now, to check the goodness of the fit, we calculated the reduced chi-square value of those three best-fit models (Table \ref{tb:timing_models}) using this formula: ${\chi}^{2}_{red}$  = ${\chi}^{2}$/n, where n= total number of degrees of freedom.  We expect the reduced chi-square value to be nearly equal to 1 for a good fit, but here, the reduced chi-square value of all three models is significantly above unity, providing not a good fit overall. From some previous studies (\citealt{2016MNRAS.457.4205S, 2015MNRAS.450.3101B, 10.1093/mnras/stad248}), we have found this is a common case in most of the transit timing analyses. As the orbital decay model provides the lowest reduced chi-squared value (1.49 with 227 degrees of freedom) in comparison to a linear model (${\chi}^{2}_{red}$ = 1.53 with 228 degrees of freedom) and apsidal precession model (${\chi}^{2}_{red}$ = 1.55 with 225 degrees of freedom), that’s why it provides a marginally better fit to the timing data of TrES-2b and this model is preferred over the other two models.
\subsubsection{The Akaike Information Criterion}
To explore this further and to determine statistically the most parsimonious model that fits the transit timing data well, astronomers have recently used two widely used metrics: the Akaike Information Criterion (AIC) and the Bayesian Information Criterion (BIC). First we calculated the Akaike Information Criterion (AIC; \citealt{1100705}) for all of those three models (see Table \ref{tb:timing_models}), defined as AIC = ${\chi}^{2} $ + 2$k_F$, where $k_F$ is the number of free parameters in the model ($k_F$ = 2 for linear model, $k_F$ = 3 for orbital decay model and $k_F$ = 5 for apsidal precession model). For a time series analysis, the Akaike information criterion (\citealt{doi:10.1080/01621459.1995.10476572}) is one of the most omnipresent tools for selecting a suitable machine learning model for an observed data set. We obtained the AIC values of 352.41, 344.66, and 358.51 for linear, orbital decay, and apsidal precession model fits, respectively. Here, the decay model is somewhat preferable over the linear model by $\bigtriangleup\mathrm{AIC} = 7.75$, with a likelihood ratio of $\exp(\bigtriangleup\mathrm{AIC}/2)= 48.18$ in representing the data.           
\subsubsection{The Bayesian Information Criterion}
Next, we calculated the Bayesian Information Criterion (BIC; \citealt{1978AnSta...6..461S, 10.1111/j.1745-3933.2007.00306.x}), a useful heuristic and a robust way for selecting an appropriate best-fit model, having different number of free parameters, to fit our transit timing dataset using  Bayesian Approach and it is expressed as, BIC= ${\chi}^{2} + k_F\log{N_P}$, where, $N_P$ is the total number of data points (230 data points used in this analysis). Usually, preference is given to a model with a lower BIC value to describe our data. The difference in the BIC between the apsidal precession and linear models is $\bigtriangleup\mathrm{BIC} = \mathrm{BIC}_{precession} - \mathrm{BIC}_{linear} = 16.41$ and the corresponding approximate Bayes factor i.e., the ratio of posterior probabilities (\citealt{2020ApJ...888L...5Y}) is $\exp(\bigtriangleup\mathrm{BIC}/2)= 3659.20$. Therefore, the linear (constant period) model is strongly favored over the apsidal precession model for the examined transit timing data of TrES-2b. However, the orbital decay model exhibits a notably superior fit compared to the linear model, as evidenced by the difference in Bayesian Information Criterion (BIC) values, $\bigtriangleup\mathrm{BIC} = 4.32$ with $\exp(\bigtriangleup\mathrm{BIC}/2)= 75.19$. This superiority renders the orbital decay model the preferred choice for representing our data, thereby weakening the possibility of a constant period in the transit timing observations of TrES-2b (see \citealt{2014ApJ...781..116B, 2016A&A...588L...6M, 2017AJ....154....4P, 2022AJ....163...77A}), despite the relatively weak distinction between the two models. Notably, \citet{2024arXiv240407339A} recently reported a $\bigtriangleup\mathrm{BIC}$ value of 4.9, consistent with the findings of this study. Based on the derived values of all three statistical quantities mentioned above, ${\chi}^{2}$, AIC, and BIC, the quadratic model gives significant evidence in detecting TTV in the TrES-2 system. Future transit and radial-velocity observational data will further confirm this. 
\subsubsection{Estimation of the Stellar Tidal Quality Factor}

Given the observed negative orbital decay rate of TrES-2b (see section \ref{sec:decay}), we assume that the decrease in orbital period is attributed to tidal dissipation occurring within the star. This assumption allows us to derive the modified stellar tidal quality factor of a star (${Q}^{'}_{\ast})$, a dimensionless parameter that characterizes the efﬁciency of energy dissipation in the host star during tidal interactions due to a perturbing body.
To calculate ${Q}^{'}_{\ast}$ for the TrES-2 system, we followed the same approach of \citet{Wilkins_2017}, \citet{2017AJ....154....4P} and \citet{2018AcA....68..371M} and we used equation (20) of the modified “constant phase lag” model of \citet{1966Icar....5..375G}:

\begin{equation}
{Q}^{'}_{\ast} = -\frac{27}{2}{\pi}\left(\frac{M_p}{M_\ast}\right)\left(\frac{a}{R_\ast}\right)^{-5}\left(\frac{1}{\dot{P}}\right),
\end{equation}

where P is the orbital period derived using the orbital decay model, $\dot{P}$ is the decay rate of the orbital period,  $\frac{M_p}{M_\ast}$ is the mass ratio of the planet to the star, and $\frac{a}{R_\ast}$  is the ratio of the semi-major axis to the stellar radius. This expression is based on the assumption that the host star’s rotational frequency is much smaller than the planet’s orbital frequency, and the ${Q}^{'}_{\ast}$ can be deﬁned from the tidal interaction model by  \citet{1977A&A....57..383Z} as ${Q}^{'}_{\ast} = \frac{3}{2} \left(\frac{Q_\ast}{K_\ast}\right)$, where ${K_\ast}$ is the stellar tidal Love number and ${Q}_{\ast}$ is the quality factor.

The estimation of ${Q}^{'}_{\ast}$ is based on these values: $\frac{M_p}{M_\ast}$ = 1.2 and $\frac{a}{R_\ast}$ = 7.63, taken from \citet{2006ApJ...651L..61O}. By substituting the derived value of $\dot{P}$ (see section \ref{sec:decay}) and the above values in equation 7, we have inferred the modiﬁed stellar tidal quality factor to be  $ \sim 9.9 \times {10}^{3}$ for TrES-2b, this is at least 2-3 orders of magnitude smaller than typically expected values reported for binary star systems (${10}^{5}–{10}^{7}$; \citealt{2015Natur.517..589M, 2018AJ....155..165P}), for the stars hosting the hot-Jupiters (${10}^{5}–{10}^{6.5}$; e.g., \citealt{2008ApJ...681.1631J, 10.1111/j.1365-2966.2012.20839.x, Barker_2020}) and for short-period hot Jupiters (${10}^{5}–{10}^{6}$; a theoretical study by \citealt{Essick_2015}. However, our value is at least consistent with the upper limit of typical modified stellar tidal quality factor value of hot Jupiter host stars (${Q}^{'}_{\ast} \leq {10}^{7}$), reported by \citet{2019AJ....158..190H}, for period range ${2} \:{days} \leq P \leq {5} \: {days}$ and the mass range ${0.5} {M}_{Jup} \leq {M}_{P} \leq 2{M}_{Jup}$. 

Since $\dot {P}$ is barely detected and there was not an appropriate error propagation from the observed $ \dot {P}$ value (see section \ref{sec:decay} ), so we also computed the $1\sigma$ uncertainty values of $ {Q}^{'}_{\ast}$. Given that $ {Q}^{'}_{\ast}$ is inversely proportional to $ \dot {P}$, any errors or uncertainties in $\dot {P}$ will have a significant impact on the uncertainties of ${Q}^{'}_{\ast}$. Consequently, even minor errors  in $\dot {P}$ can lead to substantial errors in $ {Q}^{'}_{\ast}$. By substituting the 
$1\sigma$ uncertainty values from the posterior distribution of $\frac{dP}{dE}$ (refer to Table \ref{tb:timing_models})
into the formula $\dot {P} = \frac{1}{P} \left(\frac{dP}{dE}\right)$, and subsequently substituting the resulting
$\dot {P}$ value into equation (8), we computed the lower 
(L$ {Q}^{'}_{\ast}$) and upper (U$ {Q}^{'}_{\ast}$) limits of $ {Q}^{'}_{\ast}$. Our calculated values for L${Q}^{'}_{\ast}$ and U$ {Q}^{'}_{\ast}$ are approximately $  \sim 7.5 \times {10}^{3}$ and $ \sim 14.5 \times {10}^{3}$, respectively, which appear unusually lower than the typical values observed for hot Jupiters. However, a slight overestimation of the 
$\dot {P}$ (e.g., by a $ 3\sigma $ deviation) would readily place the typical $ {Q}^{'}_{\ast}$ value within the range of ${10}^{5}–{10}^{7}$.

As the remaining lifetime (time remained for the planet to collide with its host star) of the hot Jupiter (\citealt{2009Natur.460.1098H, 2009ApJ...692L...9L, 2010ApJ...725.1995M, 2014ApJ...781..116B, 2016A&A...588L...6M, 2017AJ....154....4P})  at small eccentricity (e = 0.0014, see Table \ref{tb:timing_models}) depends on the efficiency of tidal dissipation within the host star, that’s why by substituting the value of ${Q}^{'}_{\ast}$ and the other relevant parameters (see above) in equation 5 of \citet{2009ApJ...692L...9L}, we calculated the value of ${T}_{remain}$ to be $\sim 4.9 \, Myr $, where, n = $\frac{2\pi}{P}$ is the frequency of mean orbital motion of the planet.

Moreover, TrES-2b presents ${Q}^{'}_{\ast}$ value having one order of magnitude lower than that of WASP-4b, indicating that the orbit of TrES-2b might be decaying rapidly resulting into a fast orbital evolution, which is a common case in cool stars ( ${T}_{\rm eff} < 6000 \: K$); \citealt{2020AJ....159..150P}). And the smaller value of ${Q}^{'}_{\ast}$ also indicates a strong and more efficient tidal dissipation (\citealt{1966Icar....5..375G, Wilkins_2017}). Till now, the reason behind such a low value of ${Q}^{'}_{\ast}$ is not clear; this discrepancy may appear due to adding more transit data in our study than the previous studies. Further investigation with a longer time baseline will be useful to understand better this low value of ${Q}^{'}_{\ast}$ as well as to find its dependence on planet mass (\citealt{10.1111/j.1365-2966.2010.16400.x}), orbital period (\citealt{2011AGUFMGC51F1078B}), strength of the perturbation and the internal structure of the star (\citealt{2014ARA&A..52..171O, 2018AJ....155..165P}).

\section{Concluding Remarks}         \label{sec:remarks}
In this work, we have done a combined study of ground-based photometric data along with space-based photometry to perform the transit timing analysis. We present here 64 light curves of TrES-2b observed by the space-based telescope, TESS, in sectors 26, 40, 41, 54, 55, 74 and 75. To estimate the refined values of the physical and orbital parameters as well as to do the precise transit timing analysis of our system, TrES-2b, we have now combined all TESS light curves with the best quality light curves from ETD and archival data. In total, data from 230 light curves from 2006 to 2024 have been considered here for Transit Timing Variation (TTV) analysis. Firstly, by fitting a linear ephemeris model to our transit timing data, we have derived new ephemeris for the orbital period and the mid-transit time (see Table \ref{tb:timing_models}), and these values are fully consistent with the values obtained from literature, but more precise than the previous values because of including the high-quality robust follow up of TESS. Through the frequency analysis, we have determined the absence of any short-term TTV due to lack of periodicity in the timing residuals within the system. So, we ruled out the presence of an additional planet in a close-in orbit to TrES-2b.

The absence of short-term TTV next motivates us to look for long-term TTV, which may be caused by orbital decay or apsidal precession. As the current observation time baseline covers more than a decade, we may suspect the possibility of TTV due to orbital decay for our observed system, and when we fitted the orbital decay model to our timing data using only TESS data, the decay model is slightly favoured than other two models, but when we combined all the available data, then statistically the fitting by orbital decay model gave an overwhelming response in comparison to apsidal precession, which is consistent with a previous study (\citealt{2022AJ....164..220H}). The value of the orbital decay rate, $\dot P_q$, for TrES-2b, as determined from orbital decay studies, is approximately $\sim - 5.58 \pm 1.81$~ms~yr$^{-1}$, exhibiting a declining trend. Considering this period derivative and the $ \bigtriangleup\mathrm{BIC}$ value of 4.32, it appears that the quadratic model is slightly preferred over the linear model based on the available data and the actual significance is probably reduced due to potential underestimations of uncertainties in the transit times reported in the literature. For the statistically insignificant value of e (circular orbit) and $d\omega/dE$, it appears that the apsidal precession phenomenon may not be the plausible cause of the observed TTV. Moreover, the unphysical value of the planetary love number,  $k_p$ =  $30.09~ \pm ~2.58$, is also unable to specify the interior density profile of this hot Jupiter, TrES-2b.

Assuming the negative value of $\dot P_q$ as an attribution towards orbital decay phenomenon, we estimated the value of the modified stellar tidal quality factor, $ {Q}^{'}_{\ast}\sim 9.9 \times {10}^{3}$ which is significantly smaller than the values reported by theoretical models. Therefore, the tidal decay is unlikely to be a potential cause of the observed period change here, but looking at only the BIC metric, this model appears to be the best plausible explanation for the observed TTV compared to the linear and apsidal precession model. Further investigation may give a better explanation of this finding.

We have also tried to probe other possible causes of TTV, and we have ruled out the Applegate mechanism as a possible cause of the observed TTV in TrES-2b since the largest TTV amplitude calculated by \citet{2010MNRAS.405.2037W} considering 50 yr magnetic activity cycle is an order of magnitude smaller than the observed TTV amplitude. Moreover, the observed period change of TrES-2b does not also appear to originate from the $R\phi$mer effect, as the value of orbital decay rate derived from the rate of the line of sight acceleration using RV data contradicts the decay rate obtained using transit data.  

\section{Call for Additional Observations}         \label{sec:observations}
To conclusively determine the possible cause for the change in the orbital period, we encourage further monitoring of high-precision photometric follow-up observations of primary and secondary eclipses and radial-velocity observations of the hot Jupiter system, TrES-2b. Additionally, high-precision transit observations using large or smaller aperture telescopes from different ground-based observatories would also be beneficial in studying the TTV analysis by increasing the observational time baseline.

\section{Acknowledgements}         \label{sec:Acknowledgements}
 We are grateful to the anonymous referee and editor for useful suggestions, which helped us significantly improve this paper's quality. While computing the Generalized Lomb Scargle Periodogram, the valuable suggestions given by M. Zechmeister and O{\v{g}}uz {\"O}zt{\"u}rk are acknowledged and while performing transit timing analysis the fruitful discussions with John Southworth, Kishore C. Patra, Fan Yang, Shreyas Vissapragadaare and Vineet Kumar Mannaday are also gratefully acknowledged. We are thankful to Benjamin J. Fulton for the helpful conversations with him while using the Radvel package.

This work is supported by the grant from the National Science and Technology Council, Taiwan. The grant number is 
NSTC 113-2112-M-007-030. This study incorporates data obtained
from the TESS mission's observations, accessible through the Mikulski Archive for Space Telescopes (MAST). The specific observations analyzed are retrievable via the dataset link provided:\dataset[https://doi.org/10.17909/t9-nmc8-f686]{https://doi.org/10.17909/t9-nmc8-f686}. 
The funding for this mission is provided by NASA’s Science Mission Directorate. 
This paper has made use of the publicly available transit light curves from the Exoplanet Transit Database (ETD), 
and for making the light curves publicly available, 
we are grateful to the contributors of ETD. This paper has
also used the VizieR catalog access tool, 
operated at CDS, Strasbourg, France,
the SIMBAD database, and NASA's Astrophysics Data System
Bibliographic Services.

\appendix
\restartappendixnumbering

\section{Transit fits to individual transit events} \label{app:individual_transits}
\begin{figure}
\centering
  \begin{tabular}{@{}c@{}}
    \includegraphics[width=\columnwidth]{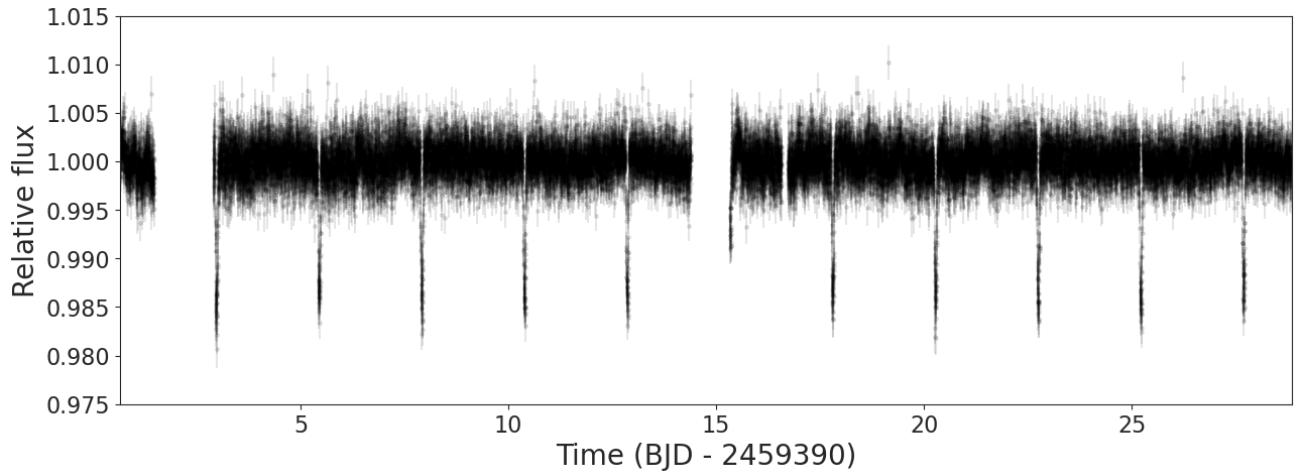}\\
    \includegraphics[width=\columnwidth]{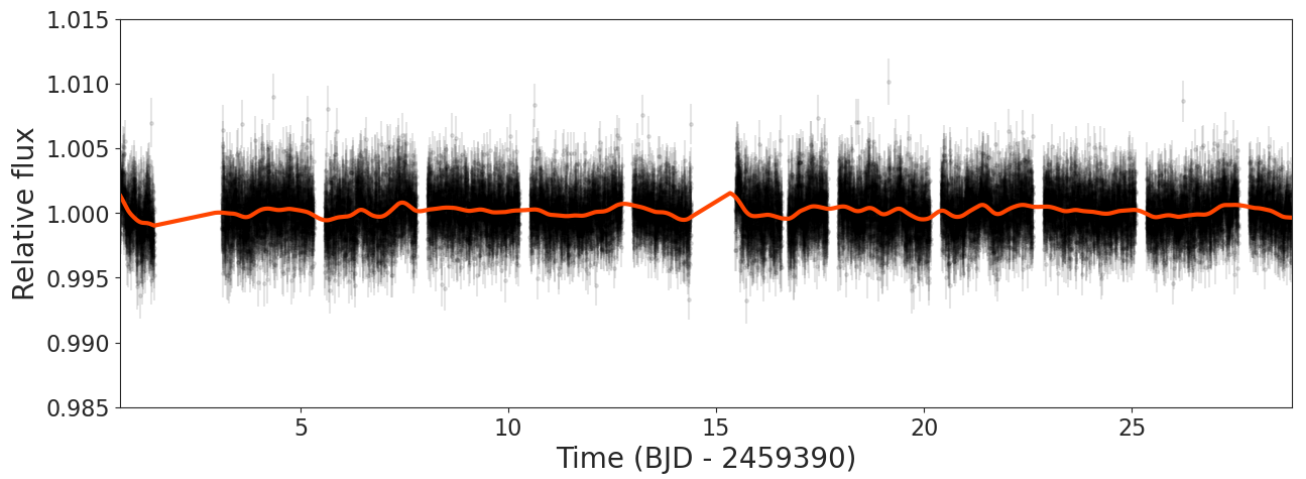}\\
    \includegraphics[width=\columnwidth]{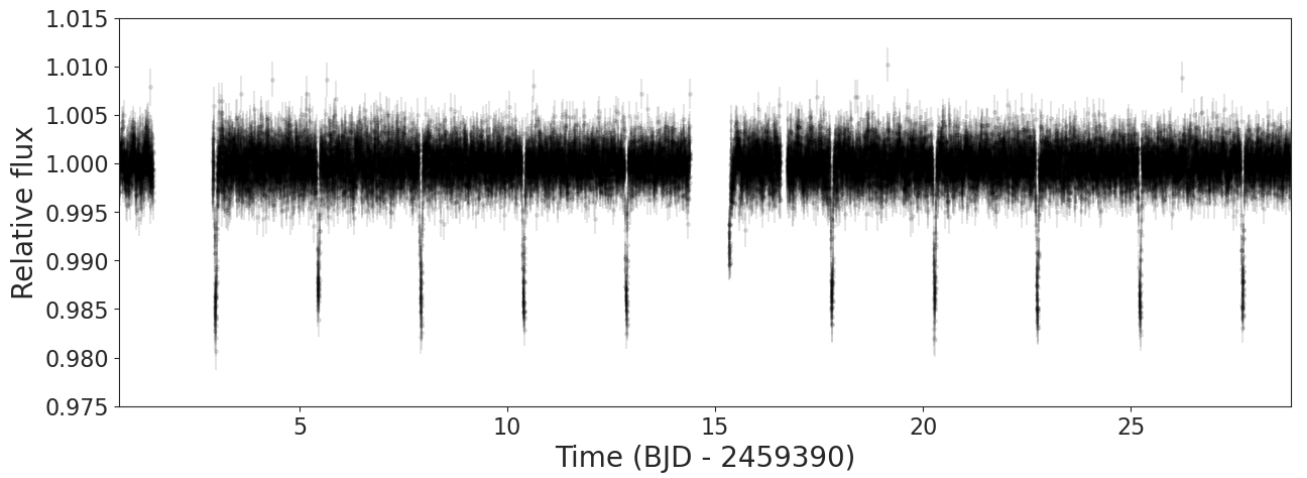} \\
  \end{tabular}
  \caption{Top panel: time series data of TrES-2b observed by TESS within sector 40. Middle panel: corresponding OOT time series data along with the best-fit GP model (red curve). Bottom panel: detrended time series data corresponding to the top panel.}
  \label{fig:ind_transits_sec40_2}
\end{figure}

\begin{figure}
\centering
  \begin{tabular}{@{}c@{}}
    \includegraphics[width=\columnwidth]{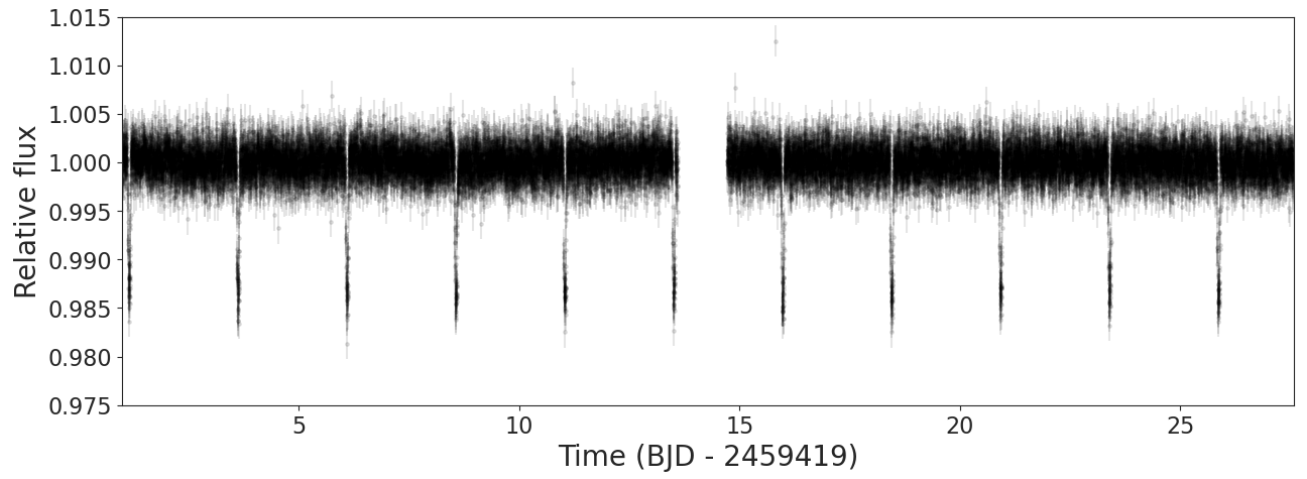}\\
    \includegraphics[width=\columnwidth]{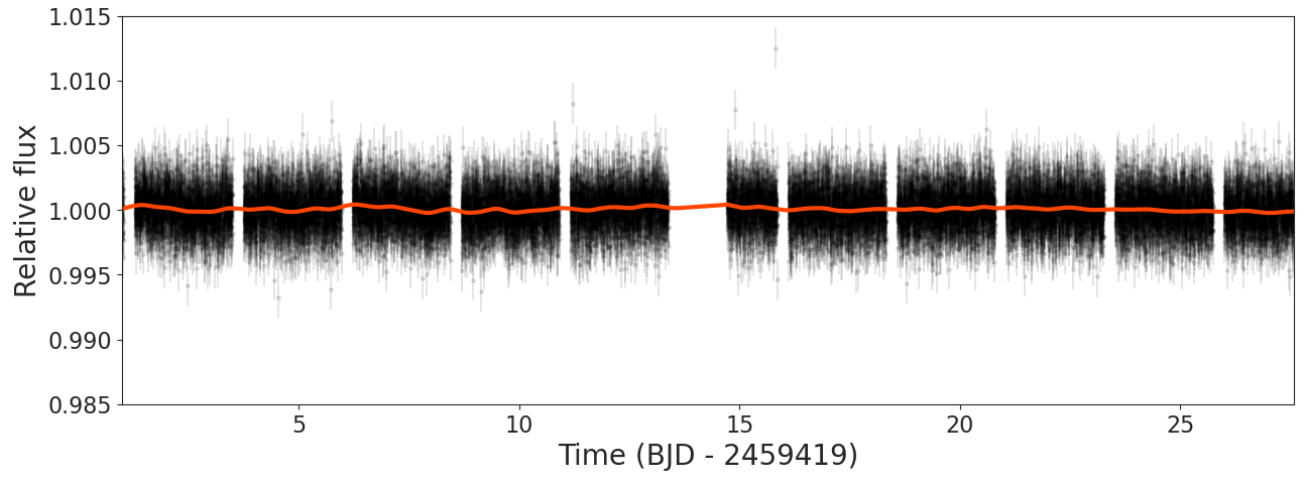}\\
    \includegraphics[width=\columnwidth]{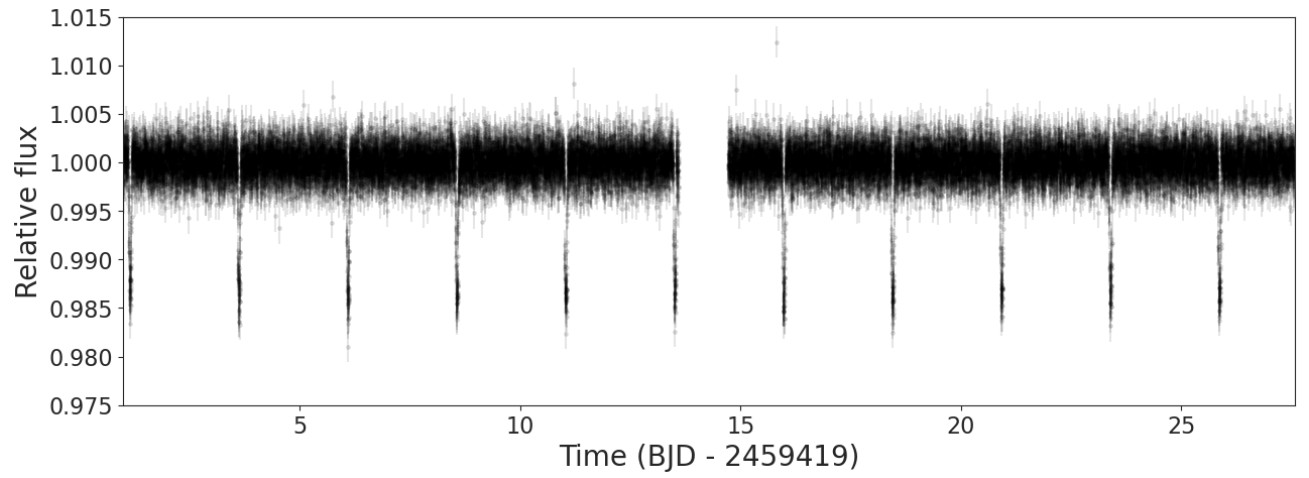} \\
  \end{tabular}
  \caption{Top panel: time series data of TrES-2b observed by TESS within sector 41. Middle panel: corresponding OOT time series data along with the best-fit GP model (red curve). Bottom panel: detrended time series data corresponding to the top panel.}
  \label{fig:ind_transits_sec40_3}
\end{figure}

\begin{figure}
\centering
  \begin{tabular}{@{}c@{}}
    \includegraphics[width=\columnwidth]{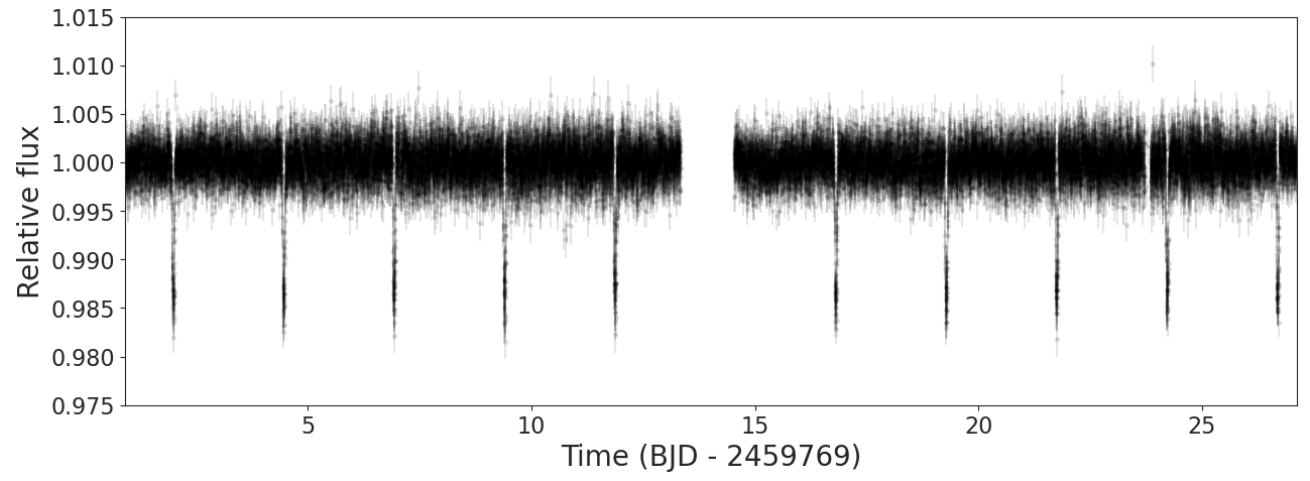}\\
    \includegraphics[width=\columnwidth]{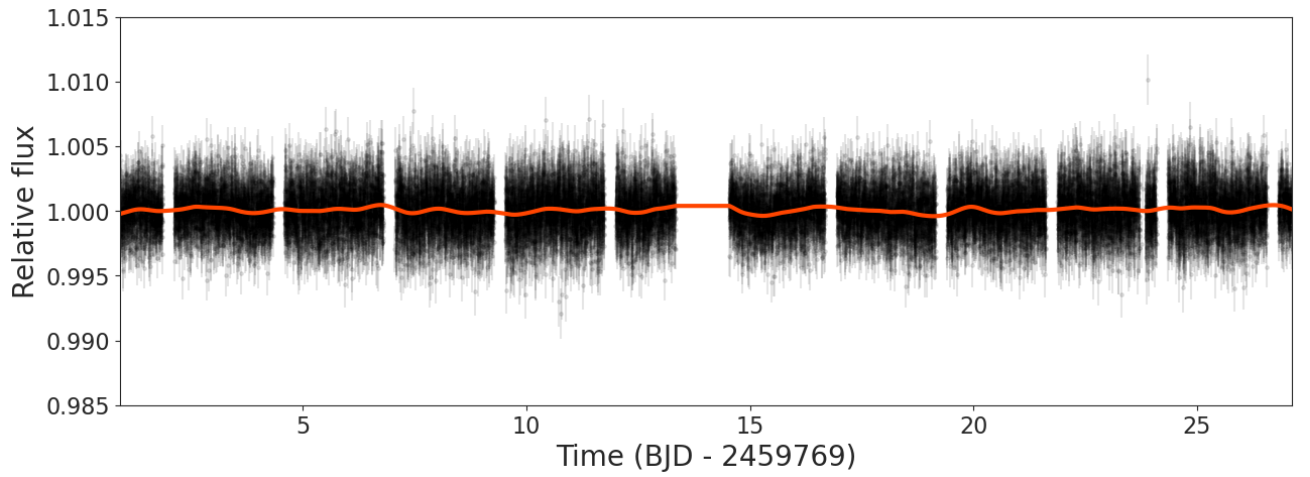}\\
    \includegraphics[width=\columnwidth]{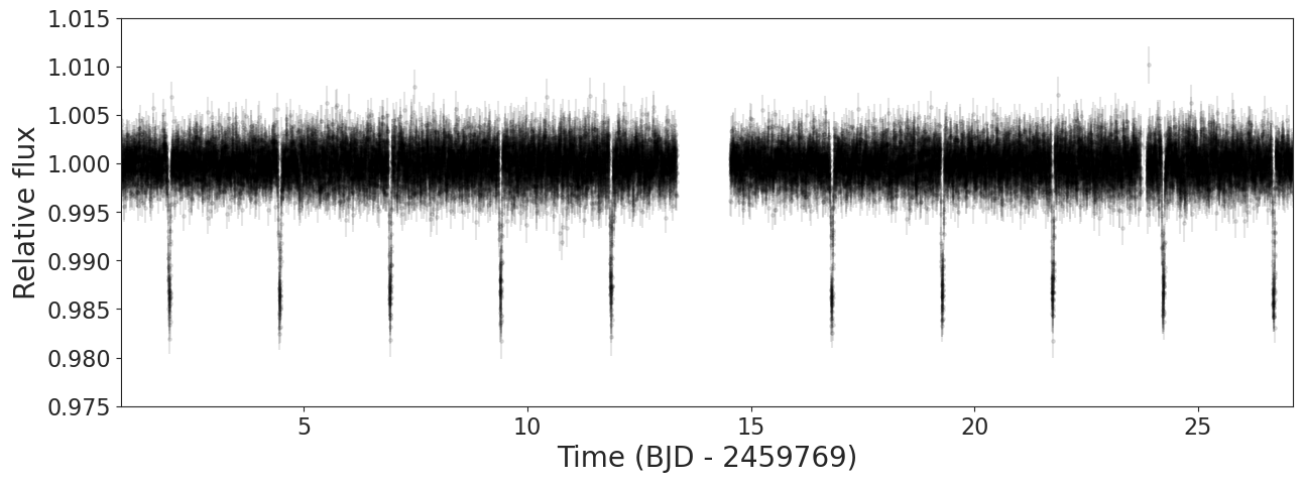} \\
  \end{tabular}
  \caption{Top panel: time series data of TrES-2b observed by TESS within sector 54. Middle panel: corresponding OOT time series data along with the best-fit GP model (red curve). Bottom panel: detrended time series data corresponding to the top panel.}
  \label{fig:ind_transits_sec40_4}
\end{figure}

\begin{figure}
\centering
  \begin{tabular}{@{}c@{}}
    \includegraphics[width=\columnwidth]{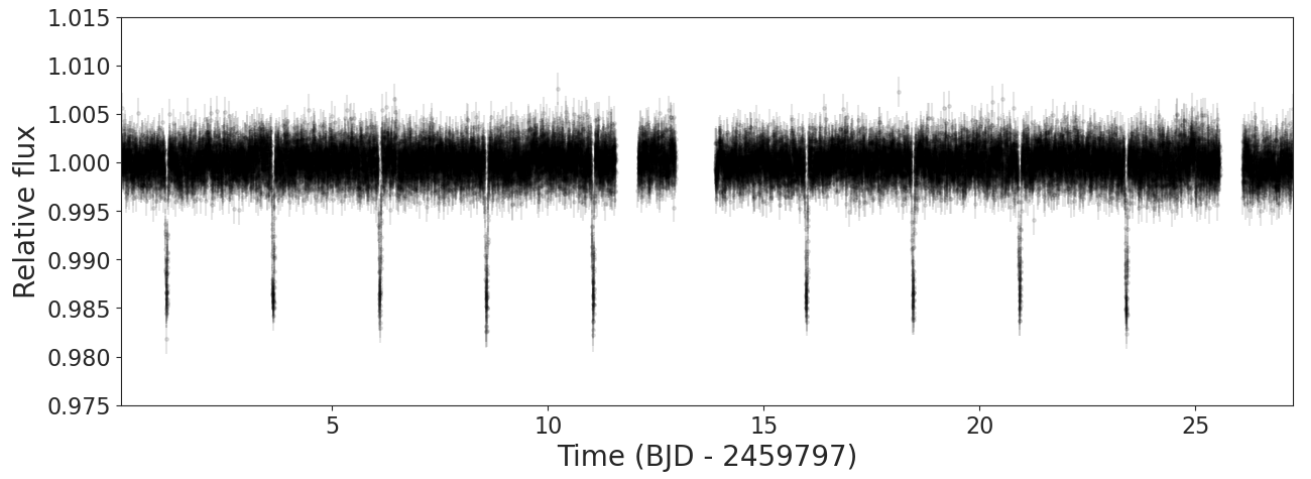}\\
    \includegraphics[width=\columnwidth]{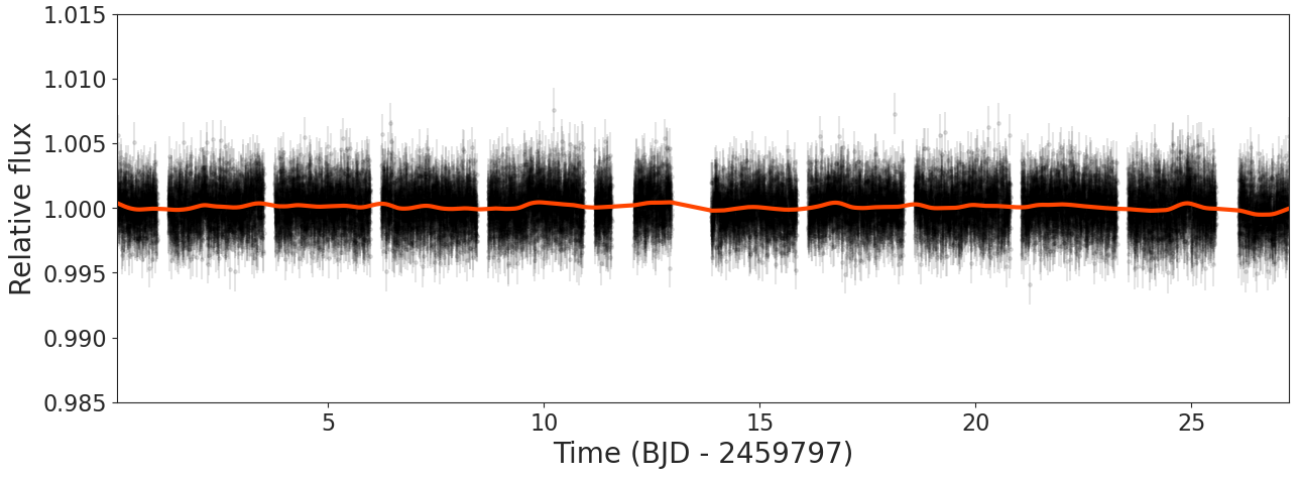}\\
    \includegraphics[width=\columnwidth]{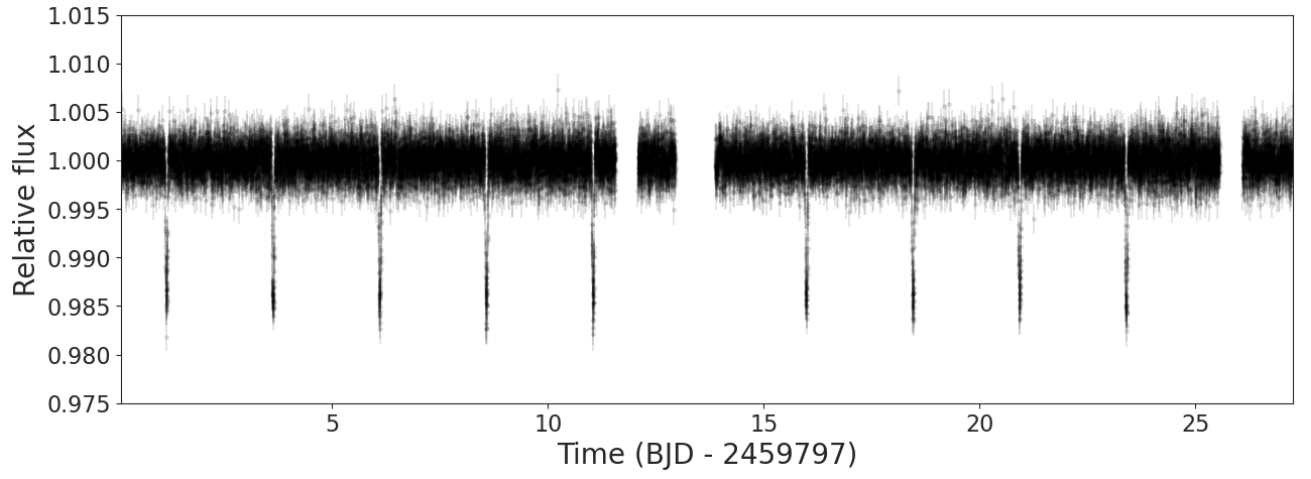} \\
  \end{tabular}
  \caption{Top panel: time series data of TrES-2b observed by TESS within sector 55. Middle panel: corresponding OOT time series data along with the best-fit GP model (red curve). Bottom panel: detrended time series data corresponding to the top panel.}
  \label{fig:ind_transits_sec40_5}
\end{figure}

\begin{figure}
\centering
  \begin{tabular}{@{}c@{}}
    \includegraphics[width=\columnwidth]{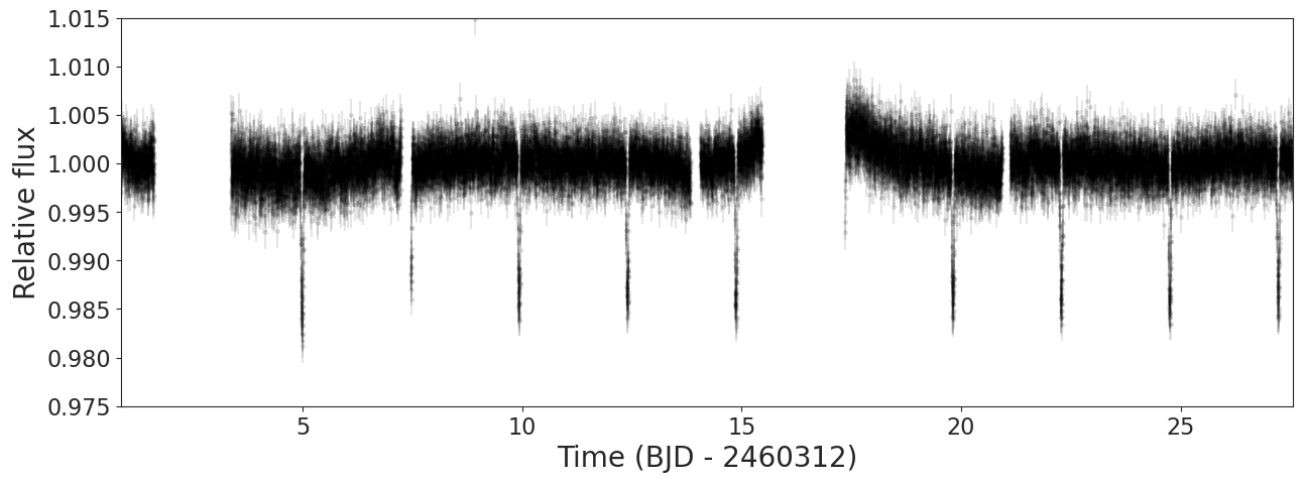}\\
    \includegraphics[width=\columnwidth]{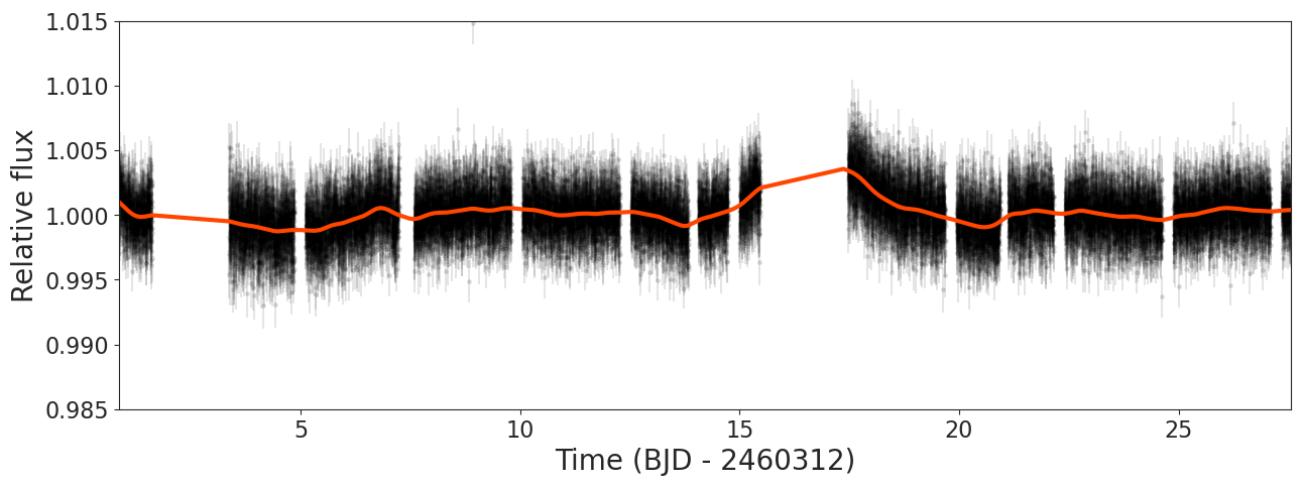}\\
    \includegraphics[width=\columnwidth]{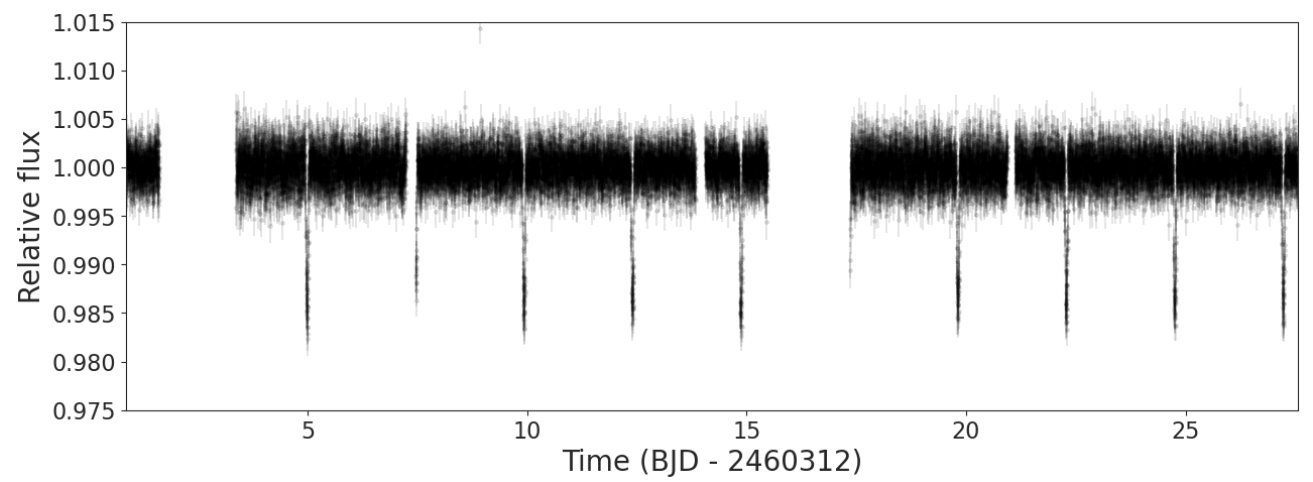} \\
  \end{tabular}
  \caption{Top panel: time series data of TrES-2b observed by TESS within sector 74. Middle panel: corresponding OOT time series data along with the best-fit GP model (red curve). Bottom panel: detrended time series data corresponding to the top panel.}
  \label{fig:ind_transits_sec40_6}
\end{figure}

\begin{figure}
\centering
  \begin{tabular}{@{}c@{}}
    \includegraphics[width=\columnwidth]{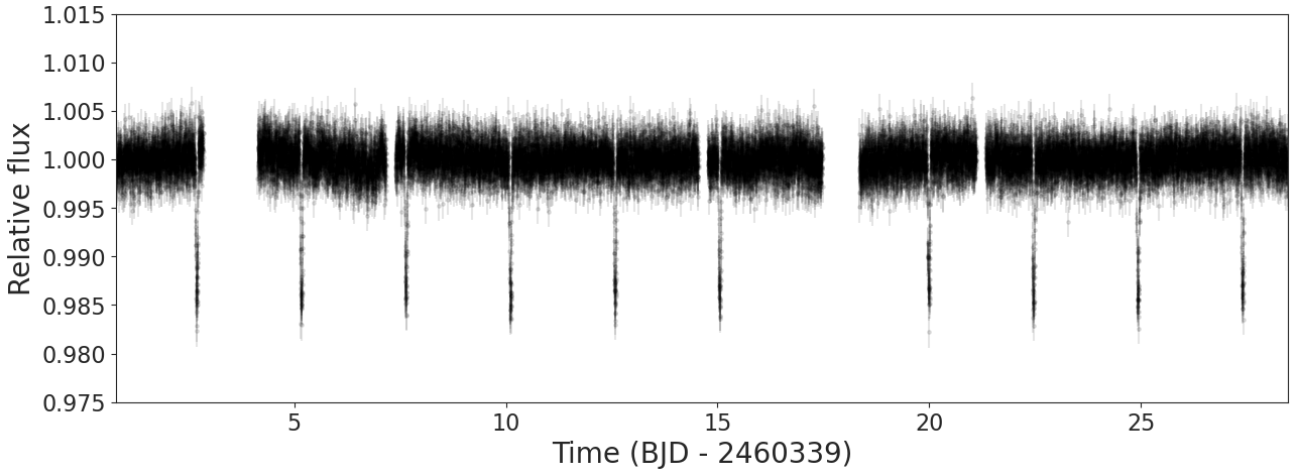}\\
    \includegraphics[width=\columnwidth]{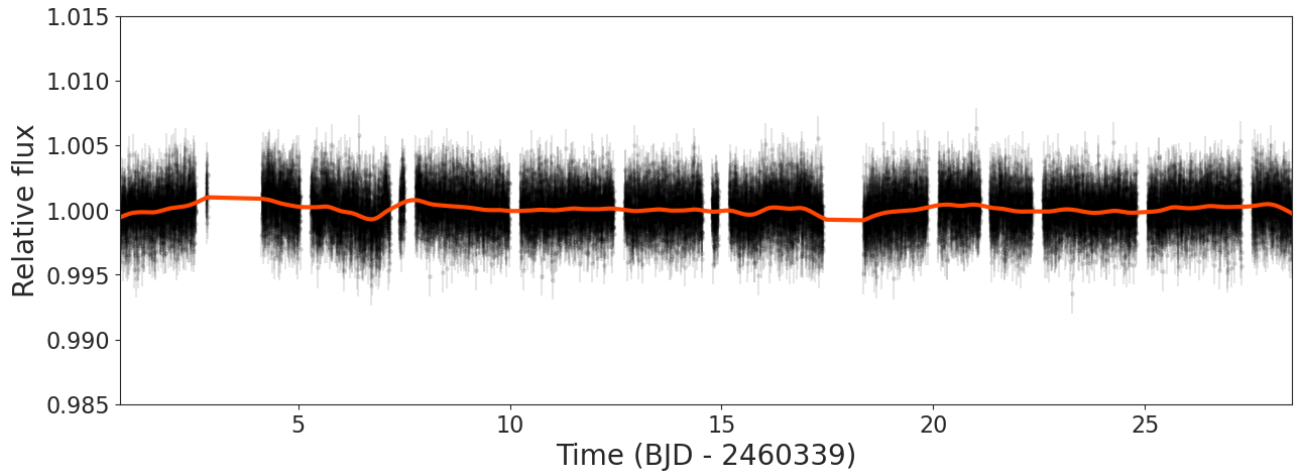}\\
    \includegraphics[width=\columnwidth]{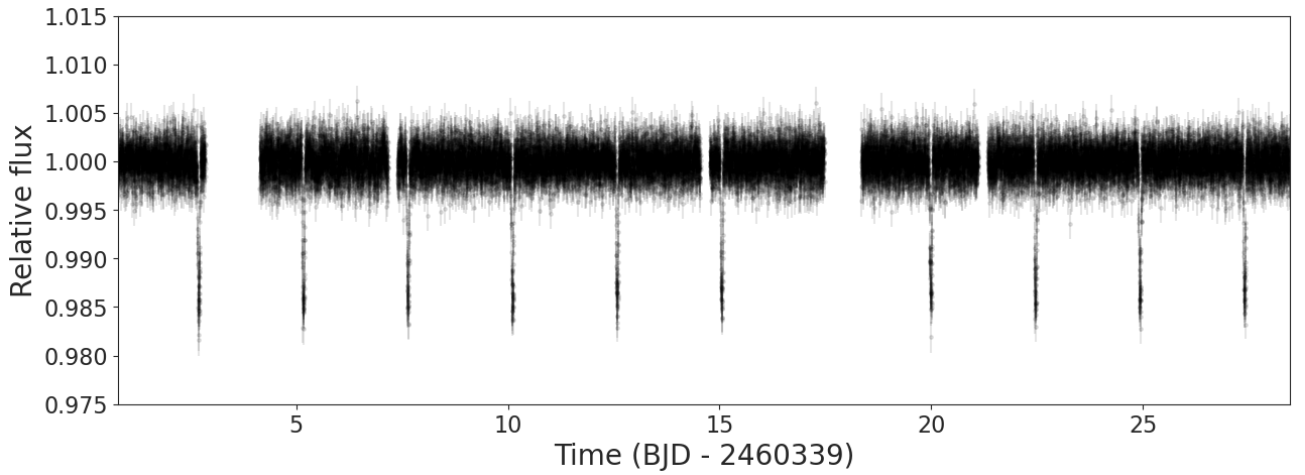} \\
  \end{tabular}
  \caption{Top panel: time series data of TrES-2b observed by TESS within sector 75. Middle panel: corresponding OOT time series data along with the best-fit GP model (red curve). Bottom panel: detrended time series data corresponding to the top panel.}
  \label{fig:ind_transits_sec40_7}
\end{figure}

\begin{figure}
    \centering
    \includegraphics[width=1.1\linewidth]{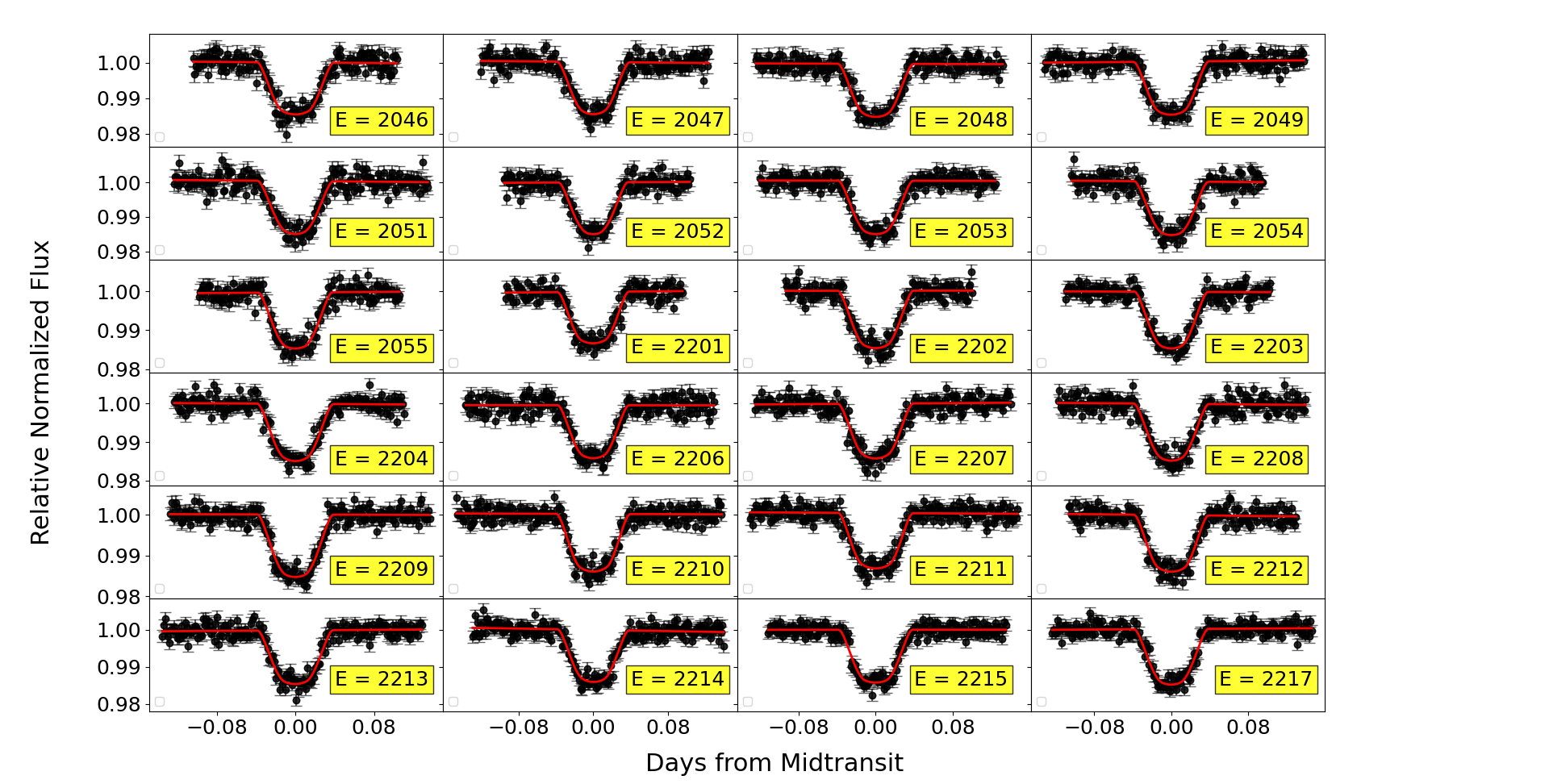}
    \caption{The normalized relative flux of TrES-2b as a function of the time (the offset from mid-transit time and in TDB-based BJD) of individual transit observed by TESS: the points are the data of raw flux, solid red lines are best-fit models for model flux, and E is the calculated epoch number.}
    \label{fig:6}
\end{figure}

\begin{figure}
    \centering
    \includegraphics[width=1.1\linewidth]{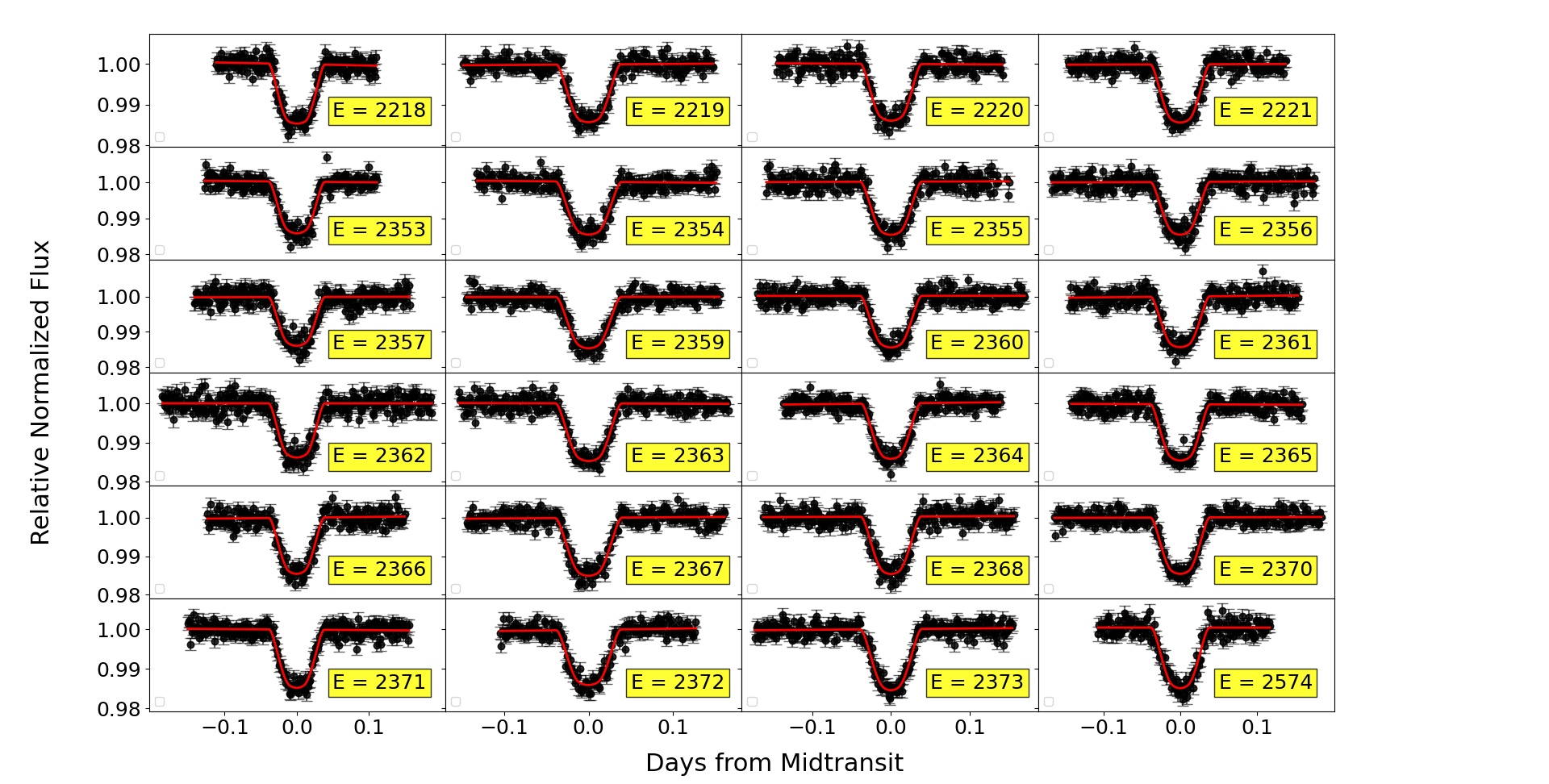}
    \caption{Same as the Figure \ref{fig:6} but for remaining epochs}
    \label{fig:7}
\end{figure}

\begin{figure}
    \centering
    \includegraphics[width=1.1\linewidth]{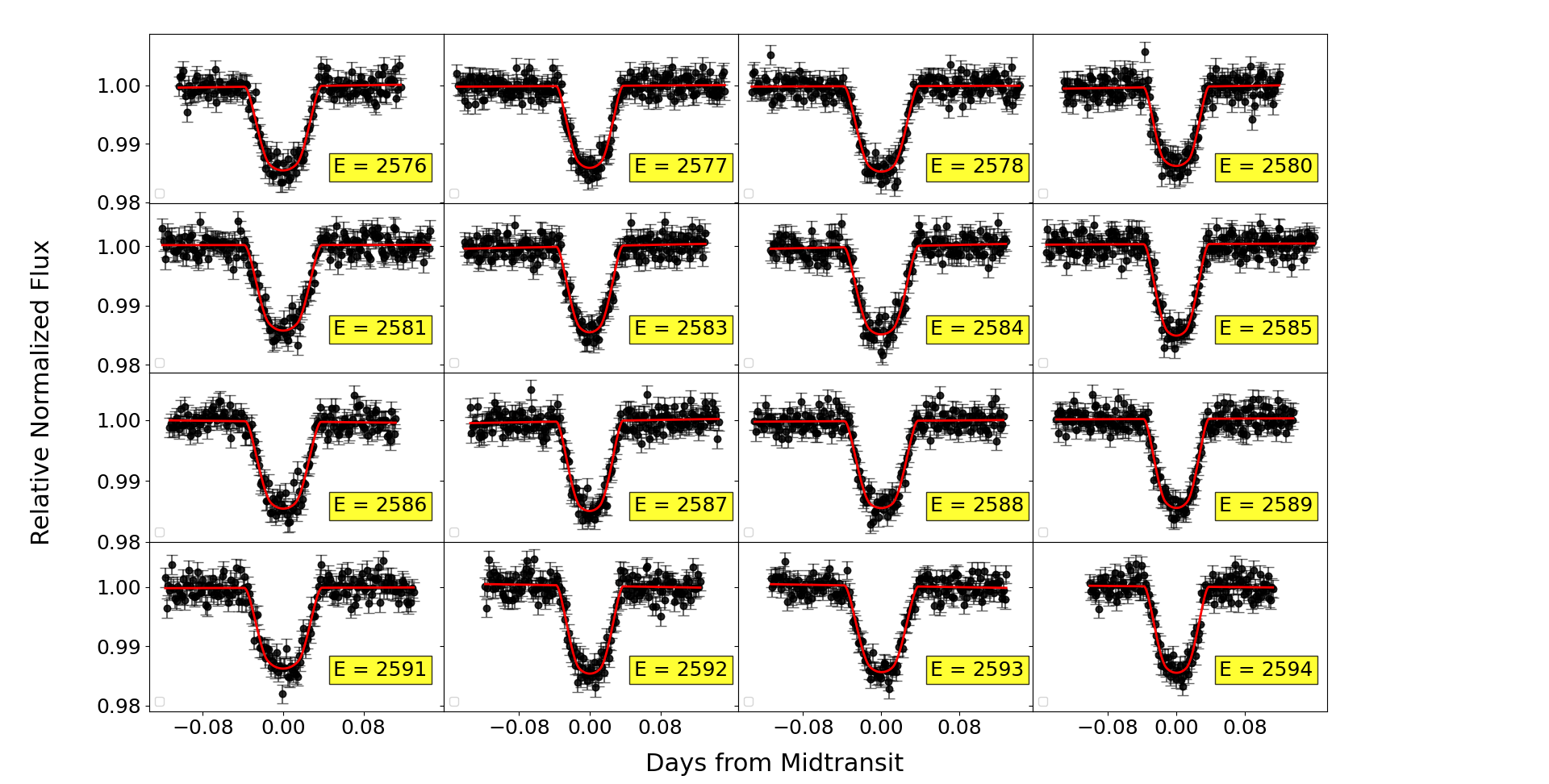}
    \caption{Same as the Figure \ref{fig:6} but for remaining epochs}
    \label{fig:8}
\end{figure}

\begin{figure}
    \centering
    \includegraphics[width=1.1\linewidth]{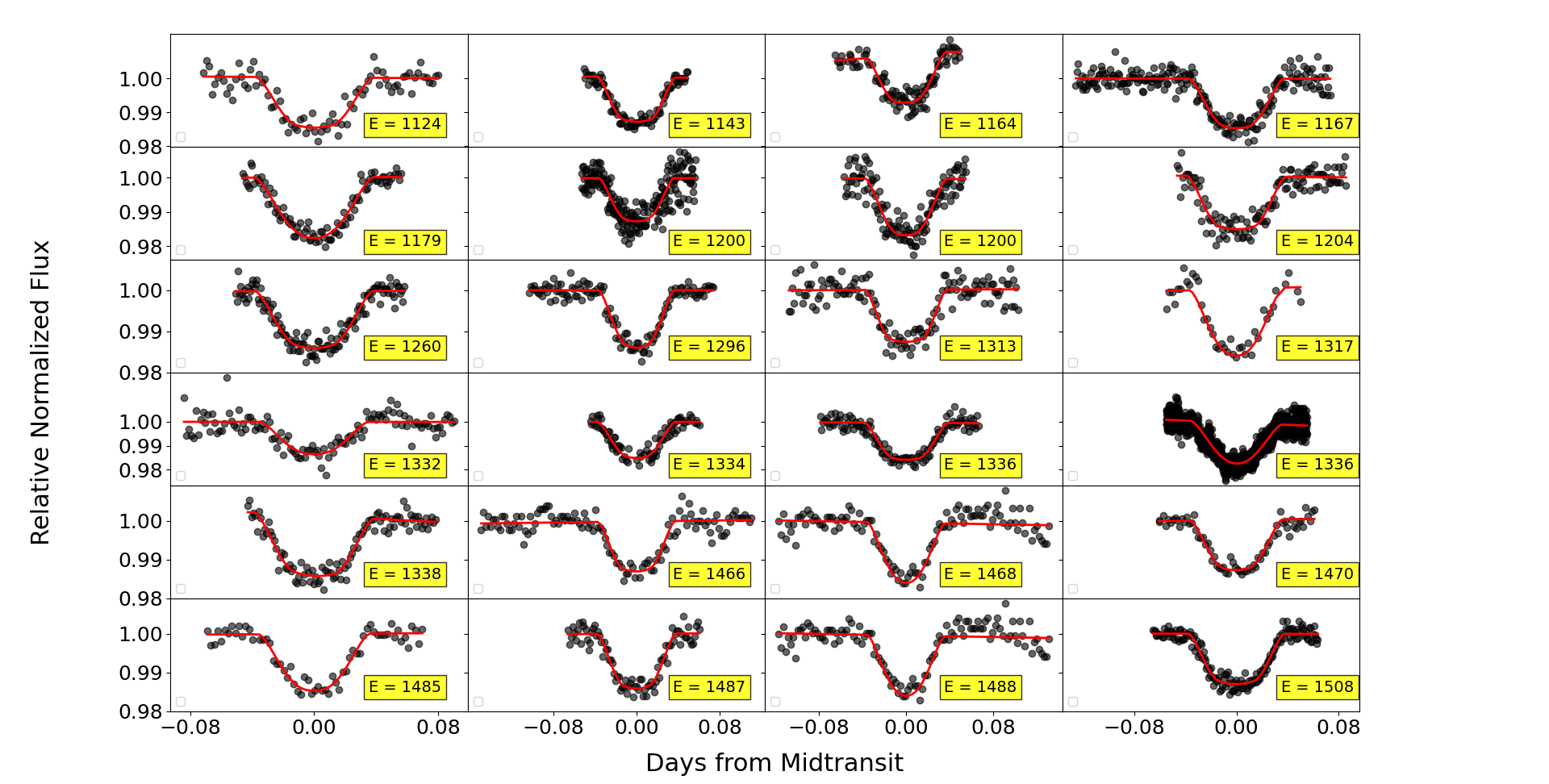}
    \caption{The normalized relative flux of TrES-2b as a function of the time (the offset from mid-transit time and in TDB-based BJD) of individual transit taken from ETD: the points are the data of raw flux, solid red lines are best-fit models for model flux, and E is the calculated epoch number.}
    \label{fig:ETD_light_curves1}
\end{figure}

\begin{figure}
    \centering
    \includegraphics[width=1.1\linewidth]{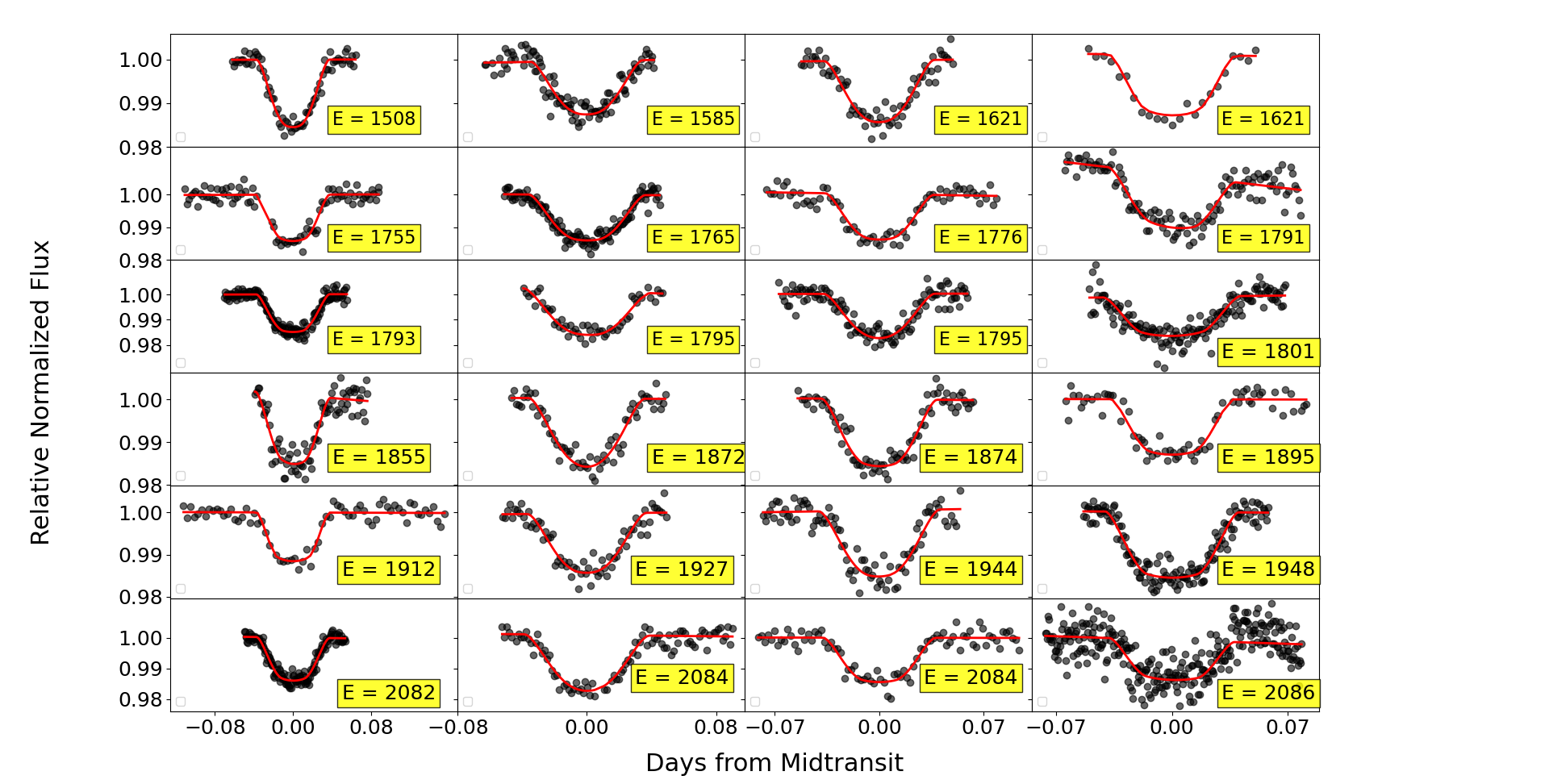}
    \caption{Same as the Figure \ref{fig:ETD_light_curves1} but for remaining epochs.}
    \label{fig:ETD_light_curves2}
\end{figure}

\begin{figure}
    \centering
    \includegraphics[width=1.1\linewidth]{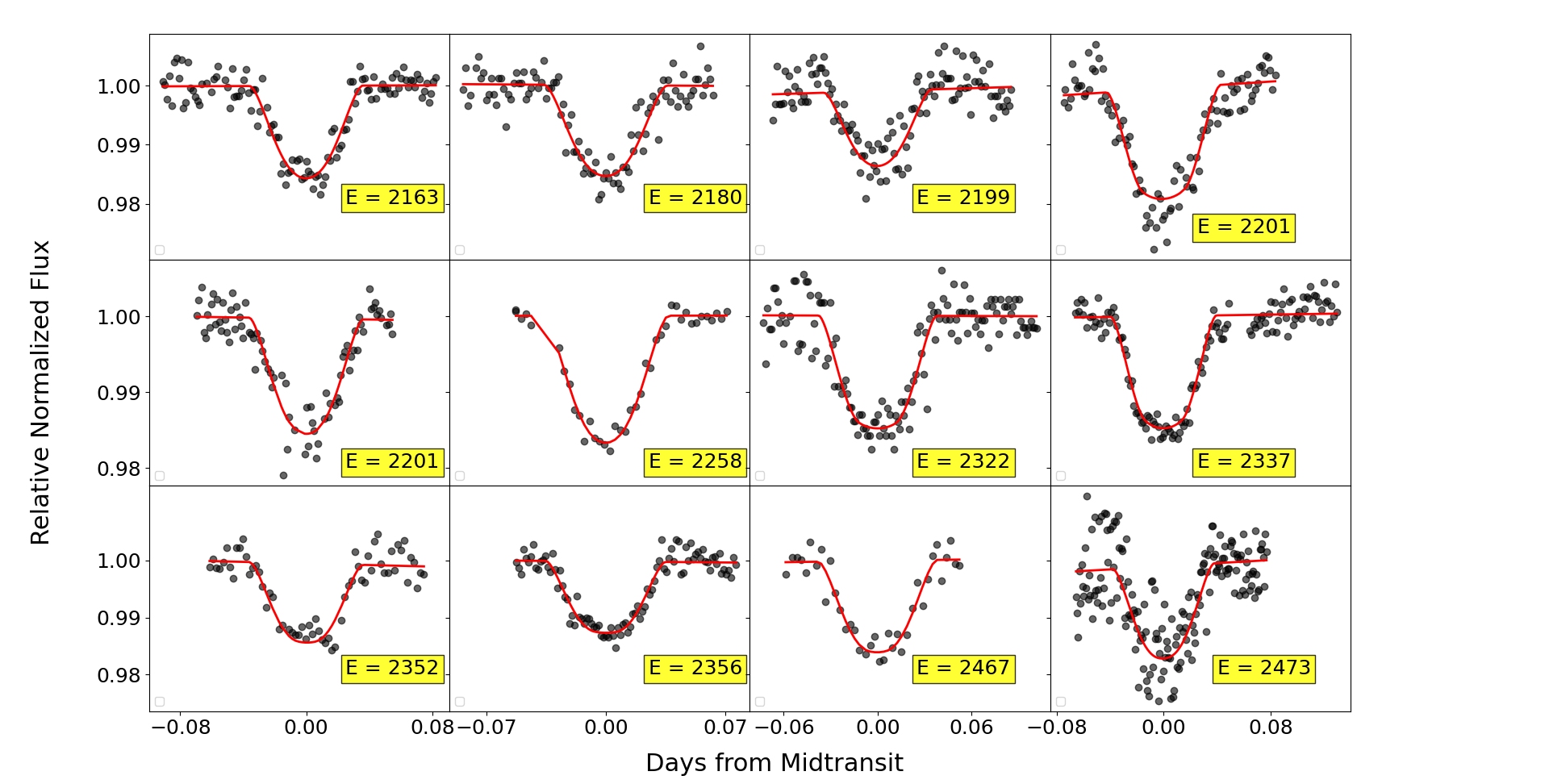}
    \caption{Same as the Figure \ref{fig:ETD_light_curves1} but for remaining epochs.}
    \label{fig:ETD_light_curves3}
\end{figure}

For all 64 TESS light curves of TrES-2b, we have represented the trend of the time series data, the OOT time-series data with the best-fitting GP model, and the detrended times series data of TrES-2 in Figures \ref{fig:ind_transits_sec40_2}--\ref{fig:ind_transits_sec40_7}. Their \texttt{TAP} model fits are shown in Figures \ref{fig:6}--\ref{fig:8}. The \texttt{TAP} model fits for all 60 light curves of ETD are shown in Figures \ref{fig:ETD_light_curves1}-\ref{fig:ETD_light_curves3} and the best-fit parameters for all these light curves of ETD i.e., $P$, ${T_m}$, ${R_p /R_\ast}$, $a/R_\ast$, {\it i}, $u_1$, $u_2$ at different epochs can be found in Table \ref{tb:lighcurve_model_TESS}.

%******************* Individual TESS Sector 2%*******************

\begin{center}
\small\addtolength{\tabcolsep}{2pt}
\begin{longtable*}{cp{2.8cm}cp{1.2cm}cp{1.8cm}cp{1.4cm}cp{1.5cm}cp{1.5cm}cp{1cm}c}
\caption{The Best-ﬁt Values of Parameters $P$, T$_{m}$, i, a/R$_\ast$, R$_p$/R$_\ast$, $u_1$, and $u_2$ for 60 ETD Transit Light Curves of TrES-2b using \texttt{TAP} \label{tb:lighcurve_model_TESS}}\\
\hline
\hline \multicolumn{1}{c}{Epoch} & \multicolumn{1}{c}{$P$} &\multicolumn{1}{c}{${T}_{m}$} & \multicolumn{1}{c}{$\it i$} & \multicolumn{1}{l}{$a/{R}_{\ast}$} & \multicolumn{1}{c}{R$_p$/R$_\ast$} & \multicolumn{1}{c}{u$_1$} & \multicolumn{1}{c}{u$_2$}\\
(E) & \ \ \ \ \ \ \ (days) &(BJD$_{\rm TDB}$) & \ \ (deg) &  &  &  & \\
\hline
\endfirsthead
		
1124 & $2.470630^{+0.000010}_{-0.0000099}$ &$2456734.6043^{+0.0017}_{-0.0016}$ & $83.97^{+0.21}_{-0.24}$ & $7.956^{+0.055}_{-0.054}$ & $0.127^{+0.013}_{-0.0098}$ &
$0.39^{+0.18}_{-0.18}$ & $0.22^{+0.20}_{-0.20}$\\
1143 & $2.4706303^{+0.0000099}_{-0.000010}$ &$2456781.5465^{+0.00098}_{-0.00097}$ & $83.98^{+0.15}_{-0.13}$ & $7.946^{+0.054}_{-0.054}$ & $0.1184^{+0.0050}_{-0.0048}$ &
$0.32^{+0.18}_{-0.17}$ & $0.24^{+0.20}_{-0.20}$\\
1164 & $2.470630^{+0.0000099}_{-0.0000099}$ &$2456833.4287^{+0.0014}_{-0.0015}$ & $83.86^{+0.19}_{-0.22}$ & $7.966^{+0.054}_{-0.055}$ & $0.124^{+0.011}_{-0.0076}$ &
$0.36^{+0.18}_{-0.18}$ & $0.24^{+0.20}_{-0.21}$\\
1167 & $2.4706302^{+0.000010}_{-0.0000099}$ &$2456840.8393^{+0.0011}_{-0.0011}$ & $83.81^{+0.17}_{-0.17}$ & $7.898^{+0.054}_{-0.054}$ & $0.1271^{+0.0073}_{-0.0057}$ &
$0.36^{+0.18}_{-0.18}$ & $0.23^{+0.20}_{-0.21}$\\
1179 & $2.470630^{+0.000010}_{-0.000010}$ &$2456870.4878^{+0.0011}_{-0.0011}$ & $83.55^{+0.19}_{-0.22}$ & $7.704^{+0.054}_{-0.055}$ & $0.142^{+0.014}_{-0.0092}$ &
$0.35^{+0.18}_{-0.17}$ & $0.28^{+0.20}_{-0.21}$\\
1200 & $2.4706301^{+0.000010}_{-0.000010}$ &$2456922.3733^{+0.0014}_{-0.0014}$ & $83.87^{+0.19}_{-0.20}$ & $7.965^{+0.055}_{-0.055}$ & $0.1179^{+0.0080}_{-0.0070}$ &
$0.37^{+0.18}_{-0.18}$ & $0.24^{+0.20}_{-0.20}$\\
1200 & $2.4706301^{+0.000010}_{-0.0000099}$ &$2456922.3734^{+0.0012}_{-0.0012}$ & $83.90^{+0.19}_{-0.20}$ & $7.967^{+0.055}_{-0.055}$ & $0.1353^{+0.0093}_{-0.0069}$ &
$0.38^{+0.18}_{-0.18}$ & $0.26^{+0.19}_{-0.20}$\\
1204 & $2.470630^{+0.000010}_{-0.0000099}$ &$2456932.2541^{+0.0015}_{-0.0015}$ & $84.07^{+0.21}_{-0.22}$ & $7.935^{+0.054}_{-0.055}$ & $0.1278^{+0.0089}_{-0.0078}$ &
$0.35^{+0.18}_{-0.18}$ & $0.22^{+0.20}_{-0.20}$\\
1260 & $2.4706303^{+0.000010}_{-0.000010}$ &$2457070.6080^{+0.0013}_{-0.0012}$ & $83.79^{+0.18}_{-0.19}$ & $7.744^{+0.054}_{-0.054}$ & $0.1231^{+0.0075}_{-0.0066}$ &
$0.40^{+0.18}_{-0.18}$ & $0.26^{+0.19}_{-0.20}$\\
1296 & $2.4706302^{+0.000010}_{-0.000010}$ &$2457159.5533^{+0.0011}_{-0.0011}$ & $83.85^{+0.18}_{-0.19}$ & $7.965^{+0.055}_{-0.055}$ & $0.1247^{+0.0076}_{-0.0056}$ &
$0.38^{+0.18}_{-0.18}$ & $0.25^{+0.20}_{-0.20}$\\
1313 & $2.4706304^{+0.000010}_{-0.0000099}$ &$2457201.55117^{+0.00093}_{-0.00093}$ & $84.05^{+0.15}_{-0.14}$ & $7.966^{+0.055}_{-0.054}$ & $0.1156^{+0.0048}_{-0.0044}$ &
$0.32^{+0.18}_{-0.17}$ & $0.20^{+0.20}_{-0.20}$\\
1317 & $2.470630^{+0.0000099}_{-0.0000099}$ &$2457211.4344^{+0.0021}_{-0.0021}$ & $83.69^{+0.26}_{-0.29}$ & $7.923^{+0.055}_{-0.054}$ & $0.137^{+0.019}_{-0.013}$ &
$0.36^{+0.18}_{-0.18}$ & $0.25^{+0.20}_{-0.21}$\\
1332 & $2.470630^{+0.0000099}_{-0.000010}$ &$2457248.4965^{+0.0037}_{-0.0034}$ & $83.68^{+0.33}_{-0.35}$ & $7.968^{+0.054}_{-0.055}$ & $0.126^{+0.026}_{-0.022}$ &
$0.41^{+0.19}_{-0.18}$ & $0.24^{+0.19}_{-0.20}$\\
1334 & $2.4706305^{+0.000010}_{-0.0000099}$ &$2457253.4357^{+0.0013}_{-0.0013}$ & $83.84^{+0.21}_{-0.25}$ & $7.915^{+0.054}_{-0.055}$ & $0.130^{+0.014}_{-0.0091}$ &
$0.40^{+0.18}_{-0.18}$ & $0.26^{+0.19}_{-0.20}$\\
1336 & $2.4706301^{+0.000010}_{-0.0000099}$ &$2457258.3748^{+0.0019}_{-0.0019}$ & $83.97^{+0.22}_{-0.24}$ & $7.894^{+0.054}_{-0.055}$ & $0.1292^{+0.0095}_{-0.0078}$ &
$0.37^{+0.18}_{-0.18}$ & $0.24^{+0.19}_{-0.20}$\\
1336 & $2.470630^{+0.000010}_{-0.0000099}$ &$2457258.3776^{+0.0010}_{-0.00098}$ & $83.40^{+0.21}_{-0.23}$ & $7.907^{+0.054}_{-0.054}$ & $0.157^{+0.019}_{-0.015}$ &
$0.38^{+0.18}_{-0.18}$ & $0.27^{+0.19}_{-0.21}$\\
1338 & $2.4706306^{+0.0000098}_{-0.0000099}$ &$2457263.3165^{+0.0012}_{-0.0012}$ & $84.06^{+0.19}_{-0.18}$ & $7.945^{+0.054}_{-0.055}$ & $0.1288^{+0.0062}_{-0.0060}$ &
$0.37^{+0.18}_{-0.18}$ & $0.25^{+0.19}_{-0.20}$\\
1466 & $2.4706301^{+0.0000098}_{-0.0000098}$ &$2457579.5559^{+0.0029}_{-0.0026}$ & $83.99^{+0.28}_{-0.31}$ & $7.963^{+0.055}_{-0.054}$ & $0.118^{+0.014}_{-0.011}$ &
$0.37^{+0.17}_{-0.18}$ & $0.24^{+0.20}_{-0.21}$\\
1468 & $2.470630^{+0.000010}_{-0.0000099}$ &$2457584.4955^{+0.0019}_{-0.0019}$ & $83.65^{+0.26}_{-0.30}$ & $7.970^{+0.055}_{-0.055}$ & $0.137^{+0.024}_{-0.017}$ &
$0.37^{+0.18}_{-0.18}$ & $0.25^{+0.19}_{-0.20}$\\
1470 & $2.4706302^{+0.0000099}_{-0.0000099}$ &$2457589.43882^{+0.00090}_{-0.00097}$ & $83.84^{+0.14}_{-0.14}$ & $7.966^{+0.055}_{-0.054}$ & $0.1195^{+0.0068}_{-0.0053}$ &
$0.35^{+0.18}_{-0.17}$ & $0.23^{+0.20}_{-0.20}$\\
1485 & $2.4706301^{+0.000010}_{-0.0000099}$ &$2457626.4957^{+0.0019}_{-0.0017}$ & $83.75^{+0.23}_{-0.28}$ & $7.968^{+0.055}_{-0.055}$ & $0.129^{+0.018}_{-0.012}$ &
$0.38^{+0.18}_{-0.18}$ & $0.25^{+0.19}_{-0.21}$\\
1487 & $2.470630^{+0.000010}_{-0.0000099}$ &$2457631.4370^{+0.0010}_{-0.0010}$ & $83.95^{+0.15}_{-0.15}$ & $7.927^{+0.054}_{-0.054}$ & $0.1237^{+0.0055}_{-0.0051}$ &
$0.36^{+0.18}_{-0.18}$ & $0.24^{+0.19}_{-0.20}$\\
1488 & $2.470630^{+0.000010}_{-0.0000099}$ &$2457584.4955^{+0.0019}_{-0.0019}$ & $83.65^{+0.26}_{-0.30}$ & $7.970^{+0.055}_{-0.055}$ & $0.137^{+0.024}_{-0.017}$ &
$0.37^{+0.18}_{-0.18}$ & $0.25^{+0.19}_{-0.20}$\\
1508 & $2.470631^{+0.000010}_{-0.0000099}$ &$2457683.32104^{+0.00068}_{-0.00076}$ & $83.79^{+0.14}_{-0.15}$ & $7.924^{+0.054}_{-0.054}$ & $0.1320^{+0.0066}_{-0.0049}$ &
$0.39^{+0.17}_{-0.18}$ & $0.25^{+0.19}_{-0.20}$\\
1508 & $2.4706306^{+0.0000098}_{-0.0000099}$ &$2457683.32114^{+0.00055}_{-0.00055}$ & $84.03^{+0.13}_{-0.12}$ & $7.936^{+0.054}_{-0.054}$ & $0.1179^{+0.0030}_{-0.0032}$ &
$0.37^{+0.17}_{-0.17}$ & $0.26^{+0.19}_{-0.20}$\\
1585 & $2.470630^{+0.000010}_{-0.000010}$ &$2457873.5594^{+0.0021}_{-0.0021}$ & $83.68^{+0.23}_{-0.29}$ & $7.976^{+0.054}_{-0.055}$ & $0.119^{+0.016}_{-0.010}$ &
$0.38^{+0.18}_{-0.18}$ & $0.25^{+0.19}_{-0.20}$\\
1621 & $2.470630^{+0.000010}_{-0.0000099}$ &$2457962.50079^{+0.00046}_{-0.00047}$ & $83.78^{+0.12}_{-0.10}$ & $7.979^{+0.054}_{-0.054}$ & $0.1198^{+0.0039}_{-0.0035}$ &
$0.23^{+0.17}_{-0.14}$ & $0.23^{+0.21}_{-0.21}$\\
1621 & $2.4706301^{+0.000010}_{-0.0000099}$ &$2457962.5012^{+0.0021}_{-0.0018}$ & $84.10^{+0.22}_{-0.23}$ & $7.951^{+0.055}_{-0.054}$ & $0.1209^{+0.0093}_{-0.0092}$ &
$0.38^{+0.18}_{-0.18}$ & $0.25^{+0.19}_{-0.20}$\\
1755 & $2.470630^{+0.0000098}_{-0.0000098}$ &$2458293.5600^{+0.0014}_{-0.0016}$ & $83.98^{+0.23}_{-0.20}$ & $7.962^{+0.055}_{-0.054}$ & $0.1238^{+0.0068}_{-0.0056}$ &
$0.36^{+0.18}_{-0.18}$ & $0.24^{+0.20}_{-0.20}$\\
1765 & $2.470630^{+0.000010}_{-0.0000099}$ &$2458318.26641^{+0.00079}_{-0.00076}$ & $83.79^{+0.15}_{-0.15}$ & $7.963^{+0.054}_{-0.054}$ & $0.1255^{+0.0058}_{-0.0051}$ &
$0.36^{+0.18}_{-0.17}$ & $0.27^{+0.19}_{-0.20}$\\
1776 & $2.470630^{+0.000010}_{-0.000010}$ &$2458345.4453^{+0.0013}_{-0.0012}$ & $83.80^{+0.17}_{-0.21}$ & $7.972^{+0.054}_{-0.055}$ & $0.125^{+0.012}_{-0.0088}$ &
$0.35^{+0.18}_{-0.18}$ & $0.23^{+0.20}_{-0.21}$\\
1791 & $2.470630^{+0.000010}_{-0.000010}$ &$2458382.5030^{+0.0015}_{-0.0016}$ & $84.06^{+0.21}_{-0.25}$ & $7.977^{+0.055}_{-0.054}$ & $0.131^{+0.012}_{-0.0093}$ &
$0.38^{+0.18}_{-0.18}$ & $0.21^{+0.20}_{-0.21}$\\
1793 & $2.4706303^{+0.000010}_{-0.0000099}$ &$2458387.44555^{+0.00051}_{-0.00054}$ & $83.79^{+0.13}_{-0.11}$ & $7.858^{+0.054}_{-0.054}$ & $0.1270^{+0.0034}_{-0.0033}$ &
$0.39^{+0.18}_{-0.18}$ & $0.23^{+0.20}_{-0.21}$\\
1795 & $2.470630^{+0.000010}_{-0.0000099}$ &$2458392.3847^{+0.0025}_{-0.0032}$ & $83.66^{+0.31}_{-0.30}$ & $7.703^{+0.054}_{-0.055}$ & $0.140^{+0.017}_{-0.013}$ &
$0.40^{+0.18}_{-0.18}$ & $0.24^{+0.19}_{-0.21}$\\
1795 & $2.470632^{+0.000010}_{-0.0000099}$ &$2458392.3869^{+0.0017}_{-0.0015}$ & $83.67^{+0.24}_{-0.28}$ & $7.924^{+0.055}_{-0.054}$ & $0.145^{+0.020}_{-0.012}$ &
$0.31^{+0.18}_{-0.17}$ & $0.20^{+0.21}_{-0.21}$\\
1801 & $2.4706296^{+0.0000099}_{-0.0000099}$ &$2458407.2101^{+0.0020}_{-0.0018}$ & $84.05^{+0.26}_{-0.28}$ & $7.639^{+0.055}_{-0.055}$ & $0.1263^{+0.0074}_{-0.0082}$ &
$0.34^{+0.18}_{-0.18}$ & $0.30^{+0.19}_{-0.20}$\\
1855 & $2.4706302^{+0.000010}_{-0.0000099}$ &$2458540.6211^{+0.0013}_{-0.0015}$ & $83.99^{+0.22}_{-0.21}$ & $7.968^{+0.055}_{-0.055}$ & $0.1327^{+0.0082}_{-0.0083}$ &
$0.40^{+0.18}_{-0.18}$ & $0.23^{+0.19}_{-0.20}$\\
1872 & $2.470630^{+0.000010}_{-0.0000099}$ &$2458582.62458^{+0.00082}_{-0.00084}$ & $83.62^{+0.17}_{-0.20}$ & $7.968^{+0.054}_{-0.054}$ & $0.140^{+0.014}_{-0.0085}$ &
$0.34^{+0.18}_{-0.17}$ & $0.25^{+0.19}_{-0.20}$\\
1874 & $2.4706299^{+0.000010}_{-0.0000099}$ &$2458587.5642^{+0.0011}_{-0.0010}$ & $83.75^{+0.17}_{-0.18}$ & $7.753^{+0.054}_{-0.054}$ & $0.1308^{+0.0080}_{-0.0069}$ &
$0.33^{+0.18}_{-0.17}$ & $0.26^{+0.20}_{-0.20}$\\
1895 & $2.470630^{+0.000010}_{-0.0000099}$ & $2458639.4487^{+0.0011}_{-0.0012}$ & $83.98^{+0.16}_{-0.15}$ & $7.967^{+0.054}_{-0.054}$ & $0.1178^{+0.0067}_{-0.0053}$ &
$0.33^{+0.18}_{-0.17}$ & $0.21^{+0.20}_{-0.20}$\\
1912 & $2.470630^{+0.000010}_{-0.0000099}$ &$2458681.4471^{+0.0013}_{-0.0013}$ & $84.07^{+0.17}_{-0.17}$ & $7.958^{+0.054}_{-0.055}$ & $0.1108^{+0.0056}_{-0.0057}$ &
$0.38^{+0.18}_{-0.18}$ & $0.20^{+0.20}_{-0.20}$\\
1927 & $2.4706302^{+0.000010}_{-0.0000099}$ &$2458718.5081^{+0.0013}_{-0.0013}$ & $83.61^{+0.19}_{-0.22}$ & $7.797^{+0.055}_{-0.054}$ & $0.129^{+0.011}_{-0.0074}$ &
$0.45^{+0.18}_{-0.18}$ & $0.26^{+0.18}_{-0.19}$\\
1944 &$2.470630^{+0.000010}_{-0.0000099}$ & $2458760.5092^{+0.0021}_{-0.0019}$ & $83.28^{+0.25}_{-0.26}$ & $7.315^{+0.054}_{-0.055}$ & $0.134^{+0.013}_{-0.010}$ &
$0.42^{+0.18}_{-0.19}$ & $0.27^{+0.19}_{-0.20}$\\
1948 & $2.470630^{+0.000010}_{-0.000010}$ &$2458770.39015^{+0.00070}_{-0.00071}$ & $83.906^{+0.040}_{-0.040}$ & $7.690^{+0.048}_{-0.049}$ & $0.1279^{+0.0039}_{-0.0035}$ &
$0.33^{+0.16}_{-0.16}$ & $0.17^{+0.18}_{-0.18}$\\
2082 & $2.470630^{+0.0000099}_{-0.0000099}$ &$2459101.45322^{+0.00070}_{-0.00070}$ & $83.95^{+0.13}_{-0.13}$ & $7.966^{+0.054}_{-0.054}$ & $0.1238^{+0.0043}_{-0.0041}$ &
$0.37^{+0.17}_{-0.17}$ & $0.26^{+0.19}_{-0.20}$\\
2084 & $2.4706301^{+0.000010}_{-0.0000099}$ &$2459106.3931^{+0.0012}_{-0.0013}$ & $83.51^{+0.21}_{-0.24}$ & $7.688^{+0.055}_{-0.054}$ & $0.146^{+0.015}_{-0.011}$ &
$0.33^{+0.18}_{-0.17}$ & $0.28^{+0.20}_{-0.21}$\\
2084 & $2.470630^{+0.0000098}_{-0.0000099}$ &$2459106.3956^{+0.0013}_{-0.0013}$ & $83.88^{+0.21}_{-0.21}$ & $7.896^{+0.054}_{-0.054}$ & $0.1256^{+0.0088}_{-0.0072}$ &
$0.36^{+0.18}_{-0.18}$ & $0.24^{+0.19}_{-0.21}$\\
2086 & $2.4706303^{+0.000010}_{-0.0000099}$ &$2459111.3331^{+0.0019}_{-0.0020}$ & $84.09^{+0.22}_{-0.23}$ & $7.964^{+0.055}_{-0.054}$ & $0.117^{+0.010}_{-0.011}$ &
$0.31^{+0.18}_{-0.17}$ & $0.23^{+0.20}_{-0.21}$\\
2163 & $2.470630^{+0.0000099}_{-0.0000098}$ &$2459301.5740^{+0.0019}_{-0.0019}$ & $83.65^{+0.24}_{-0.27}$ & $7.974^{+0.054}_{-0.054}$ & $0.138^{+0.020}_{-0.014}$ &
$0.35^{+0.18}_{-0.18}$ & $0.28^{+0.19}_{-0.21}$\\
2180 & $2.470630^{+0.000010}_{-0.0000099}$ &$2459343.5694^{+0.0019}_{-0.0019}$ & $83.703^{+0.041}_{-0.041}$ & $7.984^{+0.054}_{-0.054}$ & $0.136^{+0.011}_{-0.011}$ &
$0.37^{+0.18}_{-0.18}$ & $0.31^{+0.19}_{-0.20}$\\
		2199 & $2.4706301^{+0.0000099}_{-0.000010}$&$2459390.5136^{+0.0036}_{-0.0031}$ & $83.57^{+0.32}_{-0.33}$ & $7.971^{+0.055}_{-0.055}$ & $0.126^{+0.029}_{-0.027}$ &
$0.38^{+0.18}_{-0.18}$ & $0.26^{+0.19}_{-0.20}$\\
2201 & $2.470630^{+0.000010}_{-0.0000099}$ &$2459395.4501^{+0.0027}_{-0.0027}$ & $83.96^{+0.30}_{-0.30}$ & $7.598^{+0.055}_{-0.054}$ & $0.140^{+0.014}_{-0.014}$ &
$0.39^{+0.18}_{-0.17}$ & $0.27^{+0.19}_{-0.20}$\\
2201 & $2.4706301^{+0.000010}_{-0.0000099}$ &$2459395.4555^{+0.0018}_{-0.0015}$ & $83.712^{+0.41}_{-0.41}$ & $7.959^{+0.054}_{-0.053}$ & $0.1333^{+0.0090}_{-0.0085}$ &
$0.39^{+0.18}_{-0.18}$ & $0.25^{+0.19}_{-0.20}$\\
2258 & $2.4706302^{+0.000010}_{-0.000010}$ &$2459536.28055^{+0.00072}_{-0.00066}$ & $83.72^{+0.16}_{-0.17}$ & $7.963^{+0.054}_{-0.054}$ & $0.1404^{+0.0093}_{-0.0063}$ &
$0.36^{+0.18}_{-0.18}$ & $0.28^{+0.19}_{-0.20}$\\
2322 & $2.470630^{+0.000010}_{-0.0000099}$ &$2459694.39963^{+0.00091}_{-0.00089}$ & $83.97^{+0.16}_{-0.15}$ & $7.965^{+0.054}_{-0.054}$ & $0.1268^{+0.0053}_{-0.0049}$ &
$0.34^{+0.18}_{-0.17}$ & $0.26^{+0.20}_{-0.20}$\\
2337 & $2.470630^{+0.0000098}_{-0.0000099}$ &$2459731.4600^{+0.0012}_{-0.0013}$ & $83.855^{+0.041}_{-0.041}$ & $7.712^{+0.051}_{-0.052}$ & $0.1265^{+0.0067}_{-0.0068}$ &
$0.42^{+0.17}_{-0.17}$ & $0.27^{+0.19}_{-0.19}$\\
2352 & $2.470637^{+0.000010}_{-0.0000099}$ &$2459768.5183^{+0.0032}_{-0.0030}$ & $83.76^{+0.30}_{-0.37}$ & $7.712^{+0.055}_{-0.055}$ & $0.125^{+0.026}_{-0.020}$ &
$0.36^{+0.18}_{-0.17}$ & $0.24^{+0.20}_{-0.21}$\\
2356 & $2.4706301^{+0.000010}_{-0.0000099}$ &$2459778.4009^{+0.0013}_{-0.0012}$ & $83.715^{+0.042}_{-0.041}$ & $7.960^{+0.052}_{-0.052}$ & $0.1177^{+0.0086}_{-0.0082}$ &
$0.30^{+0.18}_{-0.16}$ & $0.21^{+0.20}_{-0.20}$\\
2467 & $2.470630^{+0.0000099}_{-0.0000099}$ &$2460052.6399^{+0.0016}_{-0.0017}$ & $83.68^{+0.22}_{-0.25}$ & $7.799^{+0.054}_{-0.054}$ & $0.134^{+0.014}_{-0.0097}$ &
$0.34^{+0.18}_{-0.17}$ & $0.27^{+0.19}_{-0.21}$\\
2473 & $2.4706302^{+0.000010}_{-0.0000099}$ &$2460067.4613^{+0.0018}_{-0.0018}$ & $83.57^{+0.25}_{-0.28}$ & $7.727^{+0.054}_{-0.054}$ & $0.136^{+0.019}_{-0.014}$ &
$0.36^{+0.18}_{-0.18}$ & $0.25^{+0.20}_{-0.21}$\\
\hline
\end{longtable*}
\end{center}

\bibliography{tres2}{}

\begin{thebibliography}{}
\expandafter\ifx\csname natexlab\endcsname\relax\def\natexlab#1{#1}\fi
\providecommand{\url}[1]{\href{#1}{#1}}
\providecommand{\dodoi}[1]{doi:~\href{http://doi.org/#1}{\nolinkurl{#1}}}
\providecommand{\doeprint}[1]{\href{http://ascl.net/#1}{\nolinkurl{http://ascl.net/#1}}}
\providecommand{\doarXiv}[1]{\href{https://arxiv.org/abs/#1}{\nolinkurl{https://arxiv.org/abs/#1}}}

\bibitem[{{A-thano} {et~al.}(2022){A-thano}, {Jiang}, {Awiphan}, {Rattanamala}, {Su}, {Hengpiya}, {Sariya}, {Yeh}, {Shlyapnikov}, {Gorbachev}, {Rublevski}, {Mannaday}, {Thakur}, {Sahu}, {Mkrtichian}, \& {Griv}}]{2022AJ....163...77A}
{A-thano}, N., {Jiang}, I.-G., {Awiphan}, S., {et~al.} 2022, \aj, 163, 77, \dodoi{10.3847/1538-3881/ac416d}

\bibitem[{{Adams} {et~al.}(2024){Adams}, {Jackson}, {Sickafoose}, {Morgenthaler}, {Worters}, {Stubbers}, {Carson}, {Bhure}, {Dekeyser}, {Huang}, \& {Weinberg}}]{2024arXiv240407339A}
{Adams}, E.~R., {Jackson}, B., {Sickafoose}, A.~A., {et~al.} 2024, arXiv e-prints, arXiv:2404.07339, \dodoi{10.48550/arXiv.2404.07339}

\bibitem[{Akaike(1974)}]{1100705}
Akaike, H. 1974, IEEE Transactions on Automatic Control, 19, 716, \dodoi{10.1109/TAC.1974.1100705}

\bibitem[{{Albrow} {et~al.}(2000){Albrow}, {Beaulieu}, {Caldwell}, {DePoy}, {Dominik}, {Gaudi}, {Gould}, {Greenhill}, {Hill}, {Kane}, {Martin}, {Menzies}, {Naber}, {Pogge}, {Pollard}, {Sackett}, {Sahu}, {Vermaak}, {Watson}, {Williams}, \& {PLANET Collaboration}}]{2000ApJ...535..176A}
{Albrow}, M.~D., {Beaulieu}, J.~P., {Caldwell}, J.~A.~R., {et~al.} 2000, \apj, 535, 176, \dodoi{10.1086/308842}

\bibitem[{{Almenara} {et~al.}(2016){Almenara}, {D{\'\i}az}, {Bonfils}, \& {Udry}}]{2016A&A...595L...5A}
{Almenara}, J.~M., {D{\'\i}az}, R.~F., {Bonfils}, X., \& {Udry}, S. 2016, \aap, 595, L5, \dodoi{10.1051/0004-6361/201629770}

\bibitem[{{Applegate}(1992)}]{1992ApJ...385..621A}
{Applegate}, J.~H. 1992, \apj, 385, 621, \dodoi{10.1086/170967}

\bibitem[{{Applegate} \& {Patterson}(1987)}]{1987ApJ...322L..99A}
{Applegate}, J.~H., \& {Patterson}, J. 1987, \apjl, 322, L99, \dodoi{10.1086/185044}

\bibitem[{Ballard {et~al.}(2010)Ballard, Christiansen, Charbonneau, Deming, Holman, Fabrycky, A'Hearn, Wellnitz, Barry, Kuchner, Livengood, Hewagama, Sunshine, Hampton, Lisse, Seager, \& Veverka}]{Ballard_2010}
Ballard, S., Christiansen, J.~L., Charbonneau, D., {et~al.} 2010, The Astrophysical Journal, 716, 1047, \dodoi{10.1088/0004-637X/716/2/1047}

\bibitem[{{Baluev} {et~al.}(2015){Baluev}, {Sokov}, {Shaidulin}, {Sokova}, {Jones}, {Tuomi}, {Anglada-Escud{\'e}}, {Benni}, {Colazo}, {Schneiter}, {D'Angelo}, {Burdanov}, {Fern{\'a}ndez-Laj{\'u}s}, {Ba{\c{s}}t{\"u}rk}, {Hentunen}, \& {Shadick}}]{2015MNRAS.450.3101B}
{Baluev}, R.~V., {Sokov}, E.~N., {Shaidulin}, V.~S., {et~al.} 2015, \mnras, 450, 3101, \dodoi{10.1093/mnras/stv788}

\bibitem[{Baluev {et~al.}(2020)Baluev, Sokov, Hoyer, Huitson, da~Silva, Evans, Sokova, Knight, \& Shaidulin}]{Baluev_2020}
Baluev, R.~V., Sokov, E.~N., Hoyer, S., {et~al.} 2020, Monthly Notices of the Royal Astronomical Society: Letters, 496, L11–L15, \dodoi{10.1093/mnrasl/slaa069}

\bibitem[{{Barker}(2019)}]{2019psce.confE..38B}
{Barker}, A. 2019, in KITP Conference: Planet-Star Connections in the Era of TESS and Gaia, 38

\bibitem[{Barker(2020)}]{Barker_2020}
Barker, A.~J. 2020, Monthly Notices of the Royal Astronomical Society, 498, 2270–2294, \dodoi{10.1093/mnras/staa2405}

\bibitem[{{Barker} {et~al.}(2011){Barker}, {Douglas}, {Jacobson}, {McClelland}, {Ilgen}, {Khosh}, {Lehn}, \& {Trainor}}]{2011AGUFMGC51F1078B}
{Barker}, A.~J., {Douglas}, T.~A., {Jacobson}, A.~D., {et~al.} 2011, in AGU Fall Meeting Abstracts, Vol. 2011, GC51F--1078

\bibitem[{Barker \& Ogilvie(2010)}]{10.1111/j.1365-2966.2010.16400.x}
Barker, A.~J., \& Ogilvie, G.~I. 2010, Monthly Notices of the Royal Astronomical Society, 404, 1849, \dodoi{10.1111/j.1365-2966.2010.16400.x}

\bibitem[{Basturk {et~al.}(2023)Basturk, Southworth, Yalcinkaya, Mancini, Esmer, Tekin, Tezcan, Evans, Tezcan, Bruni, \& Yesilyaprak}]{10.1093/mnras/stad248}
Basturk, O., Southworth, J., Yalcinkaya, S., {et~al.} 2023, Monthly Notices of the Royal Astronomical Society, 521, 1200, \dodoi{10.1093/mnras/stad248}

\bibitem[{{Bean}(2009)}]{2009A&A...506..369B}
{Bean}, J.~L. 2009, \aap, 506, 369, \dodoi{10.1051/0004-6361/200912030}

\bibitem[{{Birkby} {et~al.}(2014){Birkby}, {Cappetta}, {Cruz}, {Koppenhoefer}, {Ivanyuk}, {Mustill}, {Hodgkin}, {Pinfield}, {Sip{\H{o}}cz}, {Kov{\'a}cs}, {Saglia}, {Pavlenko}, {Barrado}, {Bayo}, {Campbell}, {Catalan}, {Fossati}, {G{\'a}lvez-Ortiz}, {Kenworthy}, {Lillo-Box}, {Mart{\'\i}n}, {Mislis}, {de Mooij}, {Nefs}, {Snellen}, {Stoev}, {Zendejas}, {del Burgo}, {Barnes}, {Goulding}, {Haswell}, {Kuznetsov}, {Lodieu}, {Murgas}, {Palle}, {Solano}, {Steele}, \& {Tata}}]{2014MNRAS.440.1470B}
{Birkby}, J.~L., {Cappetta}, M., {Cruz}, P., {et~al.} 2014, \mnras, 440, 1470, \dodoi{10.1093/mnras/stu343}

\bibitem[{{Blecic} {et~al.}(2014){Blecic}, {Harrington}, {Madhusudhan}, {Stevenson}, {Hardy}, {Cubillos}, {Hardin}, {Bowman}, {Nymeyer}, {Anderson}, {Hellier}, {Smith}, \& {Collier Cameron}}]{2014ApJ...781..116B}
{Blecic}, J., {Harrington}, J., {Madhusudhan}, N., {et~al.} 2014, \apj, 781, 116, \dodoi{10.1088/0004-637X/781/2/116}

\bibitem[{{Bonomo} {et~al.}(2017){Bonomo}, {Desidera}, {Benatti}, {Borsa}, {Crespi}, {Damasso}, {Lanza}, {Sozzetti}, {Lodato}, {Marzari}, {Boccato}, {Claudi}, {Cosentino}, {Covino}, {Gratton}, {Maggio}, {Micela}, {Molinari}, {Pagano}, {Piotto}, {Poretti}, {Smareglia}, {Affer}, {Biazzo}, {Bignamini}, {Esposito}, {Giacobbe}, {H{\'e}brard}, {Malavolta}, {Maldonado}, {Mancini}, {Martinez Fiorenzano}, {Masiero}, {Nascimbeni}, {Pedani}, {Rainer}, \& {Scandariato}}]{2017A&A...602A.107B}
{Bonomo}, A.~S., {Desidera}, S., {Benatti}, S., {et~al.} 2017, \aap, 602, A107, \dodoi{10.1051/0004-6361/201629882}

\bibitem[{{Bouchy} {et~al.}(2005){Bouchy}, {Udry}, {Mayor}, {Moutou}, {Pont}, {Iribarne}, {da Silva}, {Ilovaisky}, {Queloz}, {Santos}, {S{\'e}gransan}, \& {Zucker}}]{2005A&A...444L..15B}
{Bouchy}, F., {Udry}, S., {Mayor}, M., {et~al.} 2005, \aap, 444, L15, \dodoi{10.1051/0004-6361:200500201}

\bibitem[{Bouma {et~al.}(2020)Bouma, Winn, Howard, Howell, Isaacson, Knutson, \& Matson}]{Bouma_2020}
Bouma, L.~G., Winn, J.~N., Howard, A.~W., {et~al.} 2020, The Astrophysical Journal Letters, 893, L29, \dodoi{10.3847/2041-8213/ab8563}

\bibitem[{{Bouma} {et~al.}(2019){Bouma}, {Winn}, {Baxter}, {Bhatti}, {Dai}, {Daylan}, {D{\'e}sert}, {Hill}, {Kane}, {Stassun}, {Villasenor}, {Ricker}, {Vanderspek}, {Latham}, {Seager}, {Jenkins}, {Berta-Thompson}, {Col{\'o}n}, {Fausnaugh}, {Glidden}, {Guerrero}, {Rodriguez}, {Twicken}, \& {Wohler}}]{2019AJ....157..217B}
{Bouma}, L.~G., {Winn}, J.~N., {Baxter}, C., {et~al.} 2019, \aj, 157, 217, \dodoi{10.3847/1538-3881/ab189f}

\bibitem[{Buchner(2014)}]{Buchner_2014}
Buchner, J. 2014, Statistics and Computing, 26, 383–392, \dodoi{10.1007/s11222-014-9512-y}

\bibitem[{Caldwell {et~al.}(2020)Caldwell, Tenenbaum, Twicken, Jenkins, Ting, Smith, Hedges, Fausnaugh, Rose, \& Burke}]{Caldwell_2020}
Caldwell, D.~A., Tenenbaum, P., Twicken, J.~D., {et~al.} 2020, Research Notes of the AAS, 4, 201, \dodoi{10.3847/2515-5172/abc9b3}

\bibitem[{{Carter} \& {Winn}(2009)}]{2009ApJ...704...51C}
{Carter}, J.~A., \& {Winn}, J.~N. 2009, \apj, 704, 51, \dodoi{10.1088/0004-637X/704/1/51}

\bibitem[{{Charbonneau} {et~al.}(2005){Charbonneau}, {Allen}, {Megeath}, {Torres}, {Alonso}, {Brown}, {Gilliland}, {Latham}, {Mandushev}, {O'Donovan}, \& {Sozzetti}}]{2005ApJ...626..523C}
{Charbonneau}, D., {Allen}, L.~E., {Megeath}, S.~T., {et~al.} 2005, \apj, 626, 523, \dodoi{10.1086/429991}

\bibitem[{Christian {et~al.}(2009)Christian, Gibson, Simpson, Street, Skillen, Pollacco, Collier~Cameron, Joshi, Keenan, Stempels, Haswell, Horne, Anderson, Bentley, Bouchy, Clarkson, Enoch, Hebb, Hébrard, Hellier, Irwin, Kane, Lister, Loeillet, Maxted, Mayor, McDonald, Moutou, Norton, Parley, Pont, Queloz, Ryans, Smalley, Smith, Todd, Udry, West, Wheatley, \& Wilson}]{10.1111/j.1365-2966.2008.14164.x}
Christian, D.~J., Gibson, N.~P., Simpson, E.~K., {et~al.} 2009, Monthly Notices of the Royal Astronomical Society, 392, 1585, \dodoi{10.1111/j.1365-2966.2008.14164.x}

\bibitem[{{Christiansen} {et~al.}(2011){Christiansen}, {Ballard}, {Charbonneau}, {Deming}, {Holman}, \& {Madhusudhan}}]{2011ApJ...726...94C}
{Christiansen}, J.~L., {Ballard}, S., {Charbonneau}, D., {et~al.} 2011, \apj, 726, 94, \dodoi{10.1088/0004-637X/726/2/94}

\bibitem[{{Claret}(2017)}]{2017A&A...600A..30C}
{Claret}, A. 2017, \aap, 600, A30, \dodoi{10.1051/0004-6361/201629705}

\bibitem[{{Claret} \& {Bloemen}(2011)}]{2011A&A...529A..75C}
{Claret}, A., \& {Bloemen}, S. 2011, \aap, 529, A75, \dodoi{10.1051/0004-6361/201116451}

\bibitem[{{Cloutier} \& {Triaud}(2016)}]{2016MNRAS.462.4018C}
{Cloutier}, R., \& {Triaud}, A. H.~M.~J. 2016, \mnras, 462, 4018, \dodoi{10.1093/mnras/stw1953}

\bibitem[{{Collins} {et~al.}(2017){Collins}, {Kielkopf}, \& {Stassun}}]{2017AJ....153...78C}
{Collins}, K.~A., {Kielkopf}, J.~F., \& {Stassun}, K.~G. 2017, \aj, 153, 78, \dodoi{10.3847/1538-3881/153/2/78}

\bibitem[{{Counselman}(1973)}]{1973ApJ...180..307C}
{Counselman}, Charles~C., I. 1973, \apj, 180, 307, \dodoi{10.1086/151964}

\bibitem[{{Croll} {et~al.}(2010){Croll}, {Jayawardhana}, {Fortney}, {Lafreni{\`e}re}, \& {Albert}}]{2010ApJ...718..920C}
{Croll}, B., {Jayawardhana}, R., {Fortney}, J.~J., {Lafreni{\`e}re}, D., \& {Albert}, L. 2010, \apj, 718, 920, \dodoi{10.1088/0004-637X/718/2/920}

\bibitem[{{Czesla} {et~al.}(2019){Czesla}, {Terzenbach}, {Wichmann}, \& {Schmitt}}]{2019A&A...623A.107C}
{Czesla}, S., {Terzenbach}, S., {Wichmann}, R., \& {Schmitt}, J.~H.~M.~M. 2019, \aap, 623, A107, \dodoi{10.1051/0004-6361/201834516}

\bibitem[{Damasso {et~al.}(2015)Damasso, Biazzo, Bonomo, Desidera, Lanza, Nascimbeni, Esposito, Scandariato, Sozzetti, Cosentino, Gratton, Malavolta, Rainer, Gandolfi, Poretti, Zanmar~Sanchez, Ribas, Santos, Affer, Andreuzzi, Barbieri, Bedin, Benatti, Bernagozzi, Bertolini, Bonavita, Borsa, Borsato, Boschin, Calcidese, Carbognani, Cenadelli, Christille, Claudi, Covino, Cunial, Giacobbe, Granata, Harutyunyan, Lattanzi, Leto, Libralato, Lodato, Lorenzi, Mancini, Martinez~Fiorenzano, Marzari, Masiero, Micela, Molinari, Molinaro, Munari, Murabito, Pagano, Pedani, Piotto, Rosenberg, Silvotti, \& Southworth}]{Damasso_2015}
Damasso, M., Biazzo, K., Bonomo, A.~S., {et~al.} 2015, \aap, 575, A111, \dodoi{10.1051/0004-6361/201425332}

\bibitem[{{Dawson} \& {Johnson}(2018)}]{2018ARA&A..56..175D}
{Dawson}, R.~I., \& {Johnson}, J.~A. 2018, \araa, 56, 175, \dodoi{10.1146/annurev-astro-081817-051853}

\bibitem[{{Eastman} {et~al.}(2013){Eastman}, {Gaudi}, \& {Agol}}]{2013PASP..125...83E}
{Eastman}, J., {Gaudi}, B.~S., \& {Agol}, E. 2013, \pasp, 125, 83, \dodoi{10.1086/669497}

\bibitem[{{Eastman} {et~al.}(2010){Eastman}, {Siverd}, \& {Gaudi}}]{2010PASP..122..935E}
{Eastman}, J., {Siverd}, R., \& {Gaudi}, B.~S. 2010, \pasp, 122, 935, \dodoi{10.1086/655938}

\bibitem[{{Edwards} {et~al.}(2019){Edwards}, {Mugnai}, {Tinetti}, {Pascale}, \& {Sarkar}}]{2019AJ....157..242E}
{Edwards}, B., {Mugnai}, L., {Tinetti}, G., {Pascale}, E., \& {Sarkar}, S. 2019, \aj, 157, 242, \dodoi{10.3847/1538-3881/ab1cb9}

\bibitem[{{Enoch} {et~al.}(2012){Enoch}, {Collier Cameron}, \& {Horne}}]{2012A&A...540A..99E}
{Enoch}, B., {Collier Cameron}, A., \& {Horne}, K. 2012, \aap, 540, A99, \dodoi{10.1051/0004-6361/201117317}

\bibitem[{{Espinoza} {et~al.}(2019){Espinoza}, {Kossakowski}, \& {Brahm}}]{2019MNRAS.490.2262E}
{Espinoza}, N., {Kossakowski}, D., \& {Brahm}, R. 2019, \mnras, 490, 2262, \dodoi{10.1093/mnras/stz2688}

\bibitem[{Essick \& Weinberg(2015)}]{Essick_2015}
Essick, R., \& Weinberg, N.~N. 2015, The Astrophysical Journal, 816, 18, \dodoi{10.3847/0004-637x/816/1/18}

\bibitem[{{Feroz} {et~al.}(2009){Feroz}, {Hobson}, \& {Bridges}}]{2009MNRAS.398.1601F}
{Feroz}, F., {Hobson}, M.~P., \& {Bridges}, M. 2009, \mnras, 398, 1601, \dodoi{10.1111/j.1365-2966.2009.14548.x}

\bibitem[{Feroz {et~al.}(2019)Feroz, Hobson, Cameron, \& Pettitt}]{Feroz_2019}
Feroz, F., Hobson, M.~P., Cameron, E., \& Pettitt, A.~N. 2019, The Open Journal of Astrophysics, 2, \dodoi{10.21105/astro.1306.2144}

\bibitem[{Ford \& Tremaine(2003)}]{Ford_2003}
Ford, E., \& Tremaine, S. 2003, Publications of the Astronomical Society of the Pacific, 115, 1171–1186, \dodoi{10.1086/377594}

\bibitem[{{Ford}(2006)}]{2006ApJ...642..505F}
{Ford}, E.~B. 2006, \apj, 642, 505, \dodoi{10.1086/500802}

\bibitem[{{Foreman-Mackey} {et~al.}(2013){Foreman-Mackey}, {Hogg}, {Lang}, \& {Goodman}}]{2013PASP..125..306F}
{Foreman-Mackey}, D., {Hogg}, D.~W., {Lang}, D., \& {Goodman}, J. 2013, \pasp, 125, 306, \dodoi{10.1086/670067}

\bibitem[{{Fressin} {et~al.}(2010){Fressin}, {Knutson}, {Charbonneau}, {O'Donovan}, {Burrows}, {Deming}, {Mandushev}, \& {Spiegel}}]{2010ApJ...711..374F}
{Fressin}, F., {Knutson}, H.~A., {Charbonneau}, D., {et~al.} 2010, \apj, 711, 374, \dodoi{10.1088/0004-637X/711/1/374}

\bibitem[{{Fulton} {et~al.}(2018){Fulton}, {Petigura}, {Blunt}, \& {Sinukoff}}]{2018PASP..130d4504F}
{Fulton}, B.~J., {Petigura}, E.~A., {Blunt}, S., \& {Sinukoff}, E. 2018, \pasp, 130, 044504, \dodoi{10.1088/1538-3873/aaaaa8}

\bibitem[{Fulton {et~al.}(2011)Fulton, Shporer, Winn, Holman, Pál, \& Gazak}]{Fulton_2011}
Fulton, B.~J., Shporer, A., Winn, J.~N., {et~al.} 2011, The Astronomical Journal, 142, 84, \dodoi{10.1088/0004-6256/142/3/84}

\bibitem[{{Gavrilov} \& {Zharkov}(1977)}]{1977Icar...32..443G}
{Gavrilov}, S.~V., \& {Zharkov}, V.~N. 1977, \icarus, 32, 443, \dodoi{10.1016/0019-1035(77)90015-X}

\bibitem[{{Gazak} {et~al.}(2012){Gazak}, {Johnson}, {Tonry}, {Dragomir}, {Eastman}, {Mann}, \& {Agol}}]{2012AdAst2012E..30G}
{Gazak}, J.~Z., {Johnson}, J.~A., {Tonry}, J., {et~al.} 2012, Advances in Astronomy, 2012, 697967, \dodoi{10.1155/2012/697967}

\bibitem[{Gibson {et~al.}(2009)Gibson, Pollacco, Simpson, Barros, Joshi, Todd, Keenan, Skillen, Benn, Christian, Hrudková, \& Steele}]{Gibson_2009}
Gibson, N.~P., Pollacco, D., Simpson, E.~K., {et~al.} 2009, The Astrophysical Journal, 700, 1078, \dodoi{10.1088/0004-637X/700/2/1078}

\bibitem[{Gillon {et~al.}(2012)Gillon, Triaud, Fortney, Demory, Jehin, Lendl, Magain, Kabath, Queloz, Alonso, Anderson, Collier~Cameron, Fumel, Hebb, Hellier, Lanotte, Maxted, Mowlavi, \& Smalley}]{Gillon_2012}
Gillon, M., Triaud, A. H. M.~J., Fortney, J.~J., {et~al.} 2012, \aap, 542, A4, \dodoi{10.1051/0004-6361/201218817}

\bibitem[{{Gim{\'e}nez} \& {Bastero}(1995)}]{1995Ap&SS.226...99G}
{Gim{\'e}nez}, A., \& {Bastero}, M. 1995, \apss, 226, 99, \dodoi{10.1007/BF00626903}

\bibitem[{{Goldreich} \& {Soter}(1966)}]{1966Icar....5..375G}
{Goldreich}, P., \& {Soter}, S. 1966, \icarus, 5, 375, \dodoi{10.1016/0019-1035(66)90051-0}

\bibitem[{{Goodman} \& {Weare}(2010)}]{2010CAMCS...5...65G}
{Goodman}, J., \& {Weare}, J. 2010, Communications in Applied Mathematics and Computational Science, 5, 65, \dodoi{10.2140/camcos.2010.5.65}

\bibitem[{{Gouda} {et~al.}(2005){Gouda}, {Yano}, {Kobayashi}, {Yamada}, {Tsujimoto}, {Nakajima}, {Suganuma}, {Matsuhara}, {Ueda}, \& {Jasmine Working Group}}]{2005tvnv.conf..455G}
{Gouda}, N., {Yano}, T., {Kobayashi}, Y., {et~al.} 2005, in IAU Colloq. 196: Transits of Venus: New Views of the Solar System and Galaxy, ed. D.~W. {Kurtz}, 455--468, \dodoi{10.1017/S1743921305001614}

\bibitem[{{Hagey} {et~al.}(2022){Hagey}, {Edwards}, \& {Boley}}]{2022AJ....164..220H}
{Hagey}, S.~R., {Edwards}, B., \& {Boley}, A.~C. 2022, \aj, 164, 220, \dodoi{10.3847/1538-3881/ac959a}

\bibitem[{{Hamer} \& {Schlaufman}(2019)}]{2019AJ....158..190H}
{Hamer}, J.~H., \& {Schlaufman}, K.~C. 2019, \aj, 158, 190, \dodoi{10.3847/1538-3881/ab3c56}

\bibitem[{Harre {et~al.}(2023)Harre, Smith, Barros, Boue, Csizmadia, Ehrenreich, Floren, Fortier, Maxted, Hooton, Akinsanmi, Serrano, Rosario, Demory, Jones, Laskar, Adibekyan, Alibert, Alonso, Anderson, Anglada, Asquier, Barczy, Barrado~y Navascues, Baumjohann, Beck, Beck, Benz, Billot, Biondi, Bonfanti, Bonfils, Brandeker, Broeg, Cabrera, Cessa, Charnoz, Collier~Cameron, Davies, Deleuil, Delrez, Demangeon, Erikson, Fossati, Fridlund, Gandolfi, Gillon, Güdel, Hellier, Heng, Hoyer, Isaak, Kiss, Lecavelier~des Etangs, Lendl, Lovis, Luntzer, Magrin, Nascimbeni, Olofsson, Ottensamer, Pagano, Palle, Persson, Peter, Piotto, Pollacco, Queloz, Ragazzoni, Rando, Rauer, Ribas, Ricker, Salmon, Santos, Scandariato, Seager, Segransan, Simon, Sousa, Steller, Szabo, Thomas, Udry, Ulmer, Van~Grootel, Walton, Wilson, Winn, \& Wohler}]{Harre_2023}
Harre, J.-V., Smith, A. M.~S., Barros, S. C.~C., {et~al.} 2023, \aap, 669, A124, \dodoi{10.1051/0004-6361/202244529}

\bibitem[{{Hebb} {et~al.}(2009){Hebb}, {Collier-Cameron}, {Loeillet}, {Pollacco}, {H{\'e}brard}, {Street}, {Bouchy}, {Stempels}, {Moutou}, {Simpson}, {Udry}, {Joshi}, {West}, {Skillen}, {Wilson}, {McDonald}, {Gibson}, {Aigrain}, {Anderson}, {Benn}, {Christian}, {Enoch}, {Haswell}, {Hellier}, {Horne}, {Irwin}, {Lister}, {Maxted}, {Mayor}, {Norton}, {Parley}, {Pont}, {Queloz}, {Smalley}, \& {Wheatley}}]{2009ApJ...693.1920H}
{Hebb}, L., {Collier-Cameron}, A., {Loeillet}, B., {et~al.} 2009, \apj, 693, 1920, \dodoi{10.1088/0004-637X/693/2/1920}

\bibitem[{{Hellier} {et~al.}(2009){Hellier}, {Anderson}, {Collier Cameron}, {Gillon}, {Hebb}, {Maxted}, {Queloz}, {Smalley}, {Triaud}, {West}, {Wilson}, {Bentley}, {Enoch}, {Horne}, {Irwin}, {Lister}, {Mayor}, {Parley}, {Pepe}, {Pollacco}, {Segransan}, {Udry}, \& {Wheatley}}]{2009Natur.460.1098H}
{Hellier}, C., {Anderson}, D.~R., {Collier Cameron}, A., {et~al.} 2009, \nat, 460, 1098, \dodoi{10.1038/nature08245}

\bibitem[{Heyl \& Gladman(2007)}]{Heyl_2007}
Heyl, J.~S., \& Gladman, B.~J. 2007, Monthly Notices of the Royal Astronomical Society, 377, 1511–1519, \dodoi{10.1111/j.1365-2966.2007.11697.x}

\bibitem[{Hogg \& Foreman-Mackey(2018)}]{Hogg_2018}
Hogg, D.~W., \& Foreman-Mackey, D. 2018, The Astrophysical Journal Supplement Series, 236, 11, \dodoi{10.3847/1538-4365/aab76e}

\bibitem[{{Holman} {et~al.}(2007){Holman}, {Winn}, {Latham}, {O'Donovan}, {Charbonneau}, {Torres}, {Sozzetti}, {Fernandez}, \& {Everett}}]{2007ApJ...664.1185H}
{Holman}, M.~J., {Winn}, J.~N., {Latham}, D.~W., {et~al.} 2007, \apj, 664, 1185, \dodoi{10.1086/519077}

\bibitem[{{Hoyer} {et~al.}(2016){Hoyer}, {Pall{\'e}}, {Dragomir}, \& {Murgas}}]{2016AJ....151..137H}
{Hoyer}, S., {Pall{\'e}}, E., {Dragomir}, D., \& {Murgas}, F. 2016, \aj, 151, 137, \dodoi{10.3847/0004-6256/151/6/137}

\bibitem[{{Huber} {et~al.}(2022){Huber}, {White}, {Metcalfe}, {Chontos}, {Fausnaugh}, {Ho}, {Van Eylen}, {Ball}, {Basu}, {Bedding}, {Benomar}, {Bossini}, {Breton}, {Buzasi}, {Campante}, {Chaplin}, {Christensen-Dalsgaard}, {Cunha}, {Deal}, {Garc{\'\i}a}, {Garc{\'\i}a Mu{\~n}oz}, {Gehan}, {Gonz{\'a}lez-Cuesta}, {Jiang}, {Kayhan}, {Kjeldsen}, {Lundkvist}, {Mathis}, {Mathur}, {Monteiro}, {Nsamba}, {Ong}, {Pak{\v{s}}tien{\.{e}}}, {Serenelli}, {Silva Aguirre}, {Stassun}, {Stello}, {Norgaard Stilling}, {Lykke Winther}, {Wu}, {Barclay}, {Daylan}, {G{\"u}nther}, {Hermes}, {Jenkins}, {Latham}, {Levine}, {Ricker}, {Seager}, {Shporer}, {Twicken}, {Vanderspek}, \& {Winn}}]{2022AJ....163...79H}
{Huber}, D., {White}, T.~R., {Metcalfe}, T.~S., {et~al.} 2022, \aj, 163, 79, \dodoi{10.3847/1538-3881/ac3000}

\bibitem[{Husnoo {et~al.}(2012)Husnoo, Pont, Mazeh, Fabrycky, Hébrard, Bouchy, \& Shporer}]{10.1111/j.1365-2966.2012.20839.x}
Husnoo, N., Pont, F., Mazeh, T., {et~al.} 2012, Monthly Notices of the Royal Astronomical Society, 422, 3151, \dodoi{10.1111/j.1365-2966.2012.20839.x}

\bibitem[{{Ivshina} \& {Winn}(2022)}]{2022ApJS..259...62I}
{Ivshina}, E.~S., \& {Winn}, J.~N. 2022, \apjs, 259, 62, \dodoi{10.3847/1538-4365/ac545b}

\bibitem[{{Jackson} {et~al.}(2008){Jackson}, {Greenberg}, \& {Barnes}}]{2008ApJ...681.1631J}
{Jackson}, B., {Greenberg}, R., \& {Barnes}, R. 2008, \apj, 681, 1631, \dodoi{10.1086/587641}

\bibitem[{{Jenkins} {et~al.}(2016){Jenkins}, {Twicken}, {McCauliff}, {Campbell}, {Sanderfer}, {Lung}, {Mansouri-Samani}, {Girouard}, {Tenenbaum}, {Klaus}, {Smith}, {Caldwell}, {Chacon}, {Henze}, {Heiges}, {Latham}, {Morgan}, {Swade}, {Rinehart}, \& {Vanderspek}}]{2016SPIE.9913E..3EJ}
{Jenkins}, J.~M., {Twicken}, J.~D., {McCauliff}, S., {et~al.} 2016, in Society of Photo-Optical Instrumentation Engineers (SPIE) Conference Series, Vol. 9913, Software and Cyberinfrastructure for Astronomy IV, ed. G.~{Chiozzi} \& J.~C. {Guzman}, 99133E, \dodoi{10.1117/12.2233418}

\bibitem[{{Jiang} {et~al.}(2016){Jiang}, {Lai}, {Savushkin}, {Mkrtichian}, {Antonyuk}, {Griv}, {Hsieh}, \& {Yeh}}]{2016AJ....151...17J}
{Jiang}, I.-G., {Lai}, C.-Y., {Savushkin}, A., {et~al.} 2016, \aj, 151, 17, \dodoi{10.3847/0004-6256/151/1/17}

\bibitem[{{Jiang} {et~al.}(2013){Jiang}, {Yeh}, {Thakur}, {Wu}, {Chien}, {Lin}, {Chen}, {Hu}, {Sun}, \& {Ji}}]{2013AJ....145...68J}
{Jiang}, I.-G., {Yeh}, L.-C., {Thakur}, P., {et~al.} 2013, \aj, 145, 68, \dodoi{10.1088/0004-6256/145/3/68}

\bibitem[{Jordan \& Bakos(2008)}]{Jordán_2008}
Jordan, A., \& Bakos, G.~A. 2008, The Astrophysical Journal, 685, 543, \dodoi{10.1086/590549}

\bibitem[{Kass \& Raftery(1995)}]{doi:10.1080/01621459.1995.10476572}
Kass, R.~E., \& Raftery, A.~E. 1995, Journal of the American Statistical Association, 90, 773, \dodoi{10.1080/01621459.1995.10476572}

\bibitem[{Kipping \& Bakos(2011)}]{Kipping_2011}
Kipping, D., \& Bakos, G. 2011, The Astrophysical Journal, 733, 36, \dodoi{10.1088/0004-637X/733/1/36}

\bibitem[{Kipping(2010)}]{Kipping_2010}
Kipping, D.~M. 2010, Monthly Notices of the Royal Astronomical Society, 408, 1758–1769, \dodoi{10.1111/j.1365-2966.2010.17242.x}

\bibitem[{{Knutson} {et~al.}(2014){Knutson}, {Fulton}, {Montet}, {Kao}, {Ngo}, {Howard}, {Crepp}, {Hinkley}, {Bakos}, {Batygin}, {Johnson}, {Morton}, \& {Muirhead}}]{2014ApJ...785..126K}
{Knutson}, H.~A., {Fulton}, B.~J., {Montet}, B.~T., {et~al.} 2014, \apj, 785, 126, \dodoi{10.1088/0004-637X/785/2/126}

\bibitem[{{Kokori} {et~al.}(2022){Kokori}, {Tsiaras}, {Edwards}, {Rocchetto}, {Tinetti}, {Bewersdorff}, {Jongen}, {Lekkas}, {Pantelidou}, {Poultourtzidis}, {W{\"u}nsche}, {Aggelis}, {Agnihotri}, {Arena}, {Bachschmidt}, {Bennett}, {Benni}, {Bernacki}, {Besson}, {Betti}, {Biagini}, {Brandebourg}, {Bretton}, {Brincat}, {Cal{\'o}}, {Campos}, {Casali}, {Ciantini}, {Crow}, {Dauchet}, {Dawes}, {Deldem}, {Deligeorgopoulos}, {Dymock}, {Eenm{\"a}e}, {Evans}, {Esseiva}, {Falco}, {Ferratfiat}, {Fowler}, {Futcher}, {Gaitan}, {Horta}, {Guerra}, {Hurter}, {Jones}, {Kang}, {Kiiskinen}, {Kim}, {Laloum}, {Lee}, {Lomoz}, {Lopresti}, {Mallonn}, {Mannucci}, {Marino}, {Mario}, {Marquette}, {Michelet}, {Miller}, {Mollier}, {Molina}, {Montigiani}, {Mortari}, {Morvan}, {Mugnai}, {Naponiello}, {Nastasi}, {Neito}, {Pace}, {Papadeas}, {Paschalis}, {Pereira}, {Perroud}, {Phillips}, {Pintr}, {Pioppa}, {Popowicz}, {Raetz}, {Regembal}, {Rickard}, {Roberts}, {Rousselot}, {Rubia}, {Savage}, {Sedita}, {Shave-Wall}, {Sioulas},
  {{\v{S}}koln{\'\i}k}, {Smith}, {St-Gelais}, {Stouraitis}, {Strikis}, {Thurston}, {Tomacelli}, {Tomatis}, {Trevan}, {Valeau}, {Vignes}, {Vora}, {Vra{\v{s}}{\v{t}}{\'a}k}, {Walter}, {Wenzel}, {Wright}, \& {Z{\'\i}bar}}]{2022ApJS..258...40K}
{Kokori}, A., {Tsiaras}, A., {Edwards}, B., {et~al.} 2022, \apjs, 258, 40, \dodoi{10.3847/1538-4365/ac3a10}

\bibitem[{{Konacki} {et~al.}(2003){Konacki}, {Torres}, {Sasselov}, \& {Jha}}]{2003ApJ...597.1076K}
{Konacki}, M., {Torres}, G., {Sasselov}, D.~D., \& {Jha}, S. 2003, \apj, 597, 1076, \dodoi{10.1086/378561}

\bibitem[{{Konacki} {et~al.}(2005){Konacki}, {Torres}, {Sasselov}, \& {Jha}}]{2005ApJ...624..372K}
---. 2005, \apj, 624, 372, \dodoi{10.1086/429127}

\bibitem[{{Kundurthy} {et~al.}(2013){Kundurthy}, {Barnes}, {Becker}, {Agol}, {Williams}, {Gorelick}, \& {Rose}}]{2013ApJ...770...36K}
{Kundurthy}, P., {Barnes}, R., {Becker}, A.~C., {et~al.} 2013, \apj, 770, 36, \dodoi{10.1088/0004-637X/770/1/36}

\bibitem[{{Levrard} {et~al.}(2007){Levrard}, {Correia}, {Chabrier}, {Baraffe}, {Selsis}, \& {Laskar}}]{2007A&A...462L...5L}
{Levrard}, B., {Correia}, A.~C.~M., {Chabrier}, G., {et~al.} 2007, \aap, 462, L5, \dodoi{10.1051/0004-6361:20066487}

\bibitem[{{Levrard} {et~al.}(2009){Levrard}, {Winisdoerffer}, \& {Chabrier}}]{2009ApJ...692L...9L}
{Levrard}, B., {Winisdoerffer}, C., \& {Chabrier}, G. 2009, \apjl, 692, L9, \dodoi{10.1088/0004-637X/692/1/L9}

\bibitem[{Liddle(2007)}]{10.1111/j.1745-3933.2007.00306.x}
Liddle, A.~R. 2007, Monthly Notices of the Royal Astronomical Society: Letters, 377, L74, \dodoi{10.1111/j.1745-3933.2007.00306.x}

\bibitem[{{Love}(1911)}]{1911spge.book.....L}
{Love}, A.~E.~H. 1911, {Some Problems of Geodynamics}

\bibitem[{{Maciejewski} {et~al.}(2021){Maciejewski}, {Fern{\'a}ndez}, {Aceituno}, {Ramos}, {Dimitrov}, {Donchev}, \& {Ohlert}}]{2021A&A...656A..88M}
{Maciejewski}, G., {Fern{\'a}ndez}, M., {Aceituno}, F., {et~al.} 2021, \aap, 656, A88, \dodoi{10.1051/0004-6361/202142424}

\bibitem[{{Maciejewski} {et~al.}(2013{\natexlab{a}}){Maciejewski}, {Puchalski}, {Saral}, {Derman}, {Kitze}, {Bukowiecki}, {Seeliger}, \& {Neuhaeuser}}]{2013IBVS.6082....1M}
{Maciejewski}, G., {Puchalski}, D., {Saral}, G., {et~al.} 2013{\natexlab{a}}, Information Bulletin on Variable Stars, 6082, 1

\bibitem[{{Maciejewski} {et~al.}(2013{\natexlab{b}}){Maciejewski}, {Dimitrov}, {Seeliger}, {Raetz}, {Bukowiecki}, {Kitze}, {Errmann}, {Nowak}, {Niedzielski}, {Popov}, {Marka}, {Go{\'z}dziewski}, {Neuh{\"a}user}, {Ohlert}, {Hinse}, {Lee}, {Lee}, {Yoon}, {Berndt}, {Gilbert}, {Ginski}, {Hohle}, {Mugrauer}, {R{\"o}ll}, {Schmidt}, {Tetzlaff}, {Mancini}, {Southworth}, {Dall'Ora}, {Ciceri}, {Zambelli}, {Corfini}, {Takahashi}, {Tachihara}, {Benk{\H{o}}}, {S{\'a}rneczky}, {Szabo}, {Varga}, {Va{\v{n}}ko}, {Joshi}, \& {Chen}}]{2013A&A...551A.108M}
{Maciejewski}, G., {Dimitrov}, D., {Seeliger}, M., {et~al.} 2013{\natexlab{b}}, \aap, 551, A108, \dodoi{10.1051/0004-6361/201220739}

\bibitem[{{Maciejewski} {et~al.}(2016){Maciejewski}, {Dimitrov}, {Fern{\'a}ndez}, {Sota}, {Nowak}, {Ohlert}, {Nikolov}, {Bukowiecki}, {Hinse}, {Pall{\'e}}, {Tingley}, {Kjurkchieva}, {Lee}, \& {Lee}}]{2016A&A...588L...6M}
{Maciejewski}, G., {Dimitrov}, D., {Fern{\'a}ndez}, M., {et~al.} 2016, \aap, 588, L6, \dodoi{10.1051/0004-6361/201628312}

\bibitem[{{Maciejewski} {et~al.}(2018){Maciejewski}, {Fern{\'a}ndez}, {Aceituno}, {Mart{\'\i}n-Ruiz}, {Ohlert}, {Dimitrov}, {Szyszka}, {von Essen}, {Mugrauer}, {Bischoff}, {Michel}, {Mallonn}, {Stangret}, \& {Mo{\'z}dzierski}}]{2018AcA....68..371M}
{Maciejewski}, G., {Fern{\'a}ndez}, M., {Aceituno}, F., {et~al.} 2018, \actaa, 68, 371, \dodoi{10.32023/0001-5237/68.4.4}

\bibitem[{Mandel \& Agol(2002)}]{Mandel_2002}
Mandel, K., \& Agol, E. 2002, The Astrophysical Journal, 580, L171–L175, \dodoi{10.1086/345520}

\bibitem[{{Marcy} {et~al.}(2005){Marcy}, {Butler}, {Fischer}, {Vogt}, {Wright}, {Tinney}, \& {Jones}}]{2005PThPS.158...24M}
{Marcy}, G., {Butler}, R.~P., {Fischer}, D., {et~al.} 2005, Progress of Theoretical Physics Supplement, 158, 24, \dodoi{10.1143/PTPS.158.24}

\bibitem[{Mardling(2007)}]{10.1111/j.1365-2966.2007.12500.x}
Mardling, R.~A. 2007, Monthly Notices of the Royal Astronomical Society, 382, 1768, \dodoi{10.1111/j.1365-2966.2007.12500.x}

\bibitem[{{Matsumura} {et~al.}(2010){Matsumura}, {Peale}, \& {Rasio}}]{2010ApJ...725.1995M}
{Matsumura}, S., {Peale}, S.~J., \& {Rasio}, F.~A. 2010, \apj, 725, 1995, \dodoi{10.1088/0004-637X/725/2/1995}

\bibitem[{{Mayor} \& {Queloz}(1995)}]{1995Natur.378..355M}
{Mayor}, M., \& {Queloz}, D. 1995, \nat, 378, 355, \dodoi{10.1038/378355a0}

\bibitem[{{Meibom} {et~al.}(2015){Meibom}, {Barnes}, {Platais}, {Gilliland}, {Latham}, \& {Mathieu}}]{2015Natur.517..589M}
{Meibom}, S., {Barnes}, S.~A., {Platais}, I., {et~al.} 2015, \nat, 517, 589, \dodoi{10.1038/nature14118}

\bibitem[{{Miller-Ricci} {et~al.}(2008){Miller-Ricci}, {Rowe}, {Sasselov}, {Matthews}, {Guenther}, {Kuschnig}, {Moffat}, {Rucinski}, {Walker}, \& {Weiss}}]{2008ApJ...682..586M}
{Miller-Ricci}, E., {Rowe}, J.~F., {Sasselov}, D., {et~al.} 2008, \apj, 682, 586, \dodoi{10.1086/587446}

\bibitem[{{Miralda-Escud{\'e}}(2002)}]{2002ApJ...564.1019M}
{Miralda-Escud{\'e}}, J. 2002, \apj, 564, 1019, \dodoi{10.1086/324279}

\bibitem[{Mislis \& Schmitt(2009)}]{Mislis_2009}
Mislis, D., \& Schmitt, J. H. M.~M. 2009, \aap, 500, L45–L49, \dodoi{10.1051/0004-6361/200811424}

\bibitem[{{Mislis, D.} {et~al.}(2010){Mislis, D.}, {Schroter, S.}, {Schmitt, J. H. M. M.}, {Cordes, O.}, \& {Reif, K.}}]{refId0}
{Mislis, D.}, {Schroter, S.}, {Schmitt, J. H. M. M.}, {Cordes, O.}, \& {Reif, K.} 2010, \aap, 510, A107, \dodoi{10.1051/0004-6361/200912910}

\bibitem[{{Montalto} {et~al.}(2012){Montalto}, {Gregorio}, {Bou{\'e}}, {Mortier}, {Boisse}, {Oshagh}, {Maturi}, {Figueira}, {Sousa}, \& {Santos}}]{2012MNRAS.427.2757M}
{Montalto}, M., {Gregorio}, J., {Bou{\'e}}, G., {et~al.} 2012, \mnras, 427, 2757, \dodoi{10.1111/j.1365-2966.2012.21926.x}

\bibitem[{{Moutou} {et~al.}(2005){Moutou}, {Mayor}, {Bouchy}, {Lovis}, \& {Pepe}}]{2005A&A...439..367M}
{Moutou}, C., {Mayor}, M., {Bouchy}, F., {Lovis}, C., \& {Pepe}, F. 2005, \aap, 439, 367, \dodoi{10.1051/0004-6361:20052826}

\bibitem[{{Mo{\v{c}}nik} {et~al.}(2016){Mo{\v{c}}nik}, {Clark}, {Anderson}, {Hellier}, \& {Brown}}]{2016AJ....151..150M}
{Mo{\v{c}}nik}, T., {Clark}, B.~J.~M., {Anderson}, D.~R., {Hellier}, C., \& {Brown}, D.~J.~A. 2016, \aj, 151, 150, \dodoi{10.3847/0004-6256/151/6/150}

\bibitem[{{Murray} \& {Dermott}(1999)}]{1999ssd..book.....M}
{Murray}, C.~D., \& {Dermott}, S.~F. 1999, {Solar System Dynamics}, \dodoi{10.1017/CBO9781139174817}

\bibitem[{{Naoz} {et~al.}(2011){Naoz}, {Farr}, {Lithwick}, {Rasio}, \& {Teyssandier}}]{2011Natur.473..187N}
{Naoz}, S., {Farr}, W.~M., {Lithwick}, Y., {Rasio}, F.~A., \& {Teyssandier}, J. 2011, \nat, 473, 187, \dodoi{10.1038/nature10076}

\bibitem[{{O'Donovan} {et~al.}(2006){O'Donovan}, {Charbonneau}, {Mandushev}, {Dunham}, {Latham}, {Torres}, {Sozzetti}, {Brown}, {Trauger}, {Belmonte}, {Rabus}, {Almenara}, {Alonso}, {Deeg}, {Esquerdo}, {Falco}, {Hillenbrand}, {Roussanova}, {Stefanik}, \& {Winn}}]{2006ApJ...651L..61O}
{O'Donovan}, F.~T., {Charbonneau}, D., {Mandushev}, G., {et~al.} 2006, \apjl, 651, L61, \dodoi{10.1086/509123}

\bibitem[{{Ogilvie}(2014)}]{2014ARA&A..52..171O}
{Ogilvie}, G.~I. 2014, \araa, 52, 171, \dodoi{10.1146/annurev-astro-081913-035941}

\bibitem[{{{\"O}zt{\"u}rk} \& {Erdem}(2019)}]{2019MNRAS.486.2290O}
{{\"O}zt{\"u}rk}, O., \& {Erdem}, A. 2019, \mnras, 486, 2290, \dodoi{10.1093/mnras/stz747}

\bibitem[{Patil {et~al.}(2010)Patil, Huard, \& Fonnesbeck}]{JSSv035i04}
Patil, A., Huard, D., \& Fonnesbeck, C.~J. 2010, Journal of Statistical Software, 35, 1–81, \dodoi{10.18637/jss.v035.i04}

\bibitem[{{Patra} {et~al.}(2017){Patra}, {Winn}, {Holman}, {Yu}, {Deming}, \& {Dai}}]{2017AJ....154....4P}
{Patra}, K.~C., {Winn}, J.~N., {Holman}, M.~J., {et~al.} 2017, \aj, 154, 4, \dodoi{10.3847/1538-3881/aa6d75}

\bibitem[{{Patra} {et~al.}(2020){Patra}, {Winn}, {Holman}, {Gillon}, {Burdanov}, {Jehin}, {Delrez}, {Pozuelos}, {Barkaoui}, {Benkhaldoun}, {Narita}, {Fukui}, {Kusakabe}, {Kawauchi}, {Terada}, {Bouma}, {Weinberg}, \& {Broome}}]{2020AJ....159..150P}
---. 2020, \aj, 159, 150, \dodoi{10.3847/1538-3881/ab7374}

\bibitem[{{Penev} {et~al.}(2018){Penev}, {Bouma}, {Winn}, \& {Hartman}}]{2018AJ....155..165P}
{Penev}, K., {Bouma}, L.~G., {Winn}, J.~N., \& {Hartman}, J.~D. 2018, \aj, 155, 165, \dodoi{10.3847/1538-3881/aaaf71}

\bibitem[{Poddaný {et~al.}(2010)Poddaný, Brát, \& Pejcha}]{Poddan__2010}
Poddaný, S., Brát, L., \& Pejcha, O. 2010, New Astronomy, 15, 297–301, \dodoi{10.1016/j.newast.2009.09.001}

\bibitem[{{Pont} {et~al.}(2004){Pont}, {Bouchy}, {Queloz}, {Santos}, {Melo}, {Mayor}, \& {Udry}}]{2004A&A...426L..15P}
{Pont}, F., {Bouchy}, F., {Queloz}, D., {et~al.} 2004, \aap, 426, L15, \dodoi{10.1051/0004-6361:200400066}

\bibitem[{{Pravdo} {et~al.}(2004){Pravdo}, {Shaklan}, {Henry}, \& {Benedict}}]{2004ApJ...617.1323P}
{Pravdo}, S.~H., {Shaklan}, S.~B., {Henry}, T., \& {Benedict}, G.~F. 2004, \apj, 617, 1323, \dodoi{10.1086/425653}

\bibitem[{{Rabus} {et~al.}(2009){Rabus}, {Deeg}, {Alonso}, {Belmonte}, \& {Almenara}}]{2009A&A...508.1011R}
{Rabus}, M., {Deeg}, H.~J., {Alonso}, R., {Belmonte}, J.~A., \& {Almenara}, J.~M. 2009, \aap, 508, 1011, \dodoi{10.1051/0004-6361/200912252}

\bibitem[{{Raetz} {et~al.}(2009){Raetz}, {Mugrauer}, {Schmidt}, {Roell}, {Eisenbeiss}, {Hohle}, {Tetzlaff}, {Va{\v{n}}ko}, {Seifahrt}, {Broeg}, {Koppenhoefer}, \& {Neuh{\"a}user}}]{2009AN....330..475R}
{Raetz}, S., {Mugrauer}, M., {Schmidt}, T.~O.~B., {et~al.} 2009, Astronomische Nachrichten, 330, 475, \dodoi{10.1002/asna.200811200}

\bibitem[{Raetz {et~al.}(2014)Raetz, Maciejewski, Ginski, Mugrauer, Berndt, Eisenbeiss, Adam, Raetz, Roell, Seeliger, Marka, Vaňko, Bukowiecki, Errmann, Kitze, Ohlert, Pribulla, Schmidt, Sebastian, Puchalski, Tetzlaff, Hohle, Schmidt, \& Neuhäuser}]{Raetz_2014}
Raetz, S., Maciejewski, G., Ginski, C., {et~al.} 2014, Monthly Notices of the Royal Astronomical Society, 444, 1351–1368, \dodoi{10.1093/mnras/stu1505}

\bibitem[{{Ragozzine} \& {Wolf}(2009)}]{2009ApJ...698.1778R}
{Ragozzine}, D., \& {Wolf}, A.~S. 2009, \apj, 698, 1778, \dodoi{10.1088/0004-637X/698/2/1778}

\bibitem[{{Rasio} {et~al.}(1996){Rasio}, {Tout}, {Lubow}, \& {Livio}}]{1996ApJ...470.1187R}
{Rasio}, F.~A., {Tout}, C.~A., {Lubow}, S.~H., \& {Livio}, M. 1996, \apj, 470, 1187, \dodoi{10.1086/177941}

\bibitem[{{Ricker} {et~al.}(2014){Ricker}, {Winn}, {Vanderspek}, {Latham}, \& {Bakos}}]{2014SPIE.9143E..20R}
{Ricker}, G.~R., {Winn}, J.~N., {Vanderspek}, R., {Latham}, D.~W., \& {Bakos}, G.~A. 2014, in Society of Photo-Optical Instrumentation Engineers (SPIE) Conference Series, Vol. 9143, Space Telescopes and Instrumentation 2014: Optical, Infrared, and Millimeter Wave, ed. J.~{Oschmann}, Jacobus~M., M.~{Clampin}, G.~G. {Fazio}, \& H.~A. {MacEwen}, 914320, \dodoi{10.1117/12.2063489}

\bibitem[{Sanchis-Ojeda {et~al.}(2013)Sanchis-Ojeda, Winn, Marcy, Howard, Isaacson, Johnson, Torres, Albrecht, Campante, Chaplin, Davies, Lund, Carter, Dawson, Buchhave, Everett, Fischer, Geary, Gilliland, Horch, Howell, \& Latham}]{Sanchis-Ojeda_2013}
Sanchis-Ojeda, R., Winn, J.~N., Marcy, G.~W., {et~al.} 2013, The Astrophysical Journal, 775, 54, \dodoi{10.1088/0004-637X/775/1/54}

\bibitem[{{Sariya} {et~al.}(2021){Sariya}, {Jiang}, {Su}, {Yeh}, {Chang}, {Moskvin}, {Shlyapnikov}, {Ignatov}, {Mkrtichian}, {Griv}, {Mannaday}, {Thakur}, {Sahu}, {Chand}, {Bisht}, {Sun}, \& {Ji}}]{2021RAA....21...97S}
{Sariya}, D.~P., {Jiang}, I.-G., {Su}, L.-H., {et~al.} 2021, Research in Astronomy and Astrophysics, 21, 097, \dodoi{10.1088/1674-4527/21/4/97}

\bibitem[{{Sasselov}(2003)}]{2003ApJ...596.1327S}
{Sasselov}, D.~D. 2003, \apj, 596, 1327, \dodoi{10.1086/378145}

\bibitem[{Schröter {et~al.}(2012)Schröter, Schmitt, \& Müller}]{Schr_ter_2012}
Schröter, S., Schmitt, J. H. M.~M., \& Müller, H.~M. 2012, \aap, 539, A97, \dodoi{10.1051/0004-6361/201118536}

\bibitem[{{Schwarz}(1978)}]{1978AnSta...6..461S}
{Schwarz}, G. 1978, Annals of Statistics, 6, 461

\bibitem[{{Shan} {et~al.}(2023){Shan}, {Yang}, {Lu}, {Wei}, {Tian}, {Zhang}, {Guo}, {Cui}, {Yang}, {Zhang}, \& {Liu}}]{2023ApJS..264...37S}
{Shan}, S.-S., {Yang}, F., {Lu}, Y.-J., {et~al.} 2023, \apjs, 264, 37, \dodoi{10.3847/1538-4365/aca65f}

\bibitem[{{Smith} {et~al.}(2012){Smith}, {Stumpe}, {Van Cleve}, {Jenkins}, {Barclay}, {Fanelli}, {Girouard}, {Kolodziejczak}, {McCauliff}, {Morris}, \& {Twicken}}]{2012PASP..124.1000S}
{Smith}, J.~C., {Stumpe}, M.~C., {Van Cleve}, J.~E., {et~al.} 2012, \pasp, 124, 1000, \dodoi{10.1086/667697}

\bibitem[{{Sokov} {et~al.}(2018){Sokov}, {Sokova}, {Dyachenko}, {Rastegaev}, {Burdanov}, {Rusov}, {Benni}, {Shadick}, {Hentunen}, {Salisbury}, {Esseiva}, {Garlitz}, {Bretton}, {Ogmen}, {Karavaev}, {Ayiomamitis}, {Mazurenko}, {Alonso}, \& {Velichko}}]{2018MNRAS.480..291S}
{Sokov}, E.~N., {Sokova}, I.~A., {Dyachenko}, V.~V., {et~al.} 2018, \mnras, 480, 291, \dodoi{10.1093/mnras/sty1615}

\bibitem[{{Southworth} {et~al.}(2016){Southworth}, {Tregloan-Reed}, {Andersen}, {Calchi Novati}, {Ciceri}, {Colque}, {D'Ago}, {Dominik}, {Evans}, {Gu}, {Herrera-Cordova}, {Hinse}, {J{\o}rgensen}, {Juncher}, {Kuffmeier}, {Mancini}, {Peixinho}, {Popovas}, {Rabus}, {Skottfelt}, {Tronsgaard}, {Unda-Sanzana}, {Wang}, {Wertz}, {Alsubai}, {Andersen}, {Bozza}, {Bramich}, {Burgdorf}, {Damerdji}, {Diehl}, {Elyiv}, {Figuera Jaimes}, {Haugb{\o}lle}, {Hundertmark}, {Kains}, {Kerins}, {Korhonen}, {Liebig}, {Mathiasen}, {Penny}, {Rahvar}, {Scarpetta}, {Schmidt}, {Snodgrass}, {Starkey}, {Surdej}, {Vilela}, {von Essen}, \& {Wang}}]{2016MNRAS.457.4205S}
{Southworth}, J., {Tregloan-Reed}, J., {Andersen}, M.~I., {et~al.} 2016, \mnras, 457, 4205, \dodoi{10.1093/mnras/stw279}

\bibitem[{{Southworth} {et~al.}(2019){Southworth}, {Dominik}, {J{\o}rgensen}, {Andersen}, {Bozza}, {Burgdorf}, {D'Ago}, {Dib}, {Figuera Jaimes}, {Fujii}, {Gill}, {Haikala}, {Hinse}, {Hundertmark}, {Khalouei}, {Korhonen}, {Longa-Pe{\~n}a}, {Mancini}, {Peixinho}, {Rabus}, {Rahvar}, {Sajadian}, {Skottfelt}, {Snodgrass}, {Spyratos}, {Tregloan-Reed}, {Unda-Sanzana}, \& {von Essen}}]{2019MNRAS.490.4230S}
{Southworth}, J., {Dominik}, M., {J{\o}rgensen}, U.~G., {et~al.} 2019, \mnras, 490, 4230, \dodoi{10.1093/mnras/stz2602}

\bibitem[{Sozzetti {et~al.}(2007)Sozzetti, Torres, Charbonneau, Latham, Holman, Winn, Laird, \& O’Donovan}]{Sozzetti_2007}
Sozzetti, A., Torres, G., Charbonneau, D., {et~al.} 2007, The Astrophysical Journal, 664, 1190–1198, \dodoi{10.1086/519214}

\bibitem[{{Sozzetti} {et~al.}(2009){Sozzetti}, {Torres}, {Charbonneau}, {Winn}, {Korzennik}, {Holman}, {Latham}, {Laird}, {Fernandez}, {O'Donovan}, {Mandushev}, {Dunham}, {Everett}, {Esquerdo}, {Rabus}, {Belmonte}, {Deeg}, {Brown}, {Hidas}, \& {Baliber}}]{2009ApJ...691.1145S}
{Sozzetti}, A., {Torres}, G., {Charbonneau}, D., {et~al.} 2009, \apj, 691, 1145, \dodoi{10.1088/0004-637X/691/2/1145}

\bibitem[{{Stassun} {et~al.}(2018){Stassun}, {Oelkers}, {Pepper}, {Paegert}, {De Lee}, {Torres}, {Latham}, {Charpinet}, {Dressing}, {Huber}, {Kane}, {L{\'e}pine}, {Mann}, {Muirhead}, {Rojas-Ayala}, {Silvotti}, {Fleming}, {Levine}, \& {Plavchan}}]{2018AJ....156..102S}
{Stassun}, K.~G., {Oelkers}, R.~J., {Pepper}, J., {et~al.} 2018, \aj, 156, 102, \dodoi{10.3847/1538-3881/aad050}

\bibitem[{{Stefansson} {et~al.}(2017){Stefansson}, {Mahadevan}, {Hebb}, {Wisniewski}, {Huehnerhoff}, {Morris}, {Halverson}, {Zhao}, {Wright}, {O'rourke}, {Knutson}, {Hawley}, {Kanodia}, {Li}, {Hagen}, {Liu}, {Beatty}, {Bender}, {Robertson}, {Dembicky}, {Gray}, {Ketzeback}, {McMillan}, \& {Rudyk}}]{2017ApJ...848....9S}
{Stefansson}, G., {Mahadevan}, S., {Hebb}, L., {et~al.} 2017, \apj, 848, 9, \dodoi{10.3847/1538-4357/aa88aa}

\bibitem[{{Sterken}(2005)}]{2005ASPC..335....3S}
{Sterken}, C. 2005, in Astronomical Society of the Pacific Conference Series, Vol. 335, The Light-Time Effect in Astrophysics: Causes and cures of the O-C diagram, ed. C.~{Sterken}, 3

\bibitem[{{Stumpe} {et~al.}(2014){Stumpe}, {Smith}, {Catanzarite}, {Van Cleve}, {Jenkins}, {Twicken}, \& {Girouard}}]{2014PASP..126..100S}
{Stumpe}, M.~C., {Smith}, J.~C., {Catanzarite}, J.~H., {et~al.} 2014, \pasp, 126, 100, \dodoi{10.1086/674989}

\bibitem[{{Stumpe} {et~al.}(2012){Stumpe}, {Smith}, {Van Cleve}, {Twicken}, {Barclay}, {Fanelli}, {Girouard}, {Jenkins}, {Kolodziejczak}, {McCauliff}, \& {Morris}}]{2012PASP..124..985S}
{Stumpe}, M.~C., {Smith}, J.~C., {Van Cleve}, J.~E., {et~al.} 2012, \pasp, 124, 985, \dodoi{10.1086/667698}

\bibitem[{Su {et~al.}(2021)Su, Jiang, Sariya, Lee, Yeh, Mannaday, Thakur, Sahu, Chand, Shlyapnikov, Moskvin, Ignatov, Mkrtichian, \& Griv}]{Su_2021}
Su, L.-H., Jiang, I.-G., Sariya, D.~P., {et~al.} 2021, The Astronomical Journal, 161, 108, \dodoi{10.3847/1538-3881/abd4d8}

\bibitem[{Tsapras {et~al.}(2003)Tsapras, Horne, Kane, \& Carson}]{10.1046/j.1365-8711.2003.06720.x}
Tsapras, Y., Horne, K., Kane, S., \& Carson, R. 2003, Monthly Notices of the Royal Astronomical Society, 343, 1131, \dodoi{10.1046/j.1365-8711.2003.06720.x}

\bibitem[{Turner {et~al.}(2022)Turner, Flagg, Ridden-Harper, \& Jayawardhana}]{Turner_2022}
Turner, J.~D., Flagg, L., Ridden-Harper, A., \& Jayawardhana, R. 2022, The Astronomical Journal, 163, 281, \dodoi{10.3847/1538-3881/ac686f}

\bibitem[{{Turner} {et~al.}(2021){Turner}, {Ridden-Harper}, \& {Jayawardhana}}]{2021AJ....161...72T}
{Turner}, J.~D., {Ridden-Harper}, A., \& {Jayawardhana}, R. 2021, \aj, 161, 72, \dodoi{10.3847/1538-3881/abd178}

\bibitem[{{Vissapragada} {et~al.}(2022){Vissapragada}, {Chontos}, {Greklek-McKeon}, {Knutson}, {Dai}, {P{\'e}rez Gonz{\'a}lez}, {Grunblatt}, {Huber}, \& {Saunders}}]{2022ApJ...941L..31V}
{Vissapragada}, S., {Chontos}, A., {Greklek-McKeon}, M., {et~al.} 2022, \apjl, 941, L31, \dodoi{10.3847/2041-8213/aca47e}

\bibitem[{{Wahl} {et~al.}(2016){Wahl}, {Hubbard}, \& {Militzer}}]{2016ApJ...831...14W}
{Wahl}, S.~M., {Hubbard}, W.~B., \& {Militzer}, B. 2016, \apj, 831, 14, \dodoi{10.3847/0004-637X/831/1/14}

\bibitem[{{Watson} \& {Dhillon}(2004)}]{2004MNRAS.351..110W}
{Watson}, C.~A., \& {Dhillon}, V.~S. 2004, \mnras, 351, 110, \dodoi{10.1111/j.1365-2966.2004.07763.x}

\bibitem[{{Watson} \& {Marsh}(2010)}]{2010MNRAS.405.2037W}
{Watson}, C.~A., \& {Marsh}, T.~R. 2010, \mnras, 405, 2037, \dodoi{10.1111/j.1365-2966.2010.16602.x}

\bibitem[{Wilkins {et~al.}(2017)Wilkins, Delrez, Barker, Deming, Hamilton, Gillon, \& Jehin}]{Wilkins_2017}
Wilkins, A.~N., Delrez, L., Barker, A.~J., {et~al.} 2017, The Astrophysical Journal Letters, 836, L24, \dodoi{10.3847/2041-8213/aa5d9f}

\bibitem[{{Wong} {et~al.}(2022){Wong}, {Shporer}, {Vissapragada}, {Greklek-McKeon}, {Knutson}, {Winn}, \& {Benneke}}]{2022AJ....163..175W}
{Wong}, I., {Shporer}, A., {Vissapragada}, S., {et~al.} 2022, \aj, 163, 175, \dodoi{10.3847/1538-3881/ac5680}

\bibitem[{{Yal{\c{c}}{\i}nkaya} {et~al.}(2024){Yal{\c{c}}{\i}nkaya}, {Esmer}, {Ba{\c{s}}t{\"u}rk}, {Muhaymin}, {Kutluay}, {Silistre}, {Akar}, {Southworth}, {Mancini}, {Davoudi}, {Karamanl{\i}}, {Tezcan}, {Demir}, {Y{\i}lmaz}, {G{\"u}lero{\u{g}}lu}, {Tekin}, {Ta{\c{s}}k{\i}n}, {Alada{\u{g}}}, {Sertkan}, {Kurt}, {Fi{\c{s}}ek}, {Kaptan}, {Ali{\c{s}}}, {Aksaker}, {Yelkenci}, {Tezcan}, {Kaya}, {O{\u{g}}lakkaya}, {Ayd{\i}n}, \& {Ye{\c{s}}ilyaprak}}]{2024MNRAS.tmp..965Y}
{Yal{\c{c}}{\i}nkaya}, S., {Esmer}, E.~M., {Ba{\c{s}}t{\"u}rk}, {\"O}., {et~al.} 2024, \mnras, \dodoi{10.1093/mnras/stae854}

\bibitem[{{Yee} {et~al.}(2020){Yee}, {Winn}, {Knutson}, {Patra}, {Vissapragada}, {Zhang}, {Holman}, {Shporer}, \& {Wright}}]{2020ApJ...888L...5Y}
{Yee}, S.~W., {Winn}, J.~N., {Knutson}, H.~A., {et~al.} 2020, \apjl, 888, L5, \dodoi{10.3847/2041-8213/ab5c16}

\bibitem[{Yeh {et~al.}(2024)Yeh, Jiang, \& A-thano}]{Yeh_2024}
Yeh, L.-C., Jiang, I.-G., \& A-thano, N. 2024, New Astronomy, 106, 102130, \dodoi{10.1016/j.newast.2023.102130}

\bibitem[{{Zahn}(1977)}]{1977A&A....57..383Z}
{Zahn}, J.~P. 1977, \aap, 57, 383

\bibitem[{{Zechmeister} \& {K{\"u}rster}(2009)}]{2009A&A...496..577Z}
{Zechmeister}, M., \& {K{\"u}rster}, M. 2009, \aap, 496, 577, \dodoi{10.1051/0004-6361:200811296}

\bibitem[{{Zharkov} \& {Trubitsyn}(1978)}]{1978ppi..book.....Z}
{Zharkov}, V.~N., \& {Trubitsyn}, V.~P. 1978, {Physics of planetary interiors}

\end{thebibliography}
\bibliographystyle{aasjournal}

%% This command is needed to show the entire author+affiliation list when
%% the collaboration and author truncation commands are used.  It has to
%% go at the end of the manuscript.
%\allauthors

%% Include this line if you are using the \added, \replaced, \deleted
%% commands to see a summary list of all changes at the end of the article.
%\listofchanges

\end{document}